\documentclass[bibyear]{aa} 

\usepackage{lscape}
\usepackage{float}
\usepackage{graphicx}
\usepackage{txfonts}
\begin{document} 

\title{Dense cores in the Seahorse infrared dark cloud: physical properties from modified blackbody fits to the far-infrared--submillimetre spectral energy distributions
\thanks{Based on observations with the Atacama Pathfinder EXperiment (APEX) telescope under programmes {\tt 083.F-9302(A)} and {\tt 089.F-9310(A)}. APEX is a collaboration between the Max-Planck-Institut f{\"u}r Radioastronomie, the European Southern Observatory, and the Onsala Space Observatory.}}

   \author{O.~Miettinen}

   \institute{Academy of Finland, Hakaniemenranta 6, P.O. Box 131, FI-00531 Helsinki, Finland \\ \email{oskari.miettinen@aka.fi}}

   \date{Received ; accepted}

\authorrunning{Miettinen}
\titlerunning{SEDs of dense cores in the Seahorse IRDC}

\abstract{Infrared dark clouds (IRDCs) can be the birth sites of high-mass stars, and hence determining the physical properties of dense cores in IRDCs is useful to constrain the initial conditions and theoretical models of high-mass star formation.}{We aim to determine the physical properties of dense cores in the filamentary Seahorse IRDC G304.74+01.32.}{We used data from the \textit{Wide-field Infrared Survey Explorer} (\textit{WISE}), \textit{Infrared Astronomical Satellite} (\textit{IRAS}), and \textit{Herschel} in conjuction with our previous 350~$\mu$m and 870~$\mu$m observations with the Submillimetre APEX Bolometer Camera (SABOCA) and Large APEX BOlometer CAmera (LABOCA), and constructed the far-IR to submillimetre spectral energy distributions (SEDs) of the cores. The SEDs were fitted using single or two-temperature modified blackbody emission curves to derive the dust temperatures, masses, and luminosities of the cores.}{For the 12 analysed cores, which include two IR dark cores (no \textit{WISE} counterpart), nine IR bright cores, and one \ion{H}{ii} region, the mean dust temperature of the cold (warm) component, the mass, luminosity, H$_2$ number density, and surface density were derived to be $13.3\pm1.4$~K ($47.0\pm5.0$~K), $113\pm29$~M$_{\sun}$, $192\pm94$~L$_{\sun}$, $(4.3\pm1.2)\times10^5$~cm$^{-3}$, and $0.77\pm0.19$~g~cm$^{-3}$, respectively. The \ion{H}{ii} region IRAS~13039-6108a was found to be the most luminous source in our sample ($(1.1\pm0.4)\times10^3$~L$_{\sun}$). All the cores were found to be gravitationally bound (i.e. the virial parameter $\alpha_{\rm vir}<2$). Two out of the nine analysed IR bright cores (22\%) were found to follow an accretion luminosity track under the assumptions that the mass accretion rate is $10^{-5}$~M$_{\sun}$~yr$^{-1}$, the stellar mass is 10\% of the parent core mass, and the radius of the central star is $5$~R$_{\sun}$. Most of the remaing ten cores were found to lie within 1~dex below this accretion luminosity track. Seven out of 12 of the analysed cores (58\%) were found to lie above the mass-radius thresholds of high-mass star formation proposed in the literature. The surface densities of $\Sigma>0.4$~g~cm$^{-3}$ derived for these seven cores also exceed the corresponding threshold for high-mass star formation. Five of the analysed cores (42\%) show evidence of fragmentation into two components in the SABOCA 350~$\mu$m image.} {In addition to the \ion{H}{ii} region source IRAS~13039-6108a, some of the other cores in Seahorse also appear to be capable of giving birth to high-mass stars. The 22~$\mu$m dark core SMM~9 is likely to be the youngest source in our sample that has the potential to form a high-mass star ($96\pm23$~M$_{\sun}$ within a radius of $\sim0.1$~pc). The dense core population in the Seahorse IRDC has comparable average properties to the cores in the well-studied Snake IRDC G11.11-0.12 (e.g. $T_{\rm dust}$ and $L$ agree within a factor of $\sim1.8$); furthermore, the Seahorse, which lies $\sim60$~pc above the Galactic plane, appears to be a smaller (e.g. three times shorter in projection, $\sim100$ times less massive) version of the Snake. The Seahorse core fragmentation mechanisms appear to be heterogenous, including cases of both thermal and non-thermal Jeans instability. High-resolution follow-up studies are required to address the fragmented cores' genuine potential of forming high-mass stars.}


   \maketitle
%

\section{Introduction}

Determining the physical properties of dense cores of interstellar molecular clouds is of fundamental 
importance in the field of star formation research. A powerful method for this purpose is to construct 
the spectral energy distribution (SED) of the source from observational data obtained at multiple different 
frequencies (i.e. continuum flux density $S_{\nu}$ as a function of frequency $\nu$). In particular, an SED 
analysis can be used to derive the temperature of the dust component(s), its mass, and also the luminosity 
over a frequency or wavelength range of interest. The core properties, such as temperature and mass, are 
central to understanding the initial conditions and early stages of star formation within the parent core. 

Of the interstellar molecular clouds that have proved to be fruitful targets for the studies of Galactic star formation, 
the so-called infrared dark clouds (IRDCs; \cite{perault1996}; \cite{egan1998}; \cite{simon2006}; \cite{peretto2009}) 
have attracted a lot of interest in recent years (e.g. \cite{tang2019}; \cite{soam2019}; \cite{peretto2020}; \cite{miettinen2020}; \cite{retes2020}; \cite{moser2020} to name a few recent studies). Some of the IRDCs studied so far are found to be associated with early stages of high-mass star formation (e.g. \cite{rathborne2006}; \cite{beuther2007}; \cite{chambers2009}; \cite{battersby2010}), and even a few candidates for high-mass prestellar cores have been uncovered in IRDCs (\cite{cyganowski2014}, 2017; \cite{contreras2018}). Therefore, 
IRDCs are of particular interest in the context of high-mass star formation, where our understanding of the physical mechanisms is still incomplete (e.g. \cite{motte2018} for a review).

In this paper, we present a study of the physical properties of dense cores in the filamentary IRDC G304.74+01.32, also known as the Seahorse IRDC (\cite{miettinen2018}). The Seahorse IRDC has been fairly well-studied in the submillimetre and millimetre dust continuum emission via single-dish observations (\cite{beltran2006}; \cite{miettinenharju2010}; \cite{miettinen2018}), and through molecular spectral line observations (\cite{miettinen2012}, 2020). Miettinen (2018) derived the dust temperatures of the clumps in the Seahorse IRDC using the 
250, 350, and 500~$\mu$m peak surface brightness ratios measured with the Spectral and Photometric Imaging Receiver (SPIRE; \cite{griffin2010}) on board the \textit{Herschel} satellite (\cite{pilbratt2010})\footnote{\textit{Herschel} is an ESA space observatory with science instruments provided by European-led Principal Investigator consortia and with important participation from NASA.}. Moreover, Miettinen (2018) determined the masses of the clumps and the cores hosted by the clumps through monochromatic flux densities at 870~$\mu$m and 350~$\mu$m, respectively.

To improve the determination of the physical characteristics of the dense core population in the Seahorse IRDC compared to the aforementioned study by Miettinen (2018), the present study makes use of the source SEDs in the far-IR and submillimetre regime. Moreover, the present study focusses on the usage of data at $\sim9\arcsec-20\arcsec$ angular resolution, and does not emply the lower resolution \textit{Herschel}/SPIRE data. We note that although the SED analysis technique is standard and well established, its application to IRDC substructure samples (cores and clumps) has so far mostly concerned the Galactic plane IRDCs (e.g. \cite{rathborne2010}; \cite{beuther2010}; \cite{henning2010}; \cite{ragan2012}, 2013; \cite{veena2018}), while the Snake IRDC lies about $1\fdg3$ above the Galactic plane. 

The core sample and observational data are described in Sect.~2. The analysis and results are described in Sect.~3, and discussed in Sect.~4. Section~5 summarises our results and main conclusions. Throughout this paper, we report the magnitudes in the Vega system and adopt a kinematic distance of $d=2.54$~kpc to the Seahorse IRDC (\cite{miettinen2012}, 2018).

\section{Source sample and data}

The initial source sample of the present study was taken to be the dense cores in the Seahorse IRDC uncovered 
by Miettinen (2018) through 350~$\mu$m dust continuum observations with the Submillimetre APEX BOlometer CAmera 
(SABOCA; \cite{siringo2010}) at $9\arcsec$ (0.11~pc) resolution (full width at half maximum or FWHM). These 17 cores 
are listed in Table~\ref{table:sample}.

Miettinen (2018) employed the 3.4, 4.6, 12, and 22~$\mu$m IR data from the \textit{Wide-field Infrared Survey Explorer} (\textit{WISE}; \cite{wright2010}) to study whether the aforementioned 350~$\mu$m cores are associated with embedded young stellar objects (YSOs). We note that because the Seahorse IRDC lies about $1\fdg3$ ($\sim60$~pc) above the Galactic plane, it is outside the regions observed with the \textit{Spitzer} IR satellite (\cite{werner2004}). Five of the 17 cores (29.4\%) listed in Table~\ref{table:sample} are not associated with \textit{WISE} sources and are classified as IR dark, while the remaining 12 cores (70.6\%) are IR bright. As discussed by Miettinen (2018), the cores SMM~1b and SMM~6b are associated with \textit{WISE} sources whose IR colours suggest that they could be shock features (see the \textit{WISE} colour criteria from Koenig et al. (2012; Appendix therein)), but this would still likely be a sign of ongoing star formation activity in the cores. The \textit{WISE} source seen towards SMM~7 ($6\arcsec$ from the 350~$\mu$m maximum) was found to be so weak (only detected in the W4 band at 22~$\mu$m) that it was interpreted to be a chance projection of a background extragalactic object. However, SMM~7 shows hint of a secondary core (see Appendix~A), and the \textit{WISE} 22~$\mu$m source could be associated with it ($2\farcs8$ separation). Hence, in the present study SMM~7 is taken to be an IR bright, star-forming core. The core IRAS~13039-6108a is known to be associated with an optically thin \ion{H}{ii} region (\cite{sanchez2013}), and is hence associated with high-mass star formation. We also note that three of the clumps that were detected in the Seahorse IRDC at 870~$\mu$m and $19\farcs86$ resolution with the Large APEX BOlometer CAmera (LABOCA; \cite{siringo2009}) by Miettinen (2018), namely BLOB~1, SMM~5, and SMM~8, were not detected in our SABOCA map because the emission was resolved out at $9\arcsec$ resolution.

To construct the source SEDs, we employed the \textit{WISE} 22~$\mu$m data ($12\arcsec$ FWHM resolution), and the SABOCA 350~$\mu$m and LABOCA 870~$\mu$m data from Miettinen (2018). The \textit{WISE} W4 magnitudes of the sources in the Vega system were taken from the AllWISE catalogue\footnote{\url{https://irsa.ipac.caltech.edu/data/download/wise-allwise/}} (the total in-band brightnesses; see Table~3 in \cite{miettinen2018}), and those were used to compute the 22~$\mu$m flux densities by applying the colour corrections under the assumption of a $S_{\nu}\propto \nu^{-2}$ power-law spectrum together with an additional W4 correction (see \cite{cutri2012}). In case the LABOCA 870~$\mu$m clump was resolved into two or three cores in our higher resolution SABOCA imaging (SMM~1, SMM~4, IRAS~13037-6112, SMM~6, and IRAS~13039-6108), we used the relative SABOCA 350~$\mu$m flux densities of the cores to estimate their contribution to the LABOCA 870~$\mu$m emission.

The Seahorse IRDC was observed as part of the \textit{Herschel} Gould Belt Survey (GBS; \cite{andre2010})\footnote{\url{http://gouldbelt-herschel.cea.fr}}. The observations were done with the Photodetector Array Camera and Spectrometer (PACS; \cite{poglitsch2010}) at 70, 100, and 160~$\mu$m ($\sim9\arcsec$, $\sim10\arcsec$, and $\sim13\arcsec$ resolution, respectively) and with 
SPIRE at 250, 350, and 500~$\mu$m ($\sim18\arcsec$, $\sim24\arcsec$, and $\sim35\arcsec$ resolution, respectively). To search for \textit{Herschel} point source counterparts to the Seahorse cores, we cross-matched our source catalogue with the PACS and SPIRE Point Source Catalogues\footnote{\url{https://irsa.ipac.caltech.edu/Missions/herschel.html}} using a search radius of $9\arcsec$, that is the beam size of our SABOCA data. Five cores, namely BLOB~2, SMM~4b and 4c, and IRAS~13039-6108a and 13039-6108b were not found to have counterparts in any of the PACS catalogues, and none of the target cores were found to have counterparts in the SPIRE catalogues. However, the Seahorse IRDC as a whole is clearly detected in the SPIRE images as shown in Fig.~2 in Miettinen (2018). Nevertheless, the relatively poor resolution of the SPIRE observations makes those data less useful for the present purpose (e.g. \cite{ragan2012}, 2013), and our SABOCA observations already probed the 350~$\mu$m band at 2.7 times higher resolution than SPIRE. We note that the LABOCA clumps BLOB~1, SMM~5, and SMM~8 that were not detected in our SABOCA map also had no matches in the \textit{Herschel} point source catalogues. 

For the three \textit{IRAS} (\textit{Infrared Astronomical Satellite}; \cite{neugebauer1984}) sources in the Seahorse we also employed the 25, 60, and 100~$\mu$m data from the IRAS Point Source Catalogue v2.1\footnote{\url{https://irsa.ipac.caltech.edu/Missions/iras.html}}. The angular resolution of \textit{IRAS} at these wavelengths was about $1\arcmin-2\arcmin$ (\cite{beichman1988}). For the cores IRAS~13037-6112a and IRAS~13037-6112b we used the \textit{WISE} 22~$\mu$m and PACS 70 and 100~$\mu$m data to estimate the cores' relative contribution to the IRAS flux densities. In the case of the IRAS~13039-6108a/b core pair, all the IRAS emission was assigned to the brighter IRAS~13039-6108a component because it is associated with a 22~$\mu$m source while IRAS~13039-6108b is not. 

\begin{figure}[!htb]
\centering
\resizebox{\hsize}{!}{\includegraphics{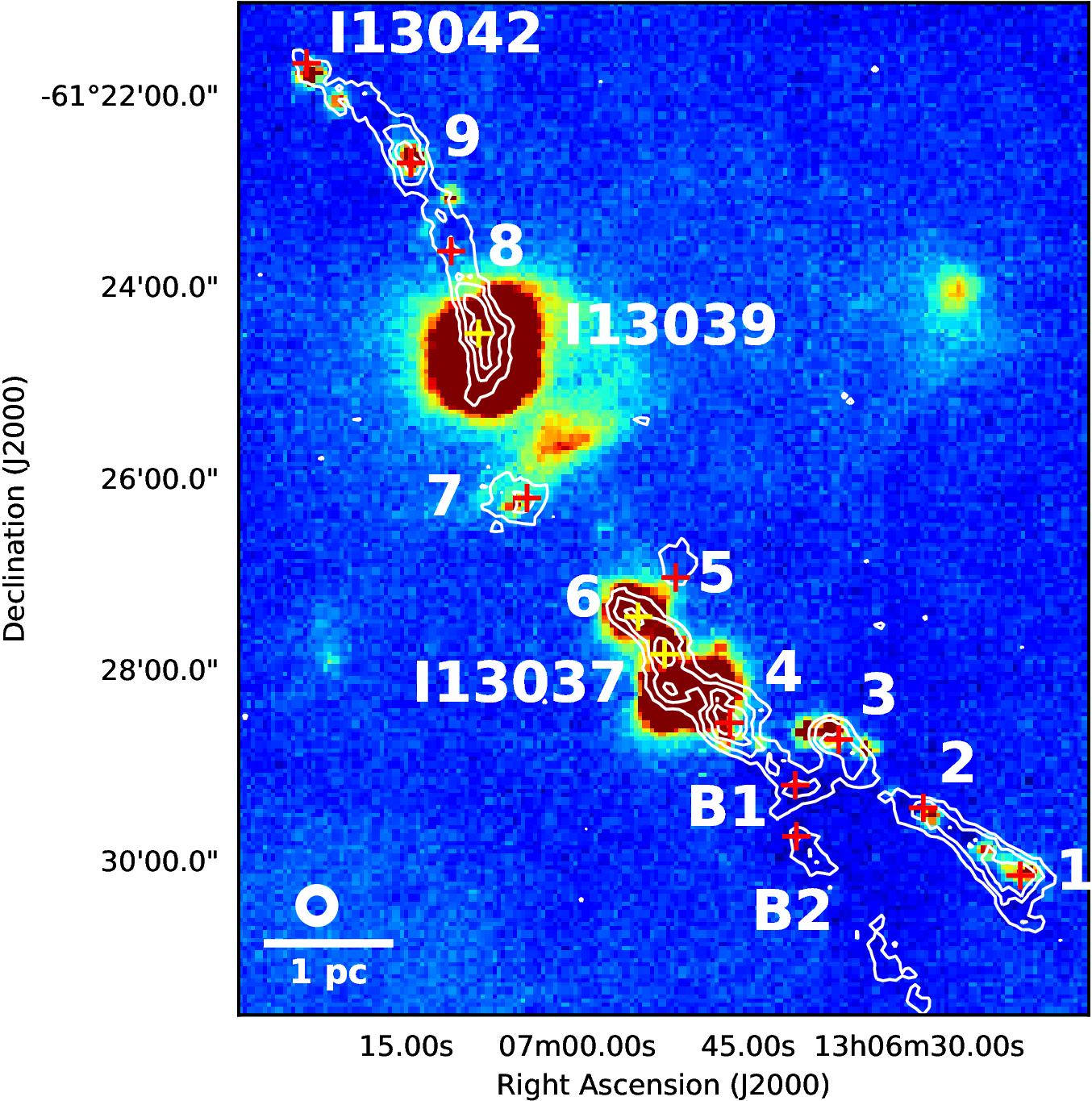}}
\caption{\textit{Herschel}/PACS 70~$\mu$m image towards the IRDC G304.74+01.32 (the Seahorse IRDC). The colour scale is displayed using 
a linear stretch. The overlaid contours represent the LABOCA 870~$\mu$m emission (\cite{miettinenharju2010}; \cite{miettinen2018}); the contours start at $3\sigma$, and increase in steps of $3\sigma$, where $3\sigma=120$~mJy~beam$^{-1}$. The 870~$\mu$m clumps are labelled so that the numbers refer to the SMM IDs (e.g. 1 refers to SMM~1), while the sources I13037, I13039, and I13042 are the
three \textit{IRAS} sources in the filament. The sources BLOB~1 and BLOB~2 are labelled as B1 and B2. The plus signs indicate the LABOCA 870~$\mu$m emission peaks of the clumps. A scale bar of 1~pc projected length, and the LABOCA beam size ($19\farcs86$ FWHM) are shown in the bottom left corner.}
\label{figure:map}
\end{figure}

We required that a source needs to have at least three detections in the far-IR and submillimetre regime in addition to the possible 
\textit{WISE} 22~$\mu$m detection in order to construct a useful source SED for the purpose of the present study. Hence,  
the final source sample is composed of 12 cores out of which two (SMM~1a and SMM~9) were not detected with \textit{WISE} and are classified as IR dark cores. The relative percentages of IR bright and IR dark cores in our final sample are therefore 83.3\% and 16.7\%. The photometric data of these sources are given in Table~\ref{table:photometry}. The PACS 70~$\mu$m image towards the Seahorse IRDC is shown in Fig.~\ref{figure:map}, while panchromatic zoom-in images towards the analysed cores are shown in Fig.~\ref{figure:images}.

\begin{table}[H]
\renewcommand{\footnoterule}{}
\caption{Initial source sample.}
{\small
\begin{minipage}{1\columnwidth}
\centering
\label{table:sample}
\begin{tabular}{c c c c}
\hline\hline 
Source & $\alpha_{2000.0}$ & $\delta_{2000.0}$ & Type\\
       & [h:m:s] & [$\degr$:$\arcmin$:$\arcsec$] &  \\ 
\hline 
SMM 1a & 13 06 19.38 & -61 30 19.42 & IR dark\\
SMM 1b & 13 06 23.36 & -61 30 10.37 & IR bright\tablefootmark{a}\\
SMM 2 &	13 06 28.82	& -61 29 43.43 & IR bright \\ 
SMM 3 &	13 06 37.00	& -61 28 51.00 & IR bright \\ 
BLOB 2 & 13 06 39.50 & -61 30 01.51 & IR dark \\ 
SMM 4a & 13 06 46.21 & -61 28 48.04 & IR bright \\ 
SMM 4b & 13 06 46.84 & -61 28 22.54 & IR bright \\ 
SMM 4c & 13 06 42.86 & -61 28 33.03 & IR dark \\ 
IRAS 13037 & 13 06 51.45 & -61 28 24.04 & IR bright \\ 
-6112a      &             &              &   \\
IRAS 13037 & 13 06 51.24 & -61 27 51.04 & IR bright \\ 
-6112b      &             &              &      \\
SMM 6a & 13 06 55.42 & -61 27 27.03 & IR bright\\ 
SMM 6b & 13 06 52.08 & -61 27 40.54 & IR bright\tablefootmark{a}\\ 
SMM 7 &	13 07 04.20	& -61 26 15.00 & IR bright\tablefootmark{b}\\ 
IRAS 13039 & 13 07 06.49 & -61 24 32.98 & IR bright/\ion{H}{ii}\tablefootmark{c} \\ 
-6108a      &             &              &  \\
IRAS 13039 & 13 07 08.58 & -61 23 56.97 & IR dark \\ 
-6108b      &             &              &  \\
SMM 9 &	13 07 12.53	& -61 22 46.43 & IR dark\\ 
IRAS 13042 & 13 07 20.66 & -61 21 52.33 & IR bright\\
-6105      &             &              & \\  	
\hline
\end{tabular} 
\tablefoot{The coordinates refer to the SABOCA 350~$\mu$m peak positions of the sources (\cite{miettinen2018}). The source type in the last column is based on the source appearance in the \textit{WISE} IR images (\cite{miettinen2018}).\tablefoottext{a}{The core is associated with a \textit{WISE} IR source that could be a shock emission knot.}\tablefoottext{b}{The weak \textit{WISE} IR source seen towards SMM~7 was considered a candidate extragalactic object in our previous studies (\cite{miettinen2018}, 2020), but it could be an embedded YSO associated with a secondary 350~$\mu$m peak position at $\alpha_{2000.0}=13^{\rm h}07^{\rm m}04\fs21$, $\delta_{2000.0}=-61\degr26\arcmin 22\farcs50$ ($2\farcs8$ or 0.03~pc offset; see Appendix~A). Hence, SMM~7 is considered an IR bright core in the present study.}\tablefoottext{c}{The core is associated with an optically thin \ion{H}{ii} region (\cite{sanchez2013}).}}
\end{minipage} }
\end{table}

\begin{table*}
\caption{Photometric data of the analysed sources.}
\begin{minipage}{2\columnwidth}
\centering
\renewcommand{\footnoterule}{}
\label{table:photometry}
\begin{tabular}{c c c c c c c c c}
\hline\hline
Source & $S_{22}$ & $S_{25}$ & $S_{60}$ & $S_{70}$ & $S_{100}$ & $S_{160}$ & $S_{350}$ & $S_{870}$\\
       & [Jy] & [Jy] & [Jy] & [Jy] & [Jy] & [Jy] & [Jy] & [Jy] \\
\hline 
SMM 1a & \ldots & \ldots & \ldots & \ldots & $2.59\pm0.11$ & $8.04\pm0.87$ & $4.04\pm1.24$ & $0.75\pm0.09$ \\
SMM 1b & $0.16^{+0.00}_{-0.01}$ & \ldots & \ldots & $0.95\pm0.05$ & $1.60\pm0.15$ & \ldots & $4.15\pm1.28$ & $0.78\pm0.09$\\
SMM 2 &	$0.35\pm0.01$ & \ldots & \ldots & $1.10\pm0.05$ & $1.65\pm0.06$ & $3.73\pm0.87$ & $2.18\pm0.69$ & $0.63\pm0.06$ \\ 
SMM 3 &	 $0.27^{+0.01}_{-0.00}$ & \ldots & \ldots & \ldots & $4.62\pm0.39$ & $10.72\pm2.38$ & $3.58\pm1.11$ & $0.82\pm0.11$ \\ 
SMM 4a & $0.03\pm0.01$ & \ldots & \ldots & \ldots & $3.32\pm1.28$ & \ldots & $4.98\pm1.53$ & $0.75\pm0.09$ \\ 
IRAS 13037 & $3.27\pm0.05$ & $5.23\pm0.42$ & $47.69\pm7.63$ & $24.08\pm1.02$ & $33.51\pm3.45$ & $56.34\pm8.01$ & $3.74\pm1.15$ & $0.77\pm0.09$\\ 
-6112a      &  & & & & $<144.25$ & \\
IRAS 13037 &  $1.36^{+0.02}_{-0.01}$ & $2.18\pm0.17$ & $17.27\pm2.76$ & $8.72\pm1.17$ & $12.14\pm1.83$ & \ldots & $2.27\pm0.72$ & $0.47\pm0.05$  \\ 
-6112b      &   & & & & $<52.25$ &  \\
SMM 6a & $0.28\pm0.01$ & \ldots & \ldots & $7.33\pm0.52$ & \ldots & \ldots & $4.97\pm0.70$ & $0.34\pm0.05$ \\ 
SMM 7 &	$0.04^{+0.001}_{-0.002}$ & \ldots & \ldots & $1.19\pm0.17$ & $2.40\pm0.46$ & \ldots & $2.12\pm0.68$ & $0.52\pm0.07$ \\ 
IRAS 13039 & $3.33\pm0.05$ & $7.43\pm0.59$ & $105.6\pm14.8$ & \ldots & \ldots & \ldots & $5.88\pm1.80$ & $1.17\pm0.13$  \\ 
-6108a      &   & & & & $196.5\pm35.4$ &    \\
SMM 9 &	\ldots & \ldots & \ldots & $3.52\pm0.07$ & $7.84\pm0.19$ & $13.10\pm1.29$ & $3.61\pm1.12$ & $0.83\pm0.10$ \\ 
IRAS 13042 & $0.46\pm0.01$ & $0.67\pm0.08$ & $<4.97$ & $2.73\pm0.03$ & $4.77\pm0.13$ & $6.58\pm0.84$ & $1.27\pm0.44$ & $0.23\pm0.05$ \\
-6105      &  & & & & $<196.5$ &\\   
\hline
\end{tabular} 
\tablefoot{The \textit{WISE} flux densities listed in the second column were computed from the AllWISE catalogue's Vega magnitudes by applying the colour corrections under the assumption of a $S_{\nu}\propto \nu^{-2}$ power-law spectrum. Moreover, an additional W4 correction was applied to calculate the final 22~$\mu$m flux density (see \cite{cutri2012}). For the \textit{IRAS} sources, the 100~$\mu$m flux density could be obtained from both \textit{Herschel}/PACS and \textit{IRAS} (given below the PACS value). However, IRAS 13039-6108a could not be found from the PACS Point Source Catalogue, and hence only the \textit{IRAS} 100~$\mu$m flux density is given in the table.}
\end{minipage} 
\end{table*}

\section{Analysis and results}

\subsection{Modified blackbody fits to the SEDs}

\begin{figure*}[!htb]
\begin{center}
\includegraphics[width=0.3\textwidth]{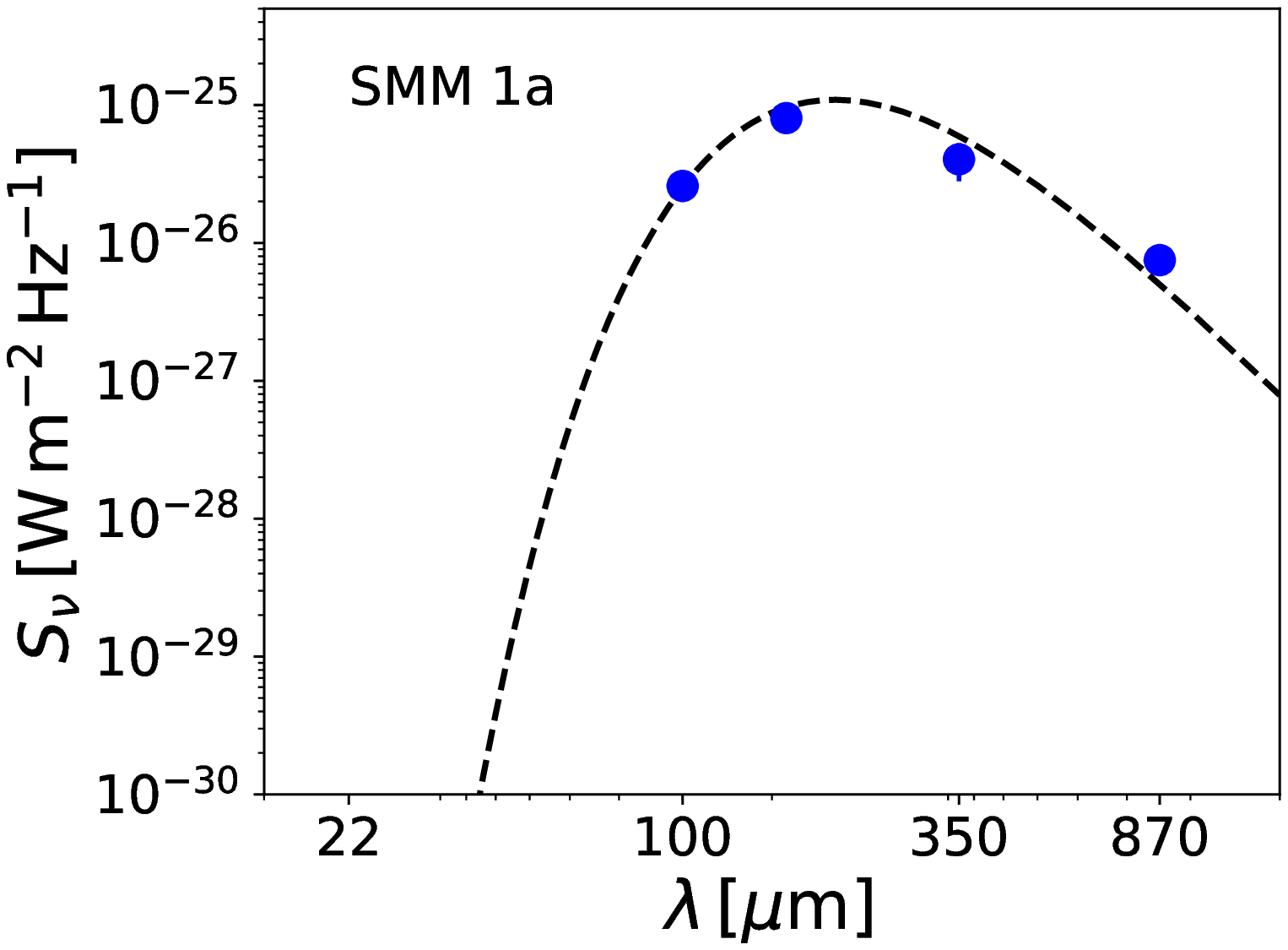}
\includegraphics[width=0.3\textwidth]{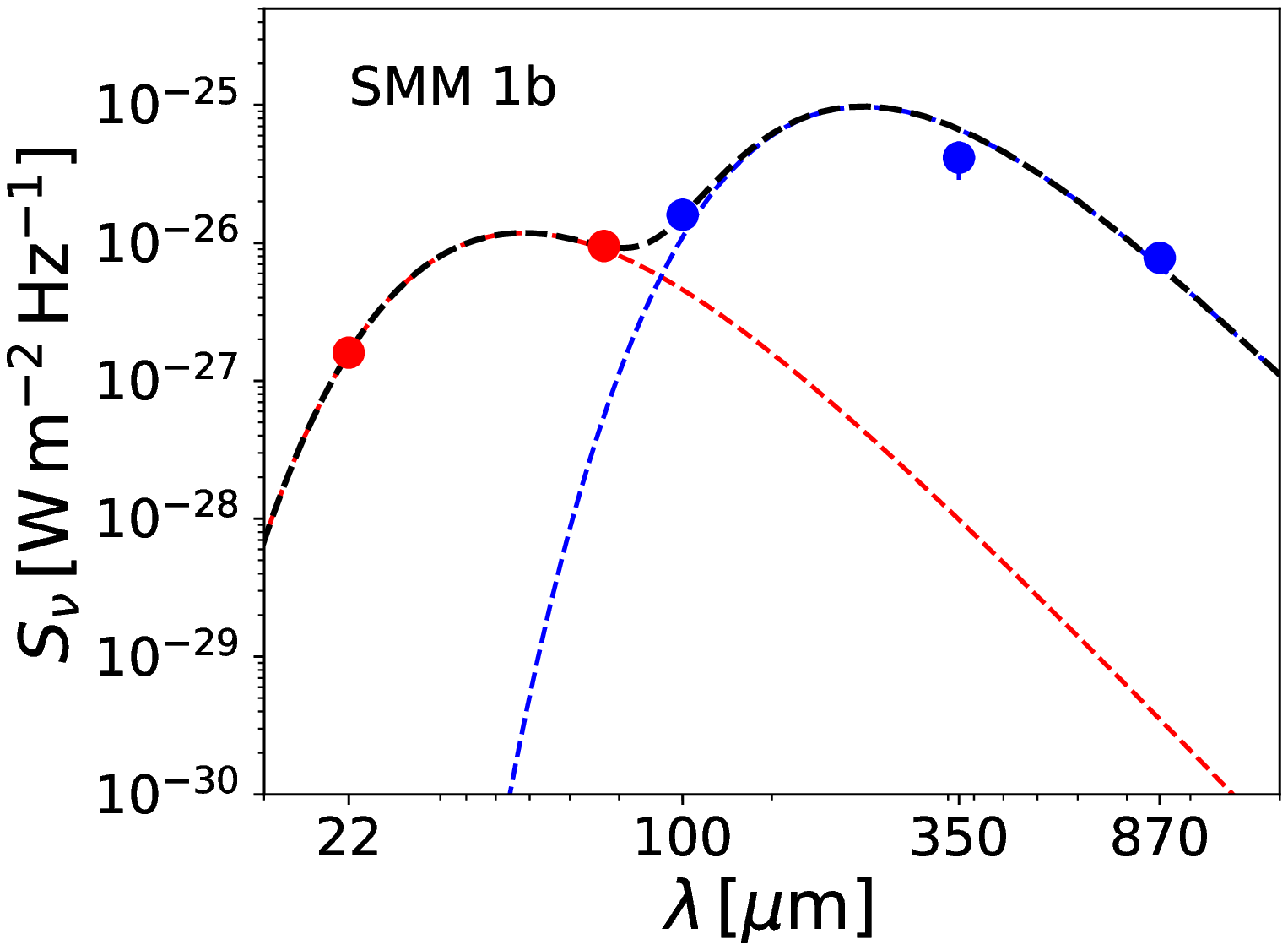}
\includegraphics[width=0.3\textwidth]{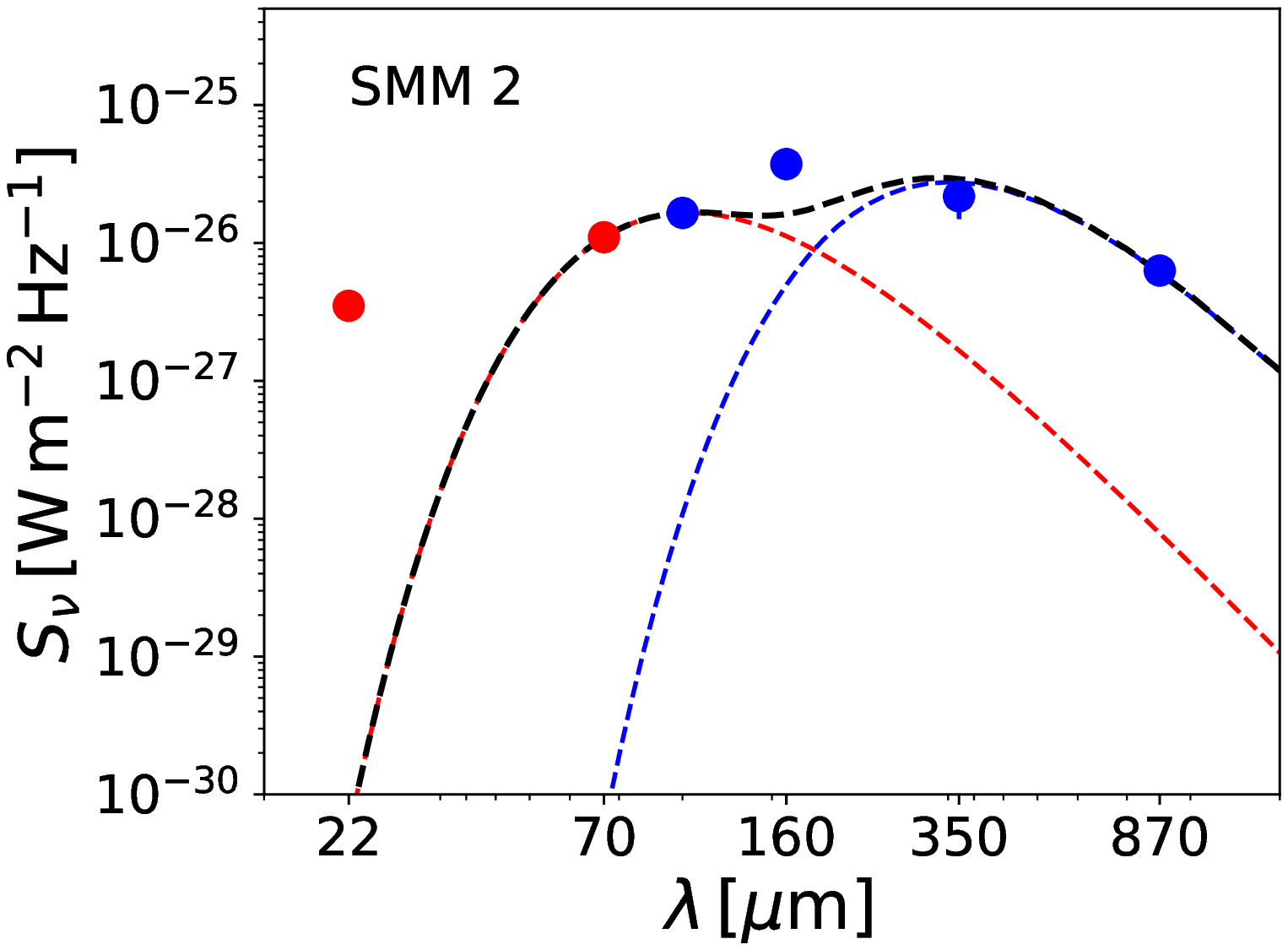}
\includegraphics[width=0.3\textwidth]{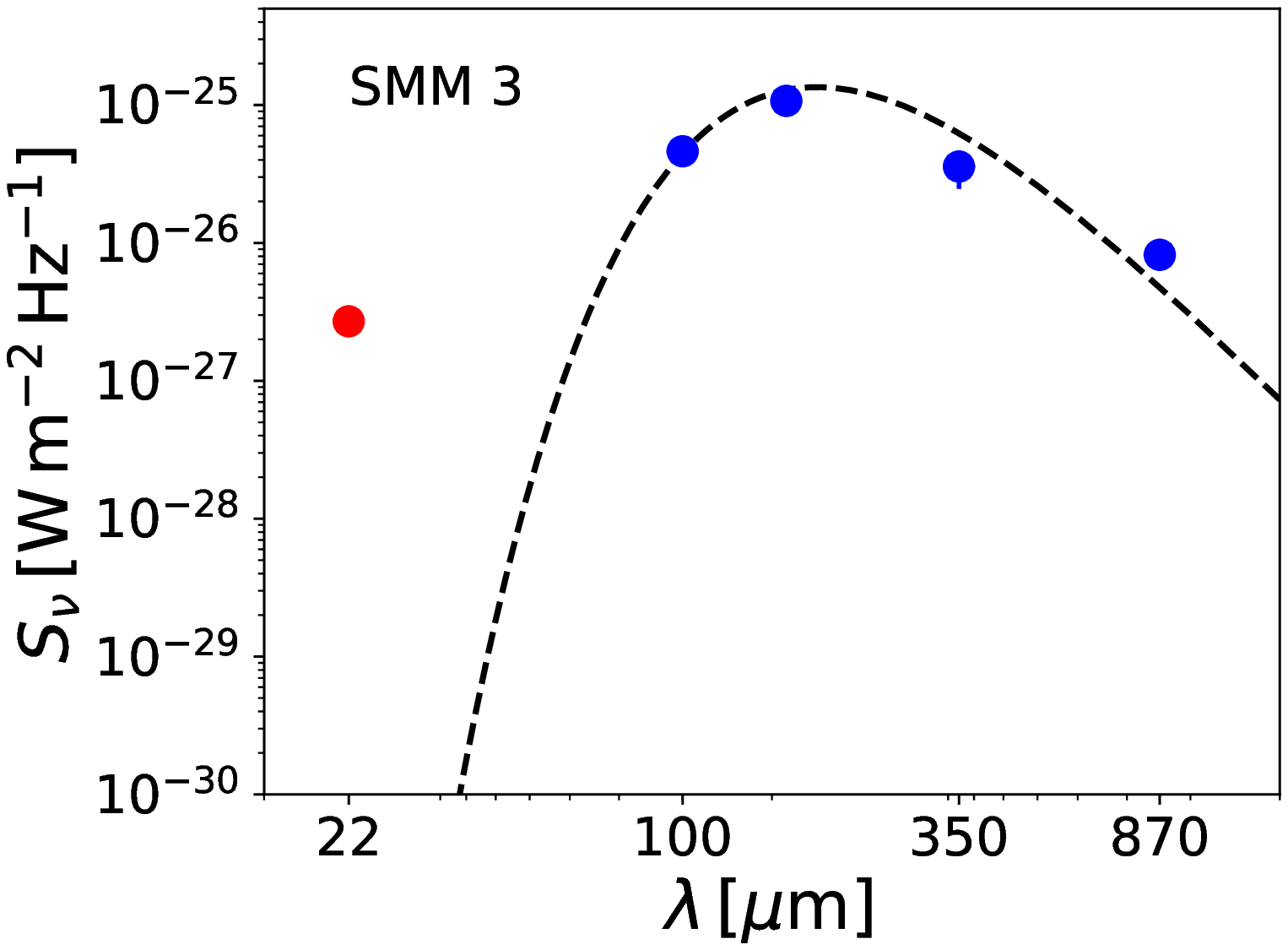}
\includegraphics[width=0.3\textwidth]{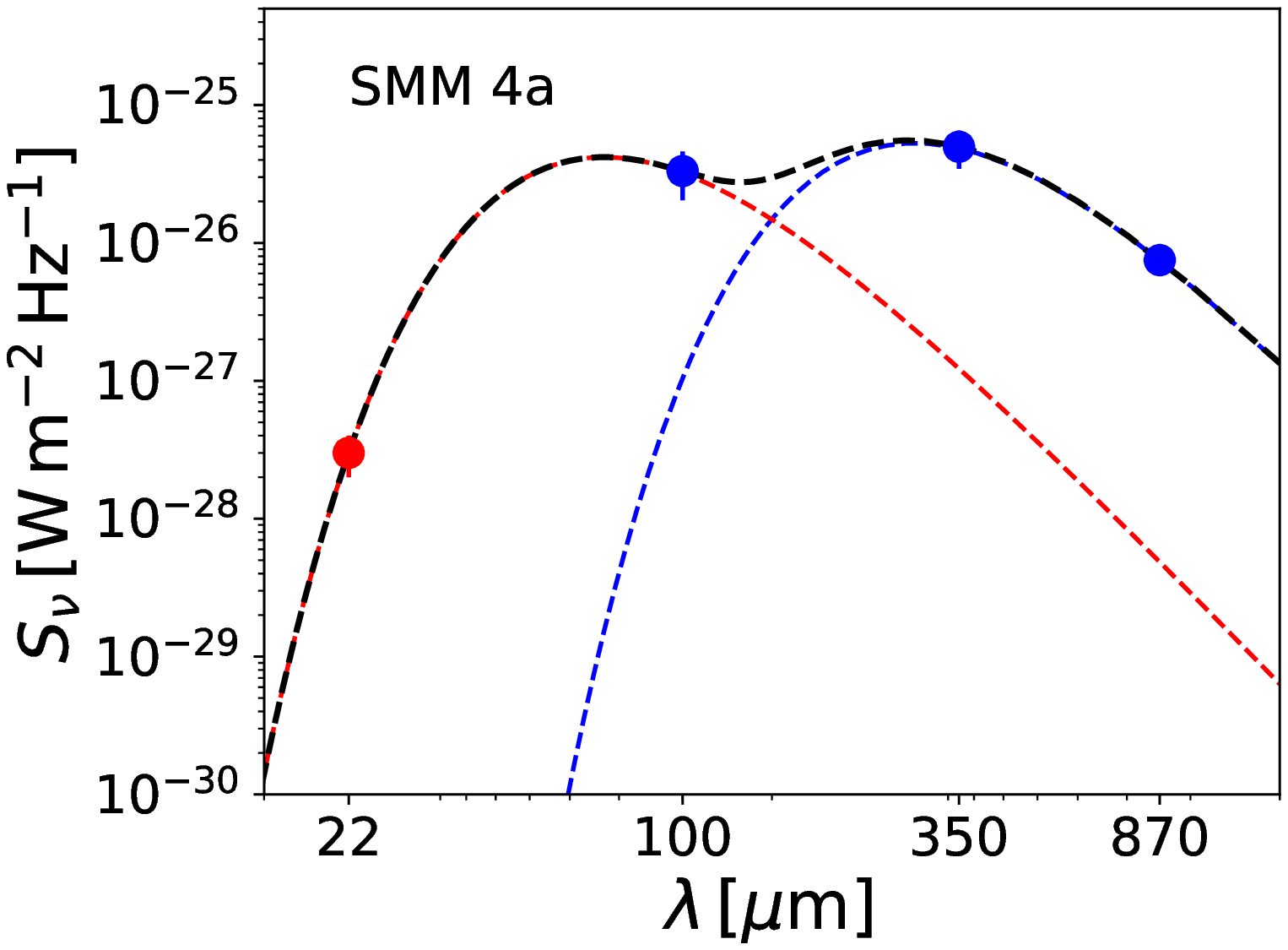}
\includegraphics[width=0.3\textwidth]{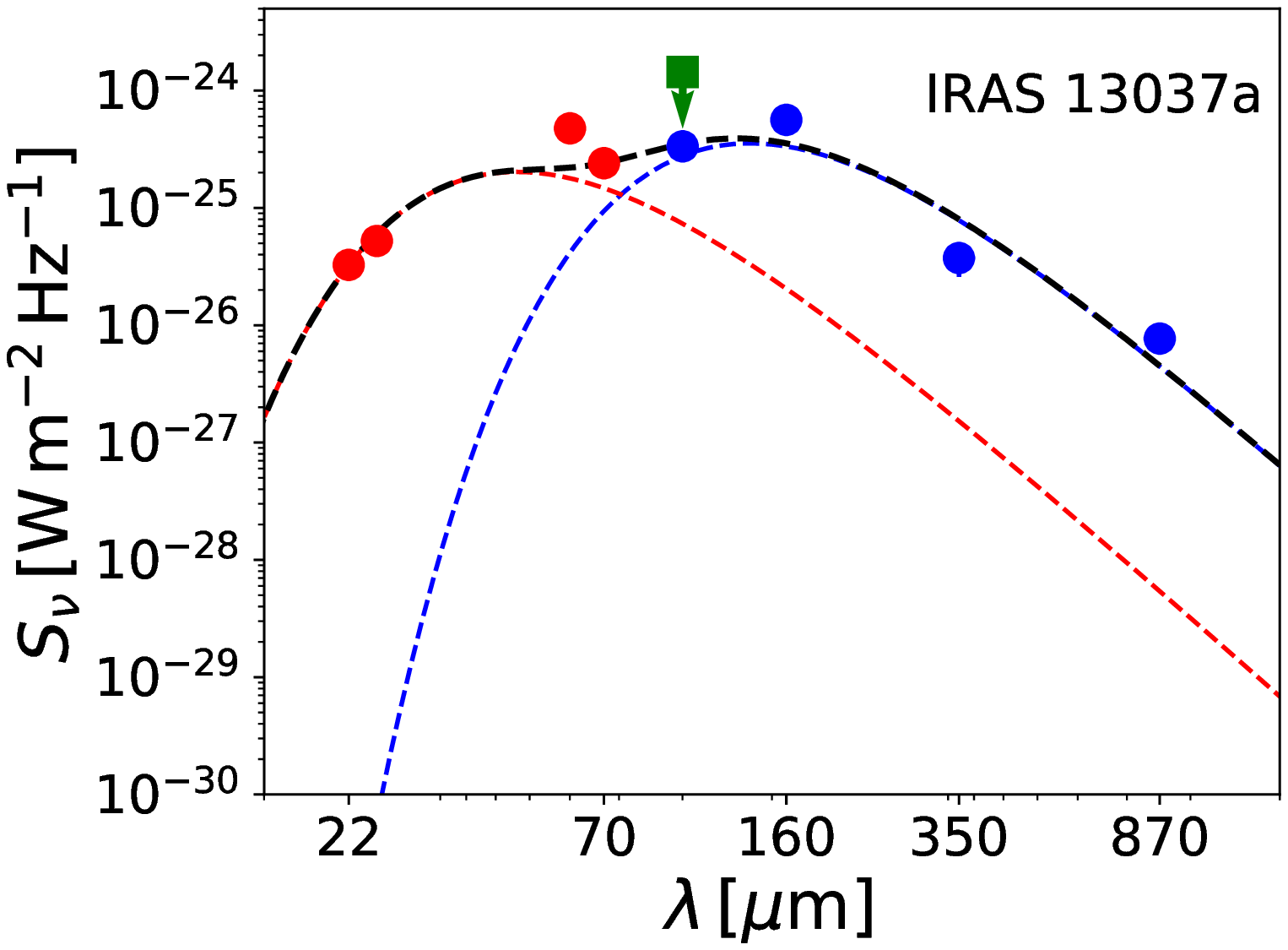}
\includegraphics[width=0.3\textwidth]{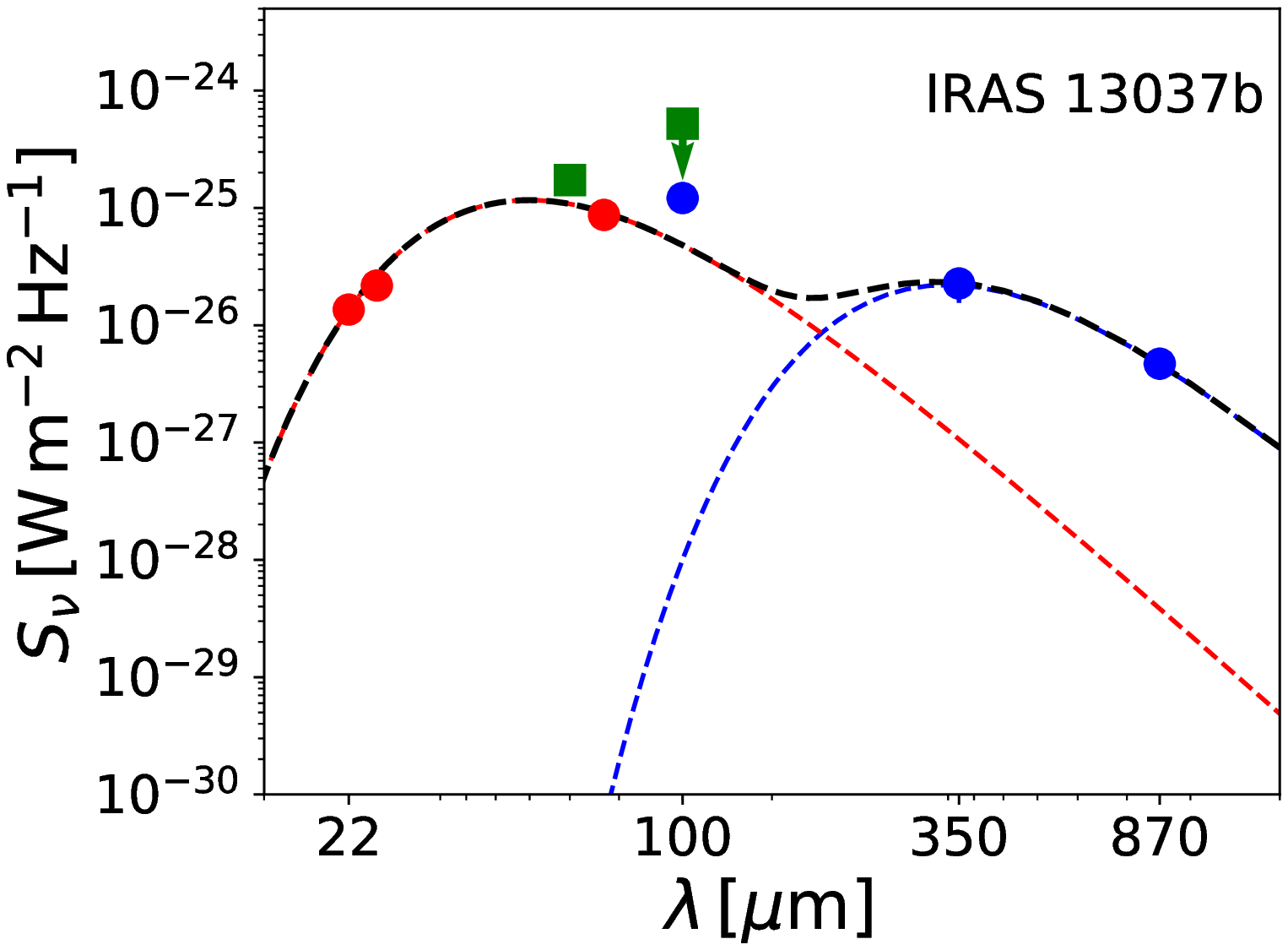}
\includegraphics[width=0.3\textwidth]{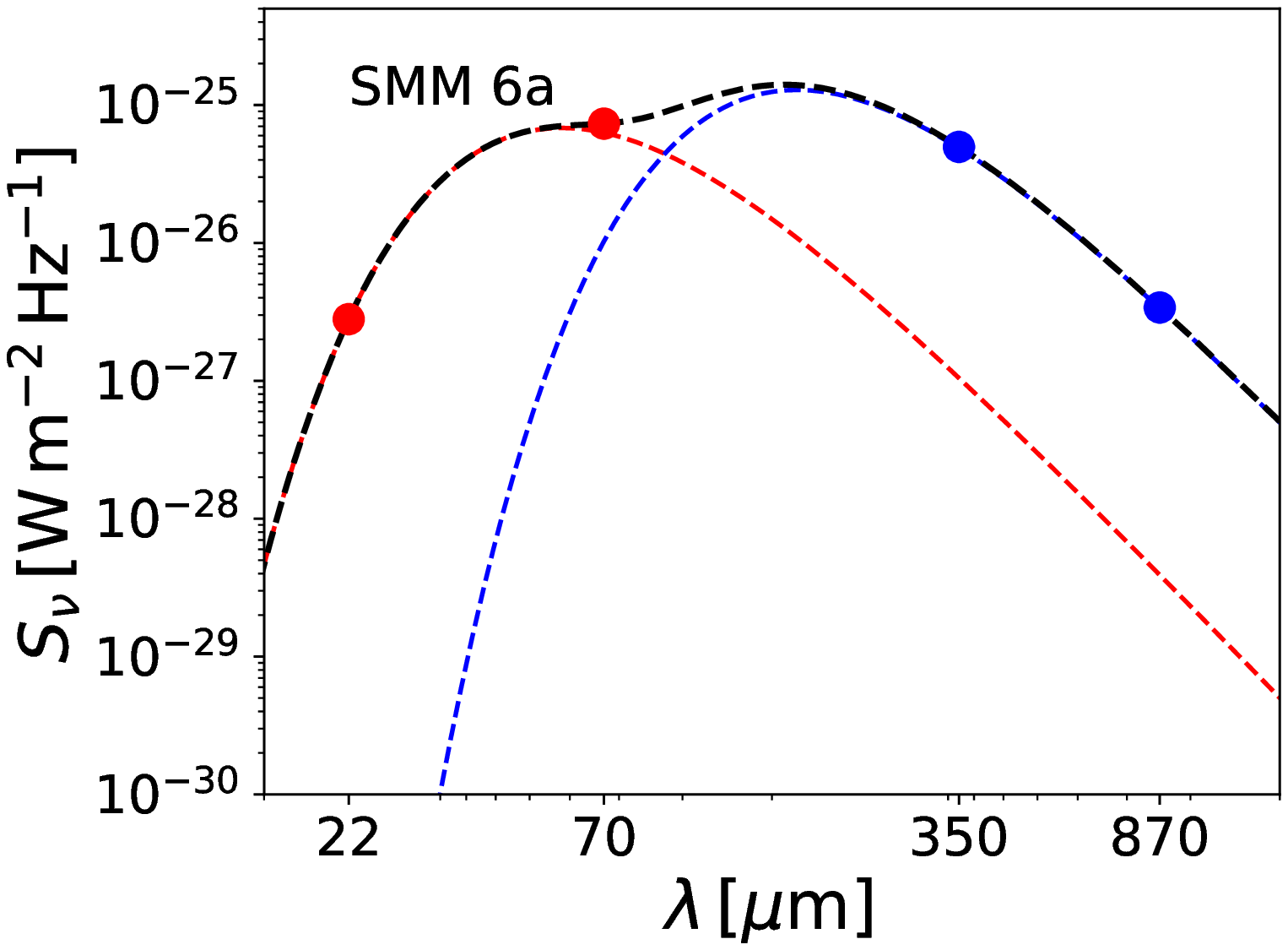}
\includegraphics[width=0.3\textwidth]{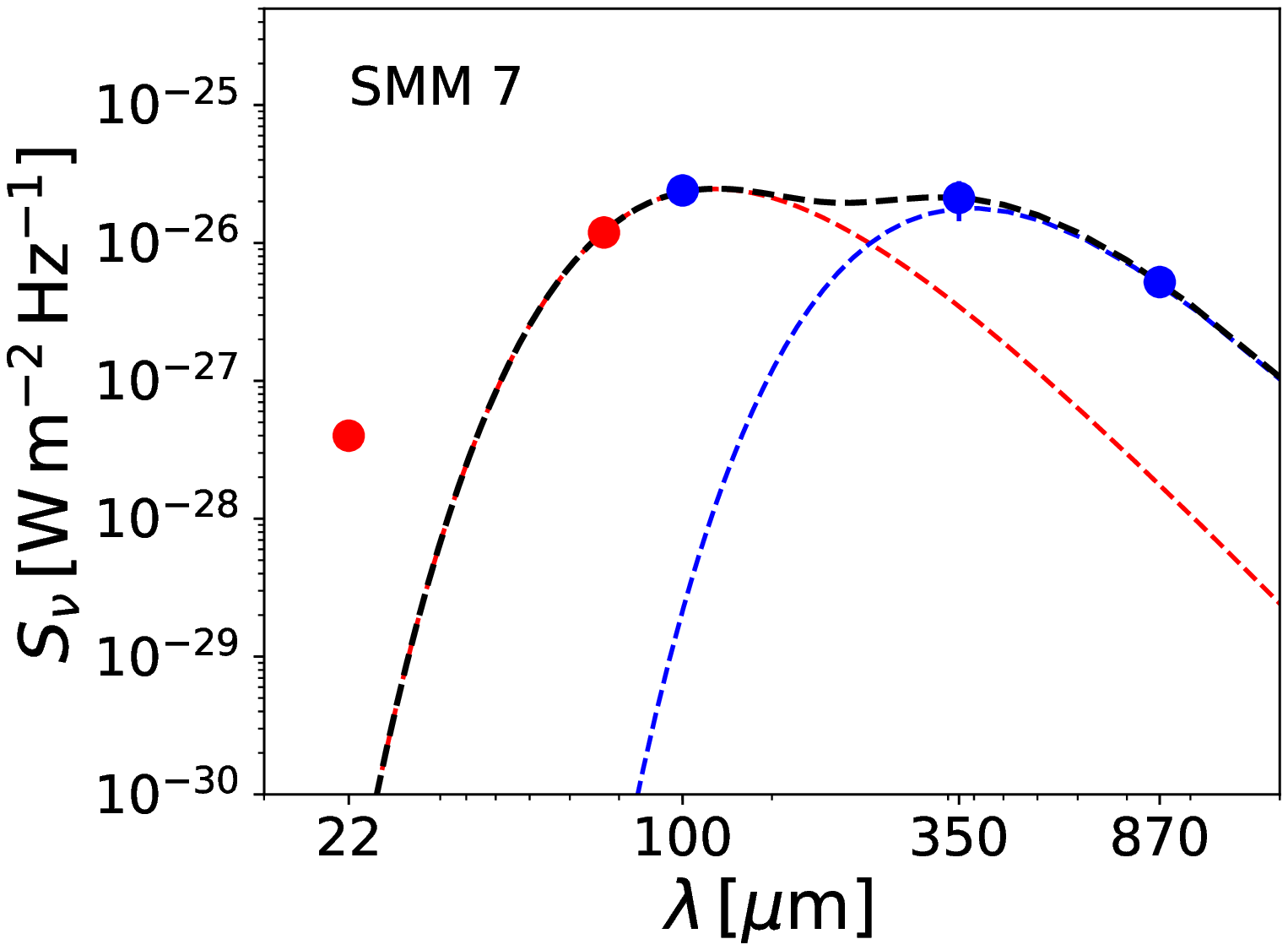}
\includegraphics[width=0.3\textwidth]{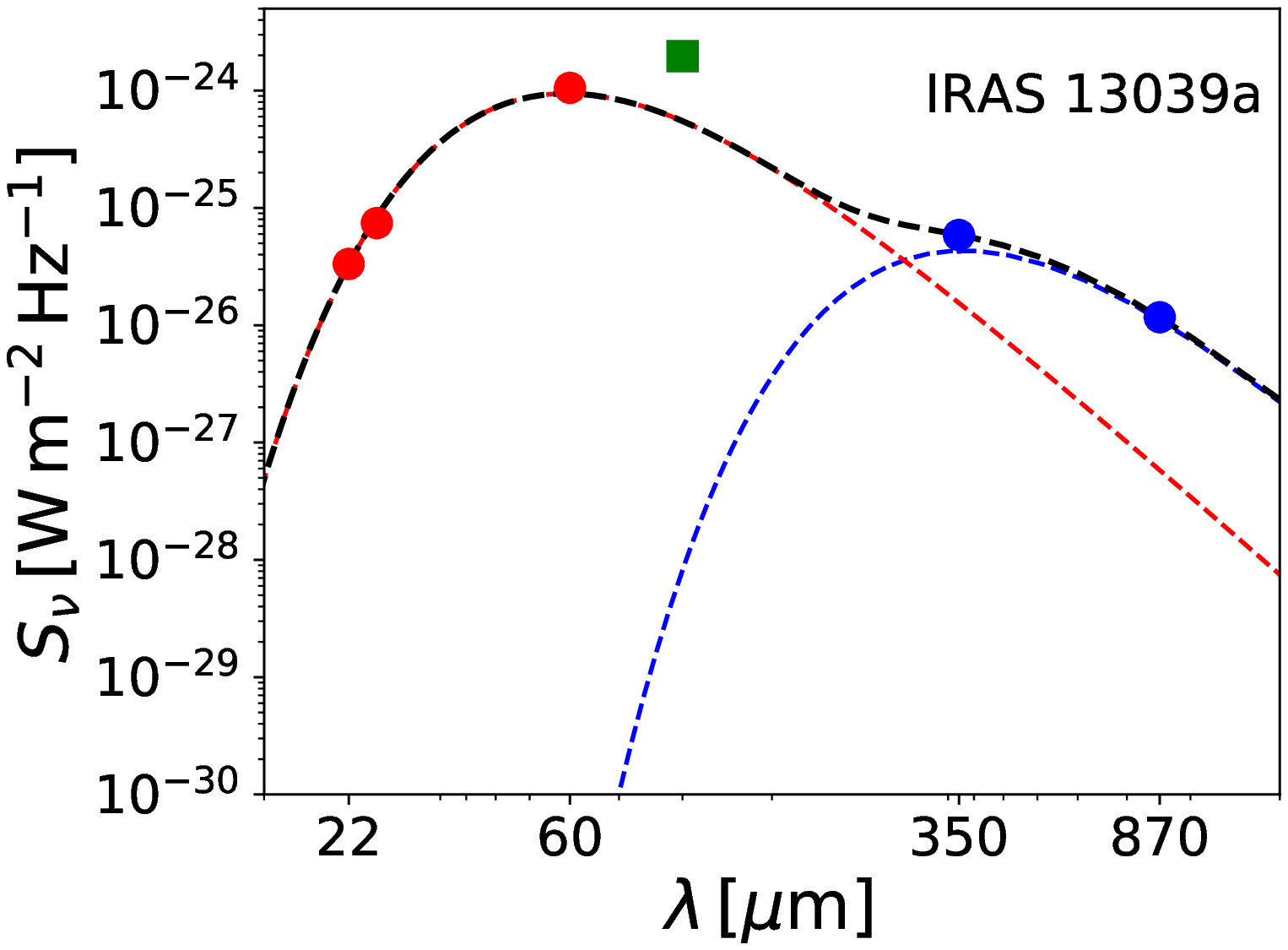}
\includegraphics[width=0.3\textwidth]{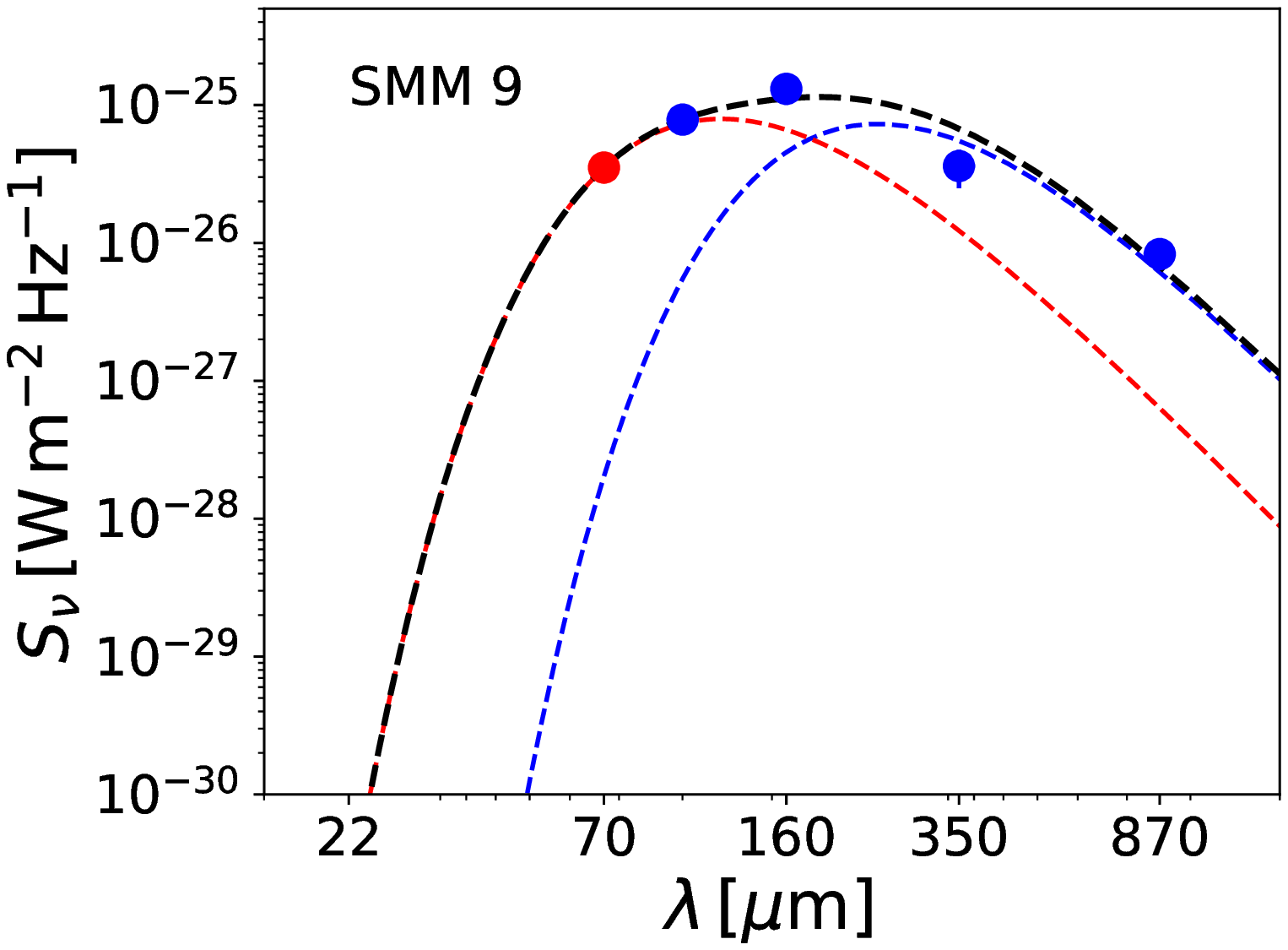}
\includegraphics[width=0.3\textwidth]{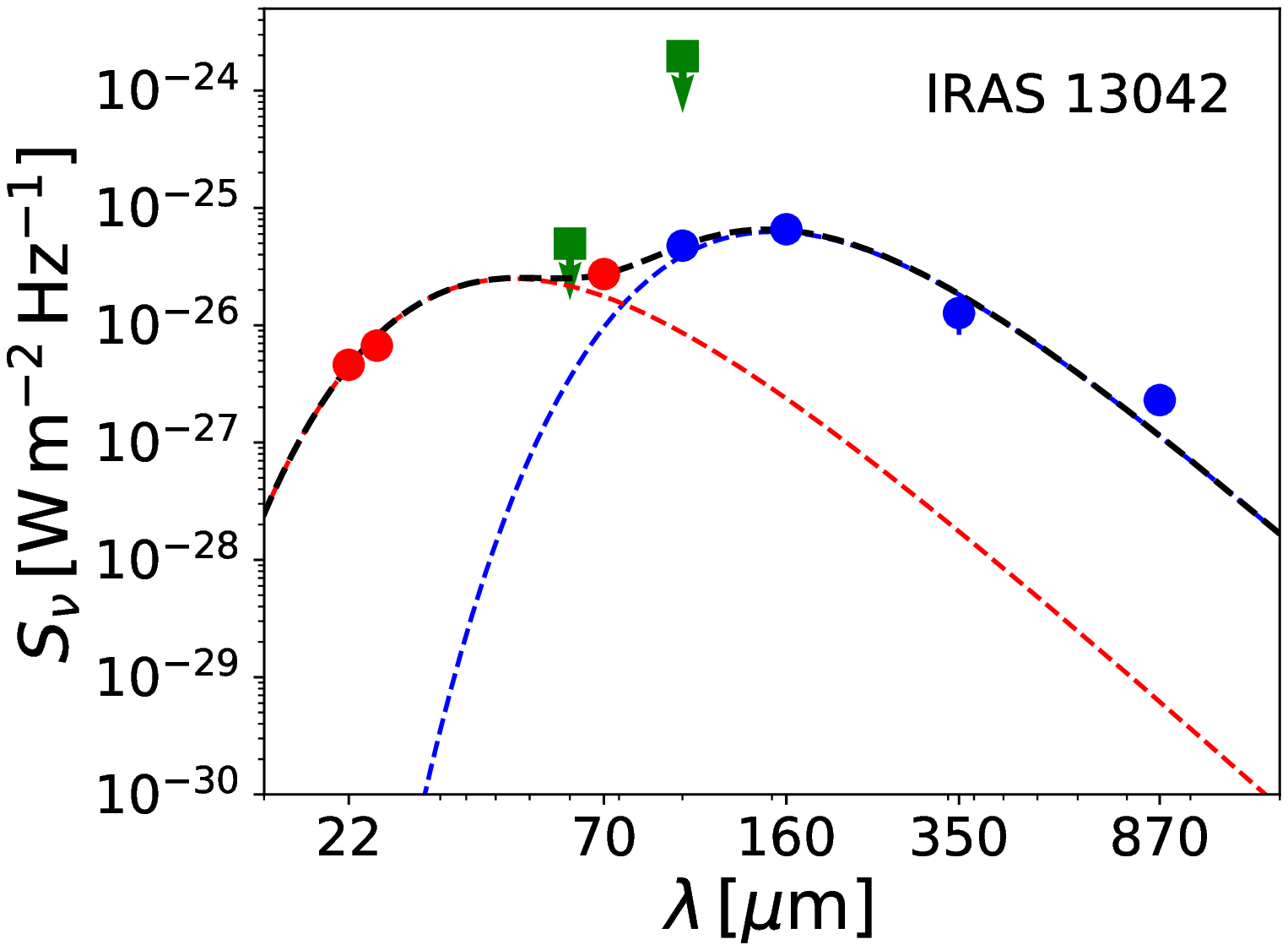}
\caption{Far-IR to submillimetre SEDs of the analysed cores. Data points at $\lambda \leq 70$~$\mu$m are highlighted in red, while the longer wavelength data are shown in blue. The data points shown by green square symbols were not included in the fit (see Sect.~4.1 for details). The downward pointing arrows indicate upper limits (relevant only for some of the \textit{IRAS} flux densities). The black, dashed lines represent the best modified blackbody fits to the data. For the two-temperature modified blackbody fits, the black, dashed line shows the sum of the two components. The blue and red dashed lines show the SED fits to the cold and warm component, respectively. We note that the $y$-axis scale is different for the bright \textit{IRAS} sources compared to the other panels.}
\label{figure:seds}
\end{center}
\end{figure*}

The SEDs of the 12 analysed cores are shown in Fig.~\ref{figure:seds}. Depending on the source, the observed far-IR to submillimetre SEDs were fitted by single or two-temperature modified blackbody functions under the assumption of 
optically thin dust emission, in which case the general function is given by (e.g. \cite{shetty2009a},b; \cite{casey2012}; \cite{bianchi2013})

\begin{equation}
\label{eqn:sed}
S_{\nu}=\frac{M_{\rm dust}}{d^2}\kappa_{\nu}B_{\nu}(T_{\rm dust})\,,
\end{equation}
where $M_{\rm dust}$ is the dust mass, $\kappa_{\nu}$ is the dust opacity, and $B_{\nu}(T_{\rm dust})$ is the Planck function 
at a dust temperature of $T_{\rm dust}$ defined as 

\begin{equation}
\label{eqn:planck}
B_{\nu}(T_{\rm dust})=\frac{2h \nu^3}{c^2}\frac{1}{e^{h\nu/k_{\rm B}T_{\rm dust}}-1}\,,
\end{equation}
where $h$ is the Planck constant, $c$ is the speed of light, and $k_{\rm B}$ is the Boltzmann constant. 
The two-temperature component SEDs were fitted with a sum of two modified Planck functions (e.g. \cite{dunne2001}).

The frequency-dependent dust opacity was assumed to follow a power-law function of the form 

\begin{equation}
\label{eqn:kappa}
\kappa_{\nu}=\kappa_0 \times \left(\frac{\nu}{\nu_0}\right)^{\beta}\,,
\end{equation}
where $\beta$ is the dust emissivity index. We assumed that the $\kappa_{\nu}$ values follow 
the Ossenkopf \& Henning (1994) dust model values for graphite-silicate dust grains that have coagulated and accreted thin ice mantles over a period of $10^5$~yr at a gas density of $n_{\rm H}=10^5$~cm$^{-3}$. For reference, the value of $\kappa_{\nu}$ is 1.38~cm$^2$~g$^{-1}$ at 870~$\mu$m and $\beta \simeq 1.9$. The non-linear least squares SED fitting was implemented in {\tt Python} using the {\tt SciPy} optimisation module (\cite{virtanen2020}).

The dust masses derived through SED fits were converted to the total gas plus dust masses using a dust-to-gas mass ratio, 
$R_{\rm dg}=M_{\rm dust}/M_{\rm gas}$. The latter value can be derived from the dust-to-hydrogen mass ratio, 
$M_{\rm dust}/M_{\rm H}$, which is about 1/100 (e.g. \cite{weingartner2001}; \cite{draine2007}). Assuming that the chemical composition of the cores is similar to the solar mixture, where the hydrogen mass percentage is $X=71\%$ and those of helium and metals are $Y=27\%$ and $Z=2\%$ (e.g. \cite{anders1989}), the ratio of total gas mass (hydrogen, helium, and metals) to hydrogen gas mass is $(X+Y+Z)/X=1.41$. Based on this assumption, we adopted a dust-to-gas mass ratio of $R_{\rm dg}=1/141$.

In addition to the dust temperature and mass that were derived through fitting the observed SEDs, we also calculated 
the bolometric luminosities of the sources by integrating over the fitted SED curves, that is (e.g. \cite{dunham2010})

\begin{equation}
\label{eqn:luminosity}
L=4\pi d^2 \int_0^{\infty} S_{\nu} {\rm d}\nu \,.
\end{equation}
To quantify the goodness of the SED fits, we calculated the reduced $\chi^2$ values defined as (e.g. \cite{dunham2006})

\begin{equation}
\label{eqn:chi}
\chi_{\rm red}^2 = \frac{1}{k}\sum\limits_{i=0}^{n}\left(\frac{S_{\nu}^{\rm obs}-S_{\nu}^{\rm model}}{\sigma(S_{\nu}^{\rm obs})}\right)^2\,,
\end{equation}
where for $n$ data points and $m$ free parameters there are $k=n-m$ degrees of freedom, and the parameters $S_{\nu}^{\rm obs}$ and $\sigma(S_{\nu}^{\rm obs})$ are the observed flux densities and their uncertainties and $S_{\nu}^{\rm model}$ is the flux density of the model fit at the corresponding frequency. 

The derived SED parameters are listed in Table~\ref{table:properties}. The quoted uncertainties were derived 
using the flux density uncertainties. The SED properties of the analysed cores and the derived physical properties of the cores 
are discussed in Sects.~4.1 and 4.2, respectively.

\subsection{Densities}

The volume-averaged H$_2$ number densities of the cores were calculated assuming a spherical geometry and using the formula

\begin{equation}
\label{eqn:density}
n({\rm H_2})=\frac{3M}{4\pi R^3 \mu_{\rm H_2}m_{\rm H}}\,,
\end{equation}
where the radii $R$ were taken from Miettinen (2018, Table~2 therein) and they refer to the 
effective core radius, which is defined as $R=\sqrt{A_{\rm proj}/\pi}$, where $A_{\rm proj}$ is the projected area of the core calculated from the total number of pixels that were assigned to the core in the SABOCA map. The quantity $\mu_{\rm H_2}$ is the mean molecular weight per H$_2$ molecule, which in our case is $\mu_{\rm H_2}=2/X=2.82$ (\cite{kauffmann2008}, Appendix~A.1 therein), and $m_{\rm H}$ is the mass of the hydrogen atom. 

We also calculated the surface densities of the cores using the formula 

\begin{equation}
\label{eqn:surface}
\Sigma=\frac{M}{\pi R^2}\,.
\end{equation} 
 
\begin{figure}[!htb]
\centering
\resizebox{\hsize}{!}{\includegraphics{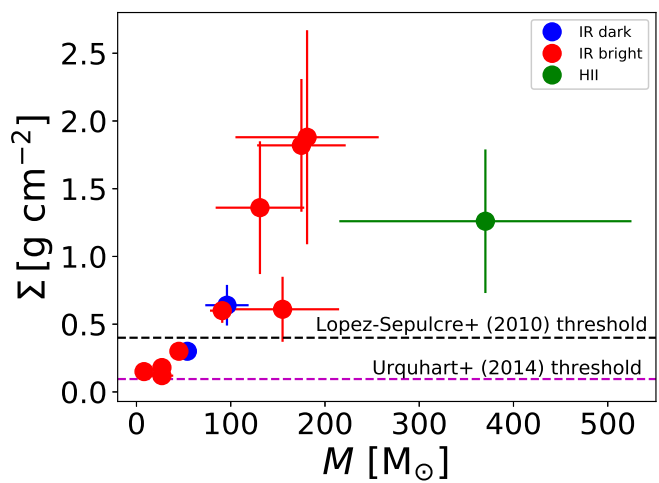}}
\caption{Surface density as a function of core mass. The IR dark cores (no \textit{WISE} counterpart) are shown by blue data points, while the red data points show the IR bright cores (cores with a \textit{WISE} counterpart). The \ion{H}{ii} region IRAS~13037-6112a is indicated by a green data point. The two horizontal dashed lines show the threshold surface densities for high-mass star formation proposed by L{\'o}pez-Sepulcre et al. (2010; black) and Urquhart et al. (2014; magenta) when scaled to the present assumptions about the dust properties (see Sect.~4.2).}
\label{figure:Sigmavsmass}
\end{figure}

The core radii and derived densities are listed in the last three columns in Table~\ref{table:properties}. To calculate the values of $n({\rm H_2})$ and $\Sigma$, we used the sum of the masses of the cold and warm component. The uncertainties in $n({\rm H_2})$ and $\Sigma$ were propagated from the mass uncertainties. The surface densities are plotted as a function of core mass in Fig.~\ref{figure:Sigmavsmass}.

\begin{table*}
\caption{SED parameters and physical properties of the sources.}
\begin{minipage}{2\columnwidth}
\centering
\renewcommand{\footnoterule}{}
\label{table:properties}
\begin{tabular}{c c c c c c c c c c}
\hline\hline
Source & $n_{\rm SED}^{\rm fit}$ & $\chi_{\rm red}^2$ & $\lambda_{\rm peak}$ & $T_{\rm dust}$ & $M$ & $L$ & $R$ & $n({\rm H_2})$ & $\Sigma$\\
       & & & [$\mu$m] & [K] & [M$_{\sun}$] & [L$_{\sun}$] & [pc] & [$10^5$ cm$^{-3}$] & [g cm$^{-2}$] \\
\hline 
SMM 1a & 4 & 6.82 & 199.3 & $14.8\pm0.2$ & $54\pm8$ & $39^{+10}_{-9}$ & 0.11 & $1.4\pm0.2$ & $0.30\pm0.04$\\
SMM 1b & 5 & 5.28 & 230.8 & $13.0\pm0.3$ & $91\pm13$ & $47^{+12}_{-10}$ & 0.10 & $3.1\pm0.4$ & $0.60\pm0.09$ \\
       &   &      &       & $60.9\pm0.7$ & $0.01\pm0.001$ & \\ 
SMM 2 &	6 & 615.90 & 333.4 & $8.8\pm0.9$ & $175\pm47$ & $17^{+11}_{-7}$ & 0.08 & $11.7\pm3.1$ & $1.82\pm0.49$\\ 
      &   &      &       & $28.8\pm0.8$ & $0.3\pm0.1$ & \\
SMM 3 &	4 & 8.21 & 183.9 & $16.1\pm0.4$ & $45\pm8$ & $52^{+19}_{-15}$ & 0.10 & $1.5\pm0.3$ & $0.30\pm0.05$\\ 
SMM 4a & 4 & \ldots\tablefootmark{a} & 272.8 & $10.3\pm1.7$ & $155\pm60$ & $55^{+73}_{-37}$ & 0.13 & $2.4\pm0.9$ & $0.61\pm0.24$\\ 
      &   &      &       & $42.3\pm1.9$ & $0.12\pm0.06$ & \\
IRAS 13037-6112a & 8 & 12.17 & 125.0 & $21.8\pm0.9$ & $27\pm5$ & $491^{+238}_{-181}$ & 0.10 & $0.9\pm0.2$ & $0.18\pm0.03$\\ 
      &   &      &       & $62.8\pm1.6$ & $0.1\pm0.02$ & \\					
IRAS 13037-6112b & 6 & 13.03 & 50.0 & $8.9\pm1.3$ & $131\pm47$ & $170^{+55}_{-43}$ & 0.08 & $8.8\pm3.1$ & $1.36\pm0.49$ \\ 
      &   &      &       & $59.3\pm0.9$ & $0.1\pm0.01$ & \\		
SMM 6a & 4 & \ldots\tablefootmark{a} & 157.9 & $17.6\pm3.6$ & $27\pm12$ & $138^{+245}_{-96}$ & 0.12 & $0.5\pm0.2$ & $0.12\pm0.06$\\ 
      &   &      &       & $51.2\pm2.0$ & $0.1\pm0.04$ & \\		
SMM 7 &	5 & 399.85 & 115.4 & $8.0\pm1.2$ & $181\pm76$ & $19^{+39}_{-16}$ & 0.08 & $12.1\pm5.1$ & $1.88\pm0.79$\\ 
      &   &      &       & $25.6\pm2.6$ & $0.8\pm0.6$ & \\
IRAS 13039-6108a & 5 & 3.77 & 58.8 & $8.3\pm1.3$ & $369\pm155$ & $1\,145^{+435}_{-356}$ & 0.14 & $4.6\pm1.9$ & $1.26\pm0.53$ \\ 
     &   &      &       & $50.2\pm0.9$ & $1.2\pm0.3$ & \\
SMM 9 &	5 & 12.34 & 187.5 & $12.2\pm1.0$ & $93\pm23$ & $68^{+41}_{-26}$ & 0.10 & $3.3\pm0.8$ & $0.64\pm0.15$\\ 
     &   &      &       & $24.8\pm0.7$ & $3\pm1$ & \\
IRAS 13042-6105 & 7 & 3.58 & 150.0 & $19.6\pm0.7$ & $8\pm2$ & $68^{+25}_{-19}$ & 0.06 & $1.3\pm0.3$ & $0.15\pm0.04$\\  
    &   &      &       & $64.0\pm0.9$ & $0.01\pm0.001$ & \\
\hline
\end{tabular} 
\tablefoot{The parameters given in the table are the number of flux density data points used in the SED fit, reduced $\chi^2$ value defined in Eq.~(\ref{eqn:chi}), wavelength of the peak position of the fitted SED, dust temperature, total (gas+dust) mass, luminosity (Eq.~(\ref{eqn:luminosity})), effective core radius taken from Miettinen (2018), volume-averaged H$_2$ number density (Eq.~(\ref{eqn:density})), and surface density (Eq.~(\ref{eqn:surface})). For the sources whose SED was fitted with a two-temperature model, the dust temperature and mass of the warm component are reported beneath the cold component values.\tablefoottext{a}{The number of flux density data points is equal to the number of free parameters of the model, and hence there are zero degrees of freedom. Therefore, the value of $\chi_{\rm red}^2$ becomes infinite.}}
\end{minipage} 
\end{table*}

\subsection{Virial parameter analysis}

To study the dynamical state of the cores, we first calculated their virial masses, $M_{\rm vir}$ (\cite{bertoldi1992}). The cores 
were assumed to be spherical with a radial density profile of the form $n(r)\propto r^{-p}$, where the density power-law index, $p$, was  used to calculate the correction factor for the effects of a non-uniform density distribution as described in Appendix~A in Bertoldi \& McKee (1992). We adopted a value of $p=1.6$, which corresponds to the mean value derived by Beuther et al. (2002) for their sample of high-mass star-forming clumps. We note that comparable values were found in other studies of high-mass star-forming clumps and cores (e.g. \cite{mueller2002}: $\langle p \rangle=1.8\pm0.4$; \cite{garay2007}: $p=1.5-2.2$; \cite{zhang2009}: $p=1.6$ and $p=2.1$ for two cores in the IRDC G28.34+0.06; \cite{li2019}: $p=0.6-2.1$ with $\langle p \rangle=1.3$). For example, for the $p$ values in the range of $1.5-2$, a given virial mass varies by 34\%.

The total (thermal plus non-thermal), one-dimensional velocity dispersion needed in the calculation of $M_{\rm vir}$
was calculated as described in Fuller \& Myers (1992; see Eq.~(3) therein). In this calculation, we assumed that the gas and dust are collisionally coupled and characterised by the same temperature, that is the gas kinetic temperature was taken to be $T_{\rm kin}=T_{\rm dust}$, which is expected to happen at densities $n({\rm H_2})\gtrsim3.2\times10^4$~cm$^{-3}$ (\cite{goldsmith2001}). Furthermore, we used the temperature of the cold dust component because it dominates the mass in each target core (see Sect.~4.2). 

In the present study, we have adopted a $[{\rm He}]/[{\rm H}]$ abundance ratio of $Y/(4X)=0.095$, which together with the assumption that all hydrogen is molecular leads to a mean molecular weight per free particle of $\mu_{\rm p}=2.37$. We note that the very often used value of $\mu_{\rm p}=2.33$ applies for a $[{\rm He}]/[{\rm H}]$ ratio of 0.1 with no metals (see \cite{kauffmann2008}). As the observed spectral linewidth we used the FWHM of the H$^{13}$CO$^+(J=2-1)$ rotational line detected by Miettinen (2020) towards all the parent clumps of the present target cores. The H$^{13}$CO$^+$ linewidths were derived through fitting the hyperfine structure of the line, and the critical density of the $J=2-1$ transition is $n_{\rm crit}=(2.6-3.1)\times10^6$~cm$^{-3}$ at temperatures between 10~K and 20~K (on the basis of the Einstein $A$ coefficient and collision rates given in the Leiden Atomic and Molecular Database (LAMDA\footnote{\url{http://home.strw.leidenuniv.nl/~moldata/}}; \cite{schoier2005})). Hence, H$^{13}$CO$^+(J=2-1)$ is well suited for our purpose as a high-density tracer. In those cases where the Seahore clump is hosting multiple cores (e.g. SMM~1), the same linewidth was adopted for all cores (e.g. for SMM~1a and 1b; see Table~\ref{table:virial}). We note, however, that the spectral lines on the core scale are expected to be narrower than on the larger clump scale owing to the dissipation of non-thermal, turbulent motions (e.g. \cite{myers1983}; \cite{pineda2010}; \cite{sokolov2018}), and a lower level of gas velocity dispersion would lead to a weaker internal kinetic pressure. The total velocity dispersion also depends on the molecular weight of the observed molecular species, which for H$^{13}$CO$^+$ is $\mu_{\rm mol}=30$.

The virial masses of the cores were used to calculate their virial parameters, $\alpha_{\rm vir}=M_{\rm vir}/M$ (\cite{bertoldi1992}; Eq.~(2.8a) therein). The value of $\alpha_{\rm vir}$ quantifies the relative importance of the core's internal kinetic energy and its gravitational energy, and it can be interpreted so that non-magnetised cores with $\alpha_{\rm vir}<2$ are gravitationally bound, those with $\alpha_{\rm vir}=1$ are in virial equilibrium, and when $\alpha_{\rm vir}<1$ the core is gravitationally unstable (e.g. \cite{russeil2010}; \cite{li2019}).

\begin{figure}[!htb]
\centering
\resizebox{\hsize}{!}{\includegraphics{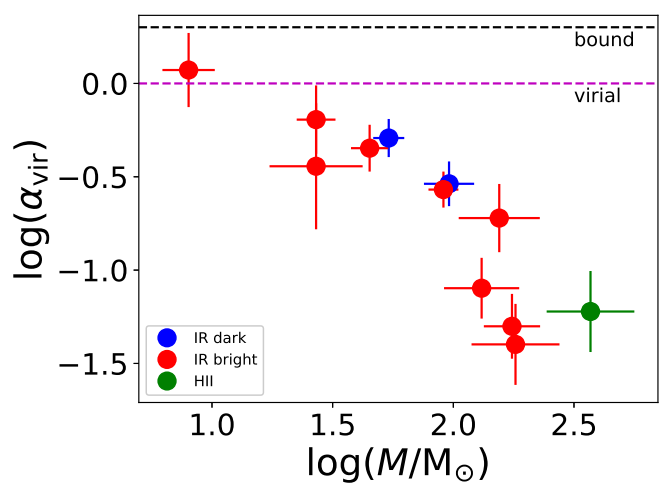}}
\caption{Virial parameter as a function of core mass. The two horizontal dashed lines show the virial equilibrium value ($\alpha_{\rm vir}=1$; magenta) and threshold of gravitational boundedness ($\alpha_{\rm vir}=2$; black).}
\label{figure:alphavsmass}
\end{figure}

The aforementioned H$^{13}$CO$^+(J=2-1)$ linewidths, and the derived virial masses and virial parameters are given in Table~\ref{table:virial}. The uncertainties in $M_{\rm vir}$ were propagated from the temperature and linewidth uncertainties, and the uncertainties in $\alpha_{\rm vir}$ were propagated from those of $M_{\rm vir}$ and $M$. The values of $\alpha_{\rm vir}$ are plotted as a function of core mass in Fig.~\ref{figure:alphavsmass}.

\begin{table}[H]
\renewcommand{\footnoterule}{}
\caption{H$^{13}$CO$^+(J=2-1)$ linewidths (FWHM), virial masses, and virial parameters of the cores.}
{\normalsize
\begin{minipage}{1\columnwidth}
\centering
\label{table:virial}
\begin{tabular}{c c c c c}
\hline\hline 
Source & $\Delta v$\tablefootmark{a} & $M_{\rm vir}$ & $\alpha_{\rm vir}$ \\
       & [km~s$^{-1}$] & [M$_{\sun}$] &  \\ 
\hline 
SMM 1a & $1.13 \pm 0.03$ & $27\pm5$ & $0.51\pm0.12$\\
SMM 1b & $1.13 \pm 0.03$ & $24\pm4$ & $0.27\pm0.06$\\
SMM 2 & $0.74 \pm 0.07$ & $9\pm2$ & $0.05\pm0.02$\\
SMM 3 &	$0.98 \pm 0.05$ & $20\pm5$ & $0.45\pm0.13$\\ 
SMM 4a & $1.11 \pm 0.03$ & $30\pm4$ & $0.19\pm0.08$\\ 
IRAS 13037-6112a & $0.83 \pm 0.05$ & $17\pm6$ & $0.64\pm0.27$\\ 
IRAS 13037-6112b & $0.83 \pm 0.05$ & $11\pm2$ & $0.08\pm0.03$\\ 
SMM 6a & $0.44 \pm 0.08$ & $10\pm6$ & $0.36\pm0.28$\\ 
SMM 7 &	$0.62 \pm 0.08$ & $7\pm2$ & $0.04\pm0.02$\\ 
IRAS 13039-6108a & $0.92 \pm 0.03$ & $23\pm4$ & $0.06\pm0.03$ \\ 
SMM 9 &	$1.22 \pm 0.01$ & $28\pm4$ & $0.29\pm0.08$ \\ 
IRAS 13042-6105 & $0.79 \pm 0.08$ & $9\pm4$ & $1.18\pm0.54$ \\	
\hline
\end{tabular} 
\tablefoot{\tablefoottext{a}{Taken from Miettinen (2020; Table~A.1 therein).} }
\end{minipage} }
\end{table}

\section{Discussion}

\subsection{Far-infrared to submillimetre SED characteristics of the Seahorse cores}

A modified blackbody composed of one or two temperature components is a simplified model for the continuum emission of a dense molecular cloud core. Emission at 22~$\mu$m is expected to originate in a warmer dust component closer to the central YSO compared to the colder envelope (see e.g. \cite{beuther2010}; \cite{ragan2012} for the context of \textit{Spitzer} 24~$\mu$m emission). Hence, a two-temperature model is expected to be a reasonable assumption for the IR bright cores in our sample. Our SED analysis was also based on the assumption of optically thin dust emission, but this assumption might be invalid at 22~$\mu$m wavelength. However, adding the dust optical thickness, $\tau$, in Eq.~(\ref{eqn:sed}) would introduce a third unknown variable (per temperature component) in the problem ($T_{\rm dust}$, $M$, $\tau$), and owing to the fairly low number of data points in our SEDs, the SED fits would start to be subject to overfitting (i.e. number of unknown parameters is comparable to the number of data points).

The angular resolution of the \textit{Herschel}/PACS data at 70, 100, and 160~$\mu$m is $9\arcsec-13\arcsec$, while  
those of SABOCA and LABOCA are $9\arcsec$ and $19\farcs86$, respectively. Hence, the resolutions from the PACS to the SABOCA wavelength 
are similar within a factor of 1.4, and even the LABOCA resolution is only 2.2 times lower than 
the highest resolution ($9\arcsec$) data used in the SED fits. However, the \textit{IRAS} data have a much lower angular resolution compared to our other data, but the corresponding flux densities still appear to be in fairly good agreement with other measurements (see below).

As shown in Fig.~\ref{figure:seds}, a one or two-temperature modified blackbody could be well-fitted to the SEDs of the 22~$\mu$m dark cores SMM~1a and SMM~9, and also to the SEDs of the 22~$\mu$m bright, YSO hosting cores SMM~1b, 4a, 6a, and IRAS~13037-6112a, 13037-6112b, 13039-6108a, and 13042-6105. Of the aforementioned cores, SMM~1a was the only source where a single-temperature model was fitted to the observed SED. In the case of SMM~4a and SMM~6a, the number of flux density data points is equal to the number of free parameters of the two-component model, and hence there are zero degrees of freedom. Hence, the value of $\chi_{\rm red}^2$ of the best-fit SED for these two sources is infinite.

For the 22~$\mu$m bright cores SMM~2 and SMM~7, the observed 22~$\mu$m flux density is not consistent with a two-temperature model fit, which could be an indication of the presence of another, still warmer dust component, or of the failure of the assumptions used in the analysis, for example that of optically thin dust emission. Moreover, for the 22~$\mu$m bright core SMM~3, a two-temperature model fit yielded two overlapping modified Planck curves that appeared similar to the single-temperature model fit shown in Fig.~\ref{figure:seds}. We opted for the single-temperature fit because it yielded a much lower $\chi_{\rm red}^2$ value (8.21) than the poor two-temperature fit ($\chi_{\rm red}^2=745.44$). Also in the case of SMM~3 a more complex model might be need to explain its 22~$\mu$m emission (more than two dust components or optically thick 22~$\mu$m emisson or both), but we note that the 70~$\mu$m data point was not available for SMM~3 so a direct comparison with the cases of SMM~2 and SMM~7 is not feasible.

The SED of IRAS~13037-6112b could not be successfully fit with a two-temperature model when the \textit{IRAS} 60~$\mu$m flux density data point was included in the fit. Hence, we decided to omit that data point from the fit, and this choice was further supported by the presence of a PACS 70~$\mu$m data point that was measured at a much higher (factor of $\sim7$) angular resolution than the \textit{IRAS} 
60~$\mu$m data point. For IRAS~13037-6112a, however, a two-temperature model could be fitted even when the 60~$\mu$m data point was included in the fit, although the best-fit model is in much better agreement with the PACS 70~$\mu$m data point. As seen in the SEDs of the \text{IRAS} sources, the \textit{WISE} 22~$\mu$m and \textit{IRAS} 25~$\mu$m flux densities are in agreement with each other, where the latter are 1.5--1.6 times higher. For IRAS~13039-6108a, a two-temperature model could not fit the \textit{IRAS} 100~$\mu$m data point, and hence we opted for an SED fit where that data point is omitted (the $\chi_{\rm red}^2$ value of the fit also dropped from 9.25 to 3.77 when the 100~$\mu$m data point was not used in the fit). We note that for IRAS~13039-6108a, which is associated with an \ion{H}{ii} region, no counterparts were found from the PACS Point Source Catalogue, which is probably the result of the extended source appearance at the PACS wavelengths (70~$\mu$m--160~$\mu$m; Fig.~\ref{figure:images}). This could explain the high \textit{IRAS} 100~$\mu$m flux density of the source. The source is also extended and associated with diffuse emission region in the \textit{WISE} 12~$\mu$m and 22~$\mu$m images as shown in Fig.~3 in Miettinen (2018) and Fig.~\ref{figure:images} herein, which is consistent with the \ion{H}{ii} region state of source evolution where a photodissociation region is surrounding the ionised gas bubble. We also note that for IRAS~13037-6112a, 13037-6112b, and 13042-6105, the SED fits are consistent with the \textit{IRAS} flux density upper limits. 

One could argue that the adopted Ossenkopf \& Henning (1994) dust model of grains with thin ice mantles is not valid for the warm dust component. For example, the adopted dust opacities in a wavelength range of 20~$\mu$m--100~$\mu$m, which typically brackets the peak of the warm SED component, are on average about $1.6\pm0.1$ times higher than those in the corresponding model of grains without ice mantles. Nevertheless, the usage of the Ossenkopf \& Henning (1994) model of bare dust particles with no icy coverages would basically only lead to about 1.6 times higher masses of the warm dust component. Because the latter masses are so low compared to the cold component ($M_{\rm warm}/M_{\rm cold}=1\times10^{-4}-0.03$ with a mean of $5\times10^{-3}$), the usage of the thin ice mantle based parameters seems justified also for the warm component.

\subsection{Physical properties of the Seahorse cores and the potential for high-mass star formation}

The mean (median) values of the derived dust temperatures of the cold component, core masses, luminosities, number densities, and surface densities are $13.3\pm1.4$~K (12.6~K), $113\pm29$~M$_{\sun}$ (94~M$_{\sun}$), $192\pm94$~L$_{\sun}$ (62~L$_{\sun}$), $(4.3\pm1.2)\times10^5$~cm$^{-3}$ ($2.8\times10^5$~cm$^{-3}$), and $0.77\pm0.19$~g~cm$^{-2}$ (0.61~g~cm$^{-2}$). The mean (median) dust temperature and mass of the warm component are $47.0\pm5.0$~K (50.7~K) and $0.6\pm0.3$~M$_{\sun}$ ($0.1$~M$_{\sun}$). As mentioned in Sect.~4.1, the mass of the warm component is only 5 per mille of the cold component's mass on average (the median $M_{\rm warm}/M_{\rm cold}$ ratio is $2.5\times10^{-3}$). As expected, the \ion{H}{ii} region source IRAS~13039-6108a was found to be the most luminous source in our sample ($L=(1.1\pm0.4)\times10^3$~L$_{\sun}$), but surprisingly the second lowest dust temperature of the cold component in our sample was derived for IRAS~13039-6108a ($T_{\rm dust}=8.3\pm1.3$~K). IRAS~13039-6108a also does not show the highest warm component temperature in our sample, but there are five cores that are still warmer ($51.2-64.0$~K compared to 50.2~K for IRAS~13039-6108a). 

\begin{figure}[!htb]
\centering
\resizebox{\hsize}{!}{\includegraphics{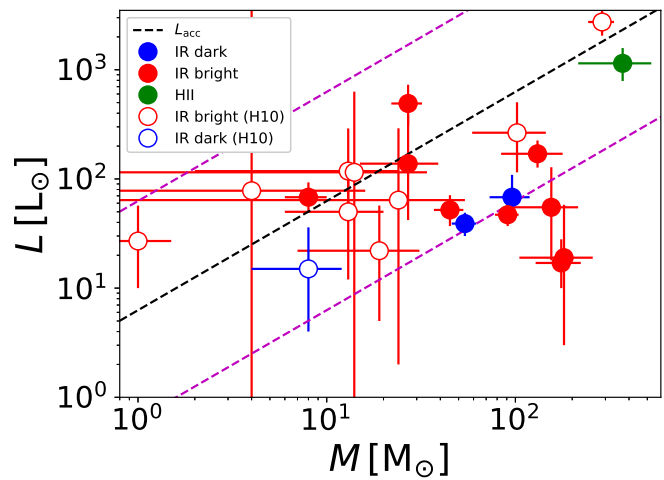}}
\caption{Core luminosity plotted against the core mass. For comparison, a sample of ten IR dark and IR bright cores in the Snake IRDC from Henning et al. (2010) are shown with empty circles (see Sect.~4.3 and Appendix~B). The black dashed line shows the accretion luminosity with a mass accretion rate of $\dot{M}_{\rm acc}=10^{-5}$~M$_{\sun}$~yr$^{-1}$ and assuming a stellar mass of $M_{\star}=0.1\times M_{\rm core}$ and stellar radius of $R_{\star}=5$~R$_{\sun}$ (see Eq.~(\ref{eqn:accretion})). The magenta dashed lines above and below the aforementioned line show a $\pm1$~dex variation of the accretion luminosity in question.}
\label{figure:luminosityvsmass}
\end{figure}

In Fig.~\ref{figure:luminosityvsmass}, we plot the core luminosity as a function of core mass. The black dashed line plotted in Fig.~\ref{figure:luminosityvsmass} indicates the accretion luminosity defined as 

\begin{equation}
\label{eqn:accretion}
L_{\rm acc}=\frac{G\dot{M}_{\rm acc}M_{\star}}{R_{\star}}\,, 
\end{equation} 
where $\dot{M}_{\rm acc}$ is the mass accretion rate and $M_{\star}$ and $R_{\star}$ are the mass and radius 
of the central star. Following Ragan et al. (2012, Fig.~14 therein), we assumed that $\dot{M}_{\rm acc}=10^{-5}$~M$_{\sun}$~yr$^{-1}$,  
the stellar mass is 10\% of the parent core mass, and that $R_{\star}=5$~R$_{\sun}$. In general, the YSO evolution in
the $L-M$ diagram shown in Fig.~\ref{figure:luminosityvsmass} occurs towards the top left corner (i.e. the parent core or envelope mass decreases while the luminosity increases; e.g. \cite{saraceno1996}; \cite{molinari2008}). Two IR bright cores in our sample lie close (within a factor of $\sim1.4$) to the aforementioned accretion luminosity trend. Also, the \ion{H}{ii} region IRAS~13039-6108a lies only a factor of 2 below that trend line. Most of the remaining cores lie clearly below that trend, for example the two 22~$\mu$m dark cores
lie approximately one dex below the trend, which could be an indication of a lower mass accretion rate (and/or lower stellar mass and larger central star). Only IRAS 13037-6112a is found to lie (by a factor of 2.9) above the $L_{\rm acc}$ line corresponding to an accretion rate of 
$10^{-5}$~M$_{\sun}$~yr$^{-1}$. The latter rate is not sufficient to overcome the radiation pressure barrier of high-mass star formation, which requires values of $\dot{M}_{\rm acc}>10^{-4}$~M$_{\sun}$~yr$^{-1}$, and indeed such high values have been derived for high-mass star-forming objects (e.g. \cite{zinnecker2007}; \cite{motte2018} for reviews, and references therein). See Sect.~4.3 for further discussion. 

\begin{figure}[!htb]
\centering
\resizebox{\hsize}{!}{\includegraphics{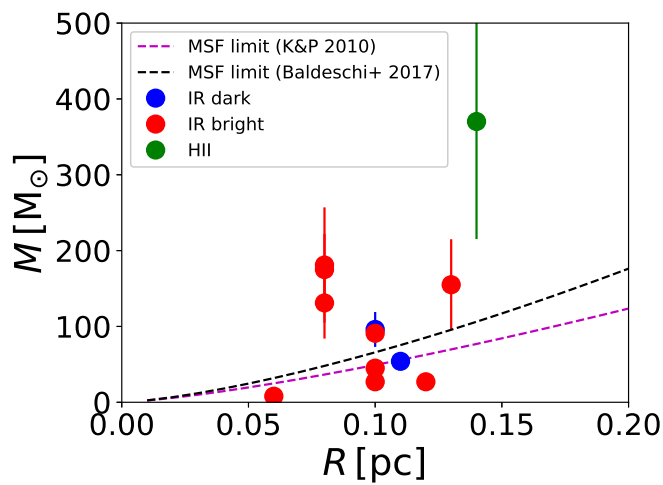}}
\caption{Core mass plotted against the effective core radius. The magenta and black dashed lines indicate the empirical mass-radius thresholds for high-mass star formation proposed by Kauffmann \& Pillai (2010) and Baldeschi et al. (2017), respectively. When scaled to the assumptions adopted in the present study, these thresholds are given by $M_{\rm thresh}(R) = 1\,051\,{\rm M}_{\sun} \times (R/{\rm pc})^{1.33}$ and $M_{\rm thresh}(R) = 1\,732\,{\rm M}_{\sun} \times (R/{\rm pc})^{1.42}$ (see Sect.~4.2 for details).}
\label{figure:massvsradius}
\end{figure}

Figure~\ref{figure:massvsradius} shows the core masses as a function of their effective radius. For comparison, the magenta dashed line plotted in the figure shows the empirical mass-radius threshold for high-mass star formation proposed by Kauffmann \& Pillai (2010); see also Kauffmann et al. (2010b). We note that the authors also used the Ossenkopf \& Henning (1994) dust opacities for grains with thin ice mantles coagulating for $10^5$~yr, but the model density was assumed to be $10^6$~cm$^{-3}$ rather than $10^5$~cm$^{-3}$ as we did. In a wavelength range of 350~$\mu$m--1~mm, the former dust opacities are on average $1.285\pm0.002$ times higher than our values. Moreover, Kauffmann \& Pillai (2010) decreased the opacities by a factor of 1.5 to calibrate between the dust extinction and dust emission based mass estimates (see \cite{kauffmann2010a} for details). We used the original Ossenkopf \& Henning (1994) opacities and a factor of 1.41 higher gas-to-dust mass ratio than what seems to have been used in the reference studies of Kauffmann \& Pillai (2010). Taking these differences into account, the Kauffmann \& Pillai (2010) threshold relationship for high-mass star formation can be written as

\begin{equation}
\label{eqn:threshold}
M_{\rm thresh}(R)=1\,051\,{\rm M}_{\sun} \times \left(\frac{R}{{\rm pc}}\right)^{1.33}\,.
\end{equation} 
Seven out of our 12 cores (58\%) lie above the Kauffmann \& Pillai (2010) threshold for high-mass star formation. These include five IR bright cores, the \ion{H}{ii} region IRAS~13039-6108a, and the 22~$\mu$m dark core SMM~9, where IRAS~13039-6108a lies furthest above the threshold, that is by a factor of $4.8\pm2.0$. Moreover, the IR bright core SMM~3 and IR dark core SMM~1a lie very close to this critical threshold, namely a factor of $1.09\pm0.19$ and $1.03\pm0.15$ below it, respectively. 

Baldeschi et al. (2017) also derived an empirical mass-radius threshold relationship for high-mass star formation (see their Eq.~(9)). The authors assumed that the opacity at 300~$\mu$m is 0.1~cm$^2$~g$^{-1}$, which includes a dust-to-gas mass ratio of 1/100. When scaled to our corresponding assumptions, the Baldeschi et al. (2017) threshold becomes 

\begin{equation}
\label{eqn:threshold2}
M_{\rm thresh}(R)=1\,732\,{\rm M}_{\sun} \times \left(\frac{R}{{\rm pc}}\right)^{1.42}\,.
\end{equation}  
As illustrated in Fig.~\ref{figure:massvsradius}, the Baldeschi et al. (2017) threshold lies above the Kauffmann \& Pillai (2010) threshold by $1.65\times (R/{\rm pc})^{1.0677}$. All the seven cores that lie above the Kauffmann \& Pillai (2010) threshold, also lie above the Baldeschi et al. (2017) threshold. For example, IRAS~13039-6108a fulfils the latter criterion by lying a factor of $3.5\pm1.5$ above it.

As illustrated in Fig.~\ref{figure:alphavsmass}, all the target cores appear to be gravitationally bound ($\alpha_{\rm vir}<2$). We note that the 22~$\mu$m dark cores SMM~1a and SMM~9 could harbour YSO(s) that are too weak to be detected with \textit{WISE} (e.g. the $5\sigma$ point-source sensitivity in the W4 band is 5.4~mJy; \cite{wright2010}), and indeed the PACS 70~$\mu$m observations suggest that the cores already have an ongoing central heating (although the 70~$\mu$m counterpart of SMM~1a could not be found from the PACS catalogue; see Appendix~A). The other way round, SMM~1a and SMM~9 are unlikely to be prestellar cores.

To quantitatively estimate the mass of the most massive star that could form in the target cores, we used the Kroupa (2001) stellar initial mass function based formula from Svoboda et al. (2016, their Eq.~(10); see also \cite{sanhueza2017}, Appendix therein), which is given by

\begin{equation}
\label{eqn:stellar}
M_{\star}^{\rm max}=20\,{\rm M}_{\sun}\times \left(\frac{\epsilon_{\rm SF}}{0.3}\times \frac{M_{\rm core}}{1\,064\,{\rm M}_{\sun}} \right)^{1/1.3}\,,
\end{equation} 
where $\epsilon_{\rm SF}$ is the star formation efficiency (SFE). Adopting a value of $\epsilon_{\rm SF}=0.3$ (e.g. \cite{lada2003} for a review; \cite{alves2007}), Eq.~(\ref{eqn:stellar}) yields maximum stellar masses of $M_{\star}^{\rm max}=0.5\pm0.1 - 5.1\pm1.7$~M$_{\sun}$ for the IR bright cores, $2.0\pm0.2$~M$_{\sun}$ and $3.1\pm0.6$~M$_{\sun}$ for the 22~$\mu$m dark cores SMM~1a and SMM~9, and $8.9\pm2.9$~M$_{\sun}$ for IRAS~13039-6108a. This suggests that the IR bright and IR dark cores in the Seahorse IRDC could give birth to multiple systems of low-mass stars ($\sim0.1-2$~M$_{\sun}$) or intermediate-mass stars ($\sim2-8$~M$_{\sun}$) rather than collapse to form a massive star of at least 8~M$_{\sun}$, which only seems to be the case in the already known high-mass star-forming object IRAS~13039-6108a. We note that according to Eq.~(\ref{eqn:stellar}), high-mass star formation requires at least a 320~M$_{\sun}$ core for $\epsilon_{\rm SF}=0.3$. On the other hand, in the competitive accretion paradigm the cores like SMM~1a and SMM~1b could competitively accrete more mass from their parent clump SMM~1, and this process has the potential to lead to the formation of a massive star at the centre of the gravitational potential well (e.g. \cite{bonnell2006}). 

To further assess our core sample's potential for high-mass star formation, in Fig.~\ref{figure:Sigmavsmass} we plot their surface densities 
as a function of core mass. For comparison, two threshold surface densities for high-mass star formation are also shown in Fig.~\ref{figure:Sigmavsmass}. L{\'o}pez-Sepulcre et al. (2010) determined an empirical surface density threshold for massive star formation of $\Sigma_{\rm thres}=0.3$~g~cm$^{-2}$ on the basis of the outflow characteristics of their sample. The authors assumed that the dust opacity is 1~cm$^2$~g$^{-1}$ at 1.2~mm and that $R_{\rm dg}=1/100$. When scaled to the present assumptions, the L{\'o}pez-Sepulcre et al. (2010) surface density threshold becomes $\Sigma_{\rm thres}=0.4$~g~cm$^{-2}$. On the other hand, on the basis of a large sample of $\sim1\,700$ massive YSOs in $\sim1\,300$ clumps drawn from the APEX Telescope Large Area Survey of the Galaxy (ATLASGAL; \cite{schuller2009}), Urquhart et al. (2014) found that a surface density of 0.05~g~cm$^{-2}$ might set a minimum threshold for high-mass star formation. Again, when taking into account that Urquhart et al. (2014) assumed different dust properties than we ($\kappa_{\rm 870\, \mu m}=1.85$~cm$^2$~g$^{-1}$ and $R_{\rm dg}=1/100$), we scaled their reported surface density threshold to 0.095~g~cm$^{-2}$. Seven (58\%) of the target cores
lie above the L{\'o}pez-Sepulcre et al. (2010) threshold by factors of $1.5\pm0.2 - 4.7\pm2.0$. The latter percentage is the same as for the cores that fulfil the Kauffmann \& Pillai (2010) and Baldeschi et al. (2017) thresholds for high-mass star formation (the cores in question are the same in both cases). All the cores are found to lie above the Urquhart et al. (2014) threshold by factors of $1.3\pm0.6 - 19.8\pm8.3$, where five cores lie in between the two $\Sigma_{\rm thres}$ values.

\subsection{Comparison with the properties of dense cores in the Snake IRDC}

As discussed in Miettinen (2018), the Seahorse IRDC shares some similarities with the Snake IRDC G11.11-0.12. Both the clouds are 
filamentary in their projected morphology, and they appear to have been hierarchically fragmented into substructures (e.g. \cite{ragan2015}). 
Altogether 17 cores were detected with SABOCA in the Seahorse IRDC, and three LABOCA clumps (SMM~5, SMM~8, and BLOB~1) were resolved out in the SABOCA map (\cite{miettinen2018}). Because 14 of these 20 sources are associated with a \textit{WISE} 22~$\mu$m source, the percentage of 22~$\mu$m bright cores is $70\%\pm19\%$, where the quoted uncertainty represents the Poisson error on counting statistics (here, SMM~7 is counted as IR bright opposite to the studies by Miettinen (2018, 2020); see Table~\ref{table:sample} and Appendix~A). Henning et al. (2010) found that 11 out of their 18 cores that are found along the Snake filament are associated with \textit{Spitzer} 24~$\mu$m emission, which makes the percentage of IR bright cores $61\%\pm18\%$, which is comparable to that in the Seahorse. Accounting for the off-filament cores in the Snake, the aforementioned percentage becomes even more similar to ours (i.e. $67\%\pm17\%$; \cite{henning2010}, Table~1 therein). The differences are that the Snake IRDC lies at a distance of 3.6~kpc (i.e. a factor of 1.4 further away than the Seahorse) and has a projected length of about 30~pc, and it is also very massive, $\sim10^5$~M$_{\sun}$ (\cite{kainulainen2013}; \cite{lin2017}). The Seahorse filament is about 9~pc long and has a mass of $\sim10^3$~M$_{\sun}$ (\cite{miettinen2018}). 

We note that Henning et al. (2010) applied a similar SED analysis to derive the Snake IRDC's core properties as in the present study (i.e. modifed blackbody fitting), and hence we took their sample as our main comparison sample of IRDC cores. However, for a better comparison with the present results, we re-analysed the SEDs of the Snake IRDC cores using the same method and assumptions as in the present study (see Appendix~B for details).

In Fig.~\ref{figure:distributions}, we show the distributions of dust temperature, mass, and luminosity of the cores in both the Seahorse IRDC and the Snake IRDC, where the latter values are listed in Table~\ref{table:snake} (the ten out of 18 on-filament sources for which SED analysis could be done). The mean (median) values of these quantities for the Snake cores are $\langle T_{\rm dust}^{\rm cold}\rangle=19.6\pm1.1$~K (19~K), $\langle T_{\rm dust}^{\rm warm}\rangle=50.9\pm3.0$~K (50~K), $\langle M\rangle = 49\pm28$~M$_{\sun}$ (14~M$_{\sun}$), and $\langle L\rangle =349\pm267$~L$_{\sun}$ (71~L$_{\sun}$). These values are $1.5\pm0.2$ (1.5), $1.1\pm0.1$ (1.0), $0.4\pm0.3$ (0.1), and $1.8\pm1.6$ (1.2) times those for the Seahorse cores (see Sect.~4.2). The Snake cores appear somewhat warmer (in the cold component), less massive (by a factor of $2.5\pm1.9$ on average), and about 80\% more luminous on average (while the median luminosities differ by only a factor of 1.2). However, at least partly these differences can be attributed to our inclusion of the longer wavelength data (350~$\mu$m and 870~$\mu$m), which leads to lower temperatures of the cold componenent (hence higher mass).

Interestingly, also the Snake IRDC contains one IR bright core, the P1 core (i.e. core no.~9), with a luminosity of about 
$2.7^{+0.8}_{-0.7}\times10^3$~L$_{\sun}$, which is comparable ($2.5^{+2.5}_{-1.2}$ times larger) to that we derived for IRAS~13039-6108a. The Snake P1 also has a mass similar to that of IRAS~13039-6108a (a factor of $1.3\pm0.6$ difference). We note that the Snake P1 is known to be associaated with 6.7~GHz Class~II methanol (CH$_3$OH) maser and 22~GHz water (H$_2$O) maser emission, which are signposts of high-mass star formation (\cite{pillai2006}; \cite{wang2014}).

In Fig.~\ref{figure:luminosityvsmass}, we also plot the luminosities and masses of the analysed core sample in the Snake IRDC filament. Three out of the nine IR bright Snake cores (33.3\%) lie close (within a factor of 1.5) to the line of accretion luminosity with $\dot{M}_{\rm acc}=10^{-5}$~M$_{\sun}$~yr$^{-1}$, although we note that the associated uncertainties are large. This is comparable to the corresponding percentage in our sample, that is two out of nine (22\%) IR bright cores lie within a factor of $\sim1.4$ of the trend; see Sect.~4.2. The one analysed IR dark core in the Snake filament lies within 1~dex below the aforementioned $L-M$ relationship, just like the two IR (22~$\mu$m) dark cores in our sample. Overall, the Seahorse and Snake cores' properties suggest that the Seahorse and Snake IRDCs are in comparable evolutionary stages, at least in terms of their star formation phase.

\begin{figure*}[!htb]
\begin{center}
\includegraphics[width=0.4\textwidth]{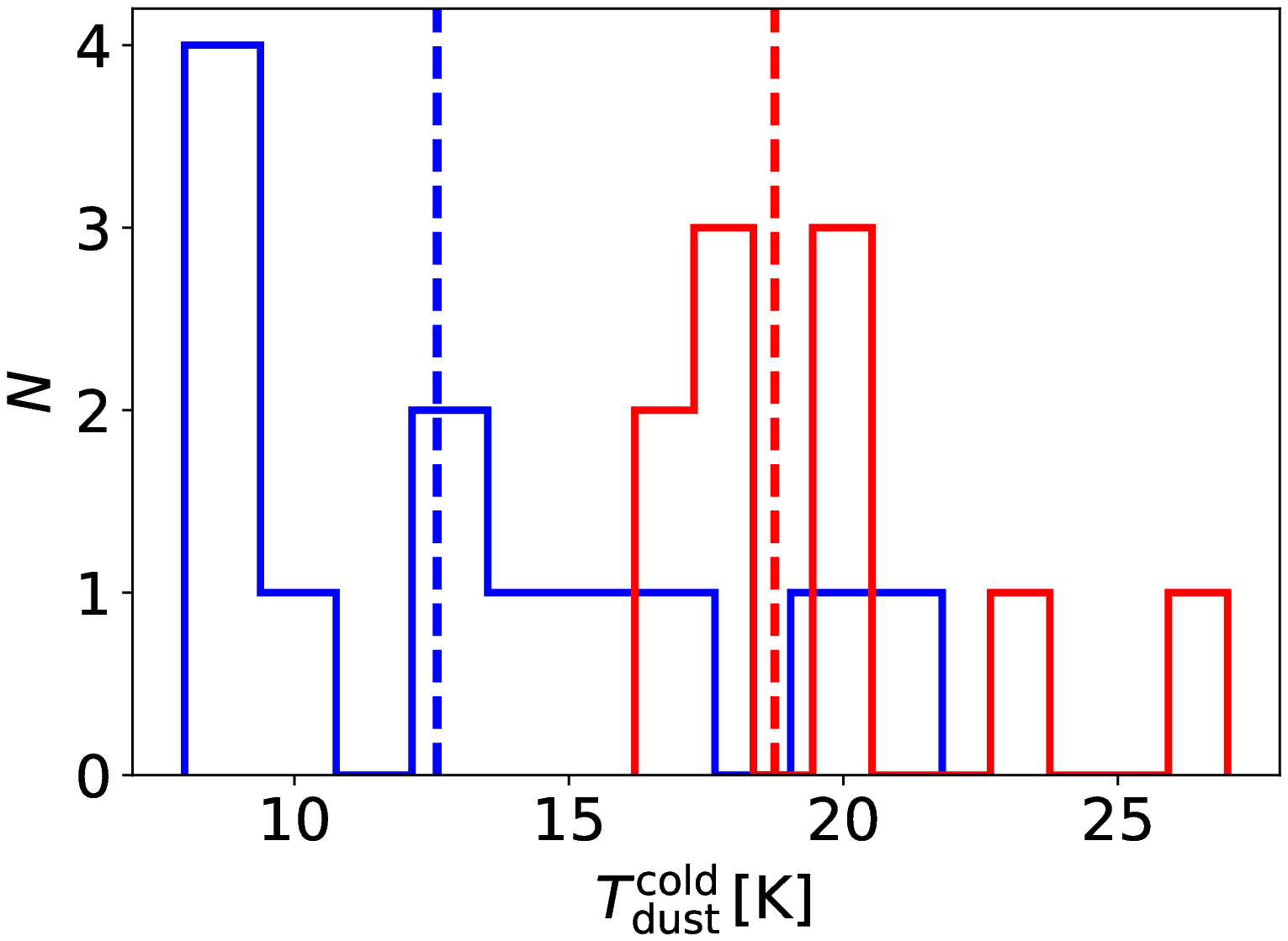}
\includegraphics[width=0.4\textwidth]{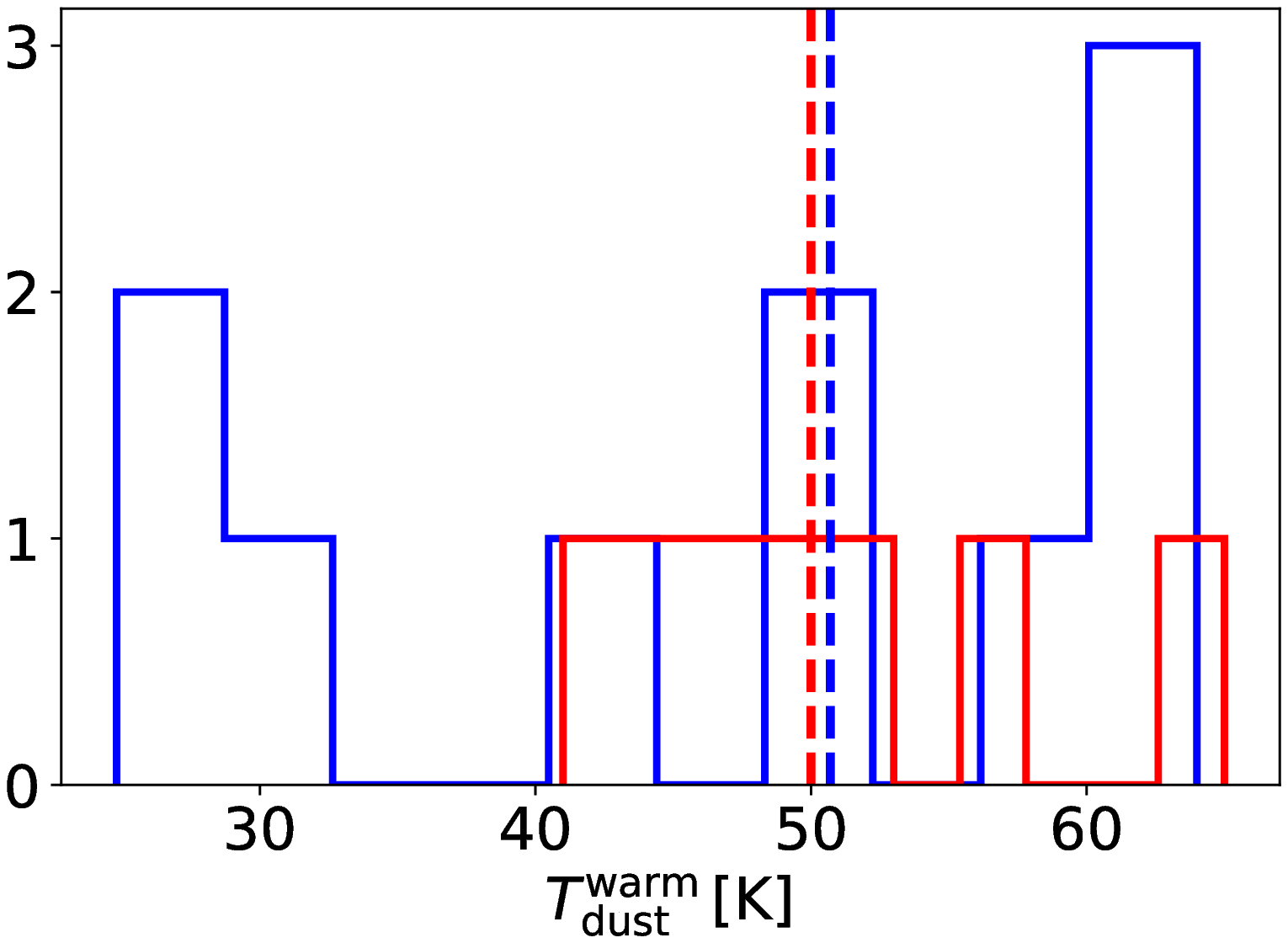}
\includegraphics[width=0.4\textwidth]{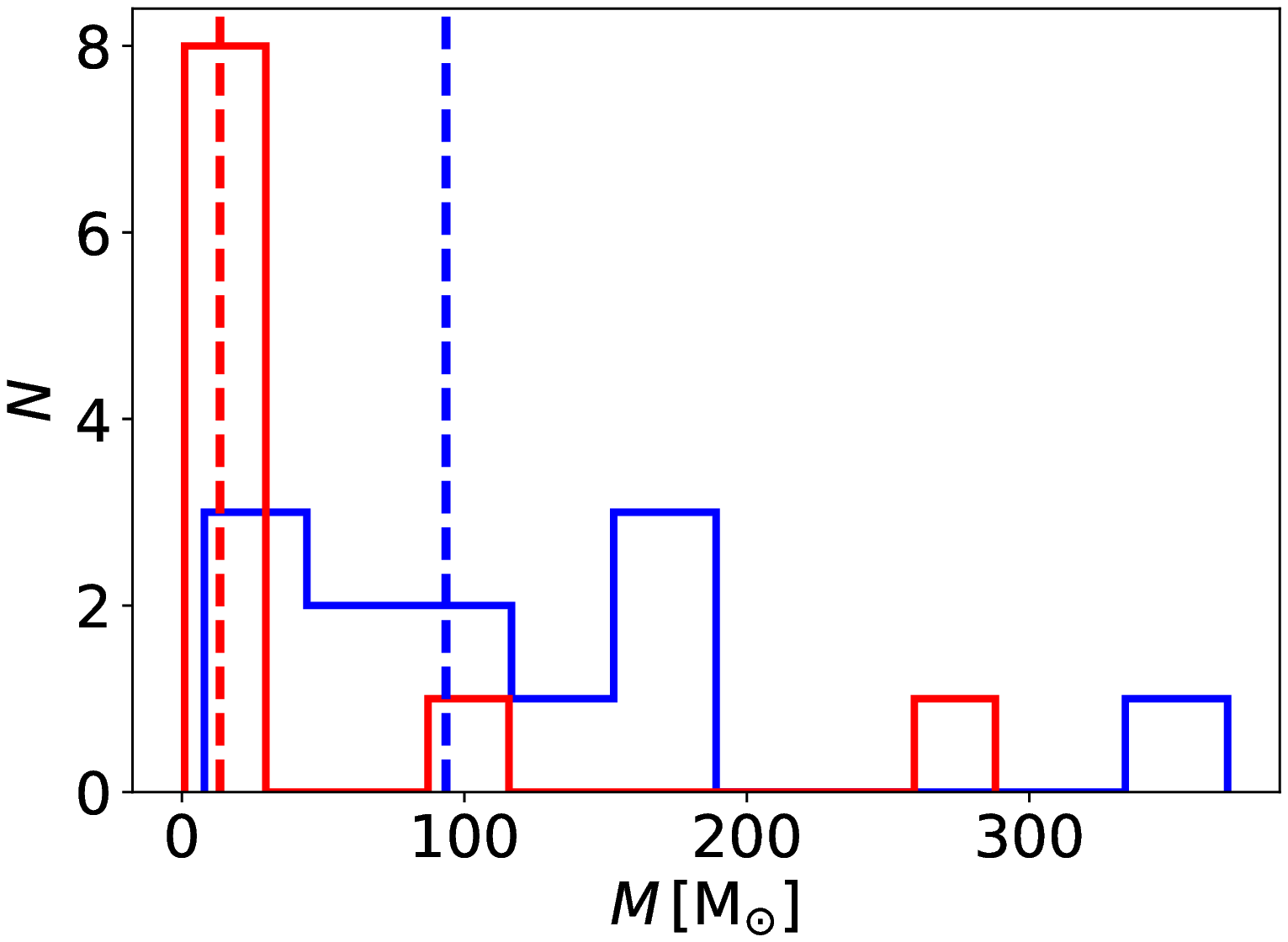}
\includegraphics[width=0.4\textwidth]{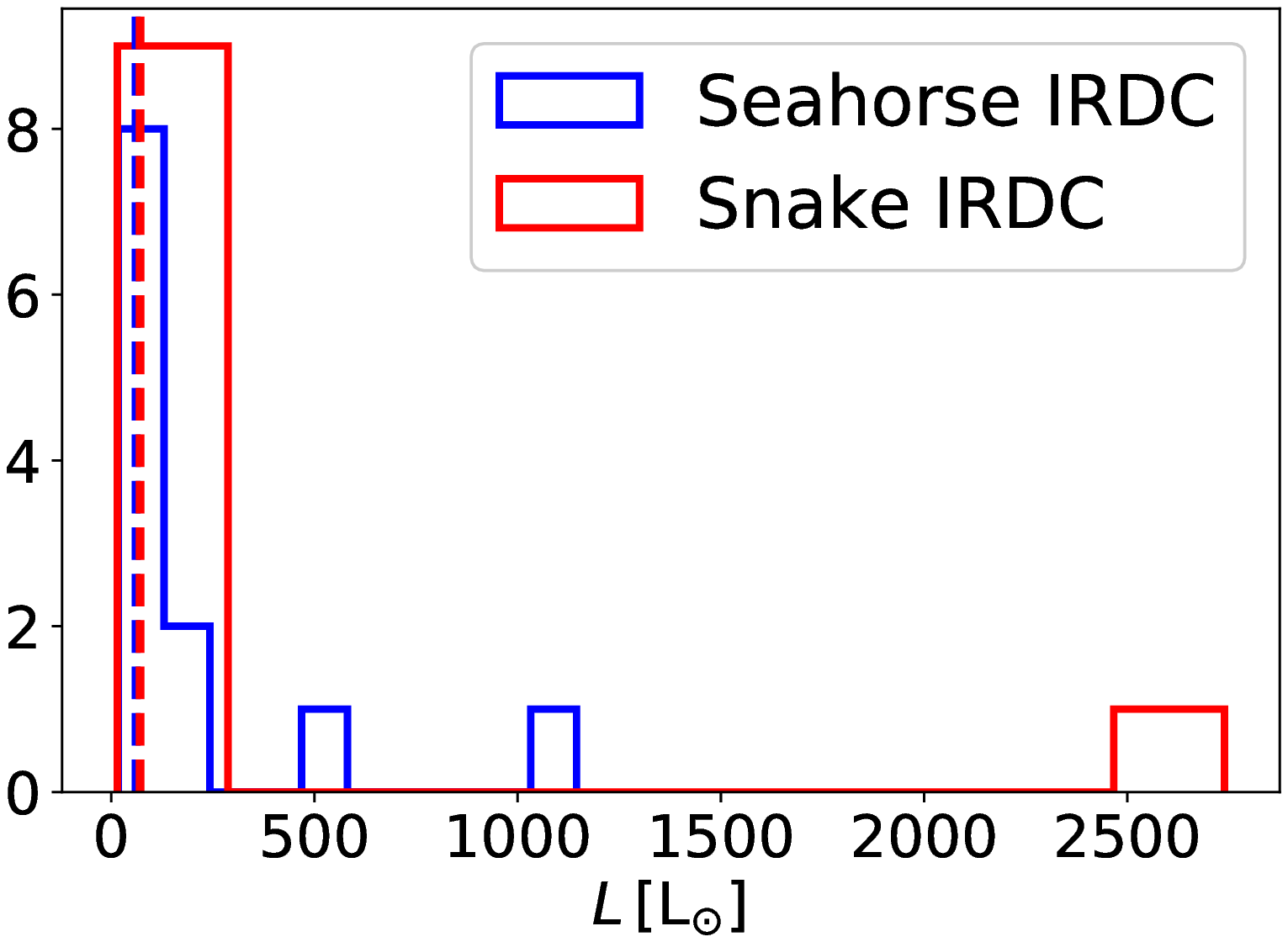}
\caption{Distributions of dust temperature, mass, and luminosity of the cores in the Seahorse IRDC (blue histogram; present study) and Snake IRDC (red histogram; \cite{henning2010}; Appendix~B herein). The vertical dashed lines indicate the sample medians.}
\label{figure:distributions}
\end{center}
\end{figure*}

\subsection{Core fragmentation}

In the paper by Miettinen (2018), we studied the fragmention of the clumps in the Seahorse IRDC into cores, and in this section 
we address the fragmention of the cores into still smaller units. As discussed in Appendix~A, five out of the 12 analysed cores (42\%) show evidence of fragmentation in the SABOCA 350~$\mu$m image. The observed, projected separations between the fragments range from 0.09~pc in SMM~7 to 0.21~pc in SMM~4a. We note that all these systems were extracted as single sources in the source extraction analysis by Miettinen (2018). 

The observed fragment separations can be compared with the thermal Jeans length of the core, $\lambda_{\rm J}\propto T_{\rm kin}^{1/2}\rho^{-1/2}$ (e.g. \cite{mckee2007} for a review, Eq.~(20) therein). Again, we assumed that $T_{\rm kin}=T_{\rm dust}$ (for the cold component), and the mass density needed in the calculation was defined as $\rho=\mu_{\rm H_2}m_{\rm H}n({\rm H_2})$ to be consistent with Eq.~(\ref{eqn:density}). The thermal Jeans mass of the cores was calculated using the definition in McKee \& Ostriker (2007), that is 
$M_{\rm J} \propto (\lambda_{\rm J}/2)^3\rho$.

The SED based core masses were compared with the Jeans mass by calculating the Jeans number, $N_{\rm J}=M/M_{\rm J}$. The aforementioned fragmentation parameters are listed in Table~\ref{table:fragmentation}. The uncertainties in the parameters were propagated from those in the temperature, density, and mass. 

As seen in Table~\ref{table:fragmentation}, the observed fragment separation in SMM~6a is comparable to its thermal Jeans length (the ratio between the two is $1.4\pm0.3$). For the remaining four cores, the observed fragment separation is $3.0\pm0.2$ to $6.4\pm1.4$ times the thermal Jeans length, which could be an indication that the derived temperature of the cold dust component underestimates the parent core temperature (e.g. $8.3\pm1.3$~K for IRAS~13039-6108a seems very low) or that also non-thermal motions contribute to the Jeans instability. For these sources, we also calculated the Jeans length where the thermal sound speed, $c_{\rm s}$, is replaced by the effective sound speed defined as $c_{\rm s,\, eff}^2=c_{\rm s}^2+\sigma_{\rm NT}^2$ (e.g. \cite{bonazzola1987}; \cite{maclow2004} for a review). The one-dimensional non-thermal velocity dispersion is often written under the assumption of isotropicity as $\sigma_{\rm NT}^2=1/3 \times {\rm v}_{\rm rms}^2$ with ${\rm v}_{\rm rms}$ the three-dimensional, rms turbulent velocity. Using the H$^{13}$CO$^+(J=2-1)$ linewidths (column~(2) in Table~\ref{table:virial}), the effective sound speed increases the Jeans lenghts in SMM~1a, 4a, and 7 and IRAS~13039-6108a to 0.14~pc, 0.11~pc, 0.03~pc, and 0.06~pc, respectively. These are 1.5--3 times the corresponding thermal Jeans lengths, and 33\%--74\% of the observed fragment separation. We note that magnetic pressure would also contribute to the effective sound speed (e.g. \cite{mckee2003}), and could also contribute to the core fragmentation.

As seen in column~(6) in Table~\ref{table:fragmentation}, the cores contain multiple thermal Jeans masses, although the Jeans numbers are associated with significant uncertainties. The highest Jeans numbers are found for SMM~1a, 4a, and 7 and IRAS~13039-6108a, but those would be decreased by factors of 3.4--21 (e.g. to $N_{\rm J}=163\pm145$ for IRAS~13039-6108a) if the non-thermal Jeans lengths are used to derive the corresponding Jeans masses. Nevertheless, it is possible that the studied cores are fragmented into multiple condensations and this hypothesis can be tested by high-resolution observations with telescopes like Atacama Large Millimetre/submillimetre Array (ALMA; \cite{wootten2009}). 

In conclusions, the analysed cores exhibit heterogeneous fragmentation properties so that SMM~6a is consistent with pure thermal Jeans fragmentation, while the other four cores listed in Table~\ref{table:fragmentation} appear to require additional mechanism(s) in their fragmentation in addition to gravity and thermal pressure support. Despite the physical mechanisms of how the cores were fragmented, their physical properties derived from the SEDs should be interpreted as those of the systems of at least two fragments, and this has direct consequence to the cores' ability to form either massive stars or multiple stellar systems.  

\begin{table*}
\caption{Core fragment separations and the Jeans analysis parameters.}
\begin{minipage}{2\columnwidth}
\centering
\renewcommand{\footnoterule}{}
\label{table:fragmentation}
\begin{tabular}{c c c c c c}
\hline\hline
Source & $d_{\rm sep}$ & $\lambda_{\rm J}$ & $d_{\rm sep}/\lambda_{\rm J}$ & $M_{\rm J}$ & $N_{\rm J}$\\
       & [pc] & [pc] &  & [M$_{\sun}$] & \\
\hline 
SMM 1a & 0.19 & $0.06\pm0.004$ & $3.0\pm0.2$ & $1.2\pm0.3$ & $44\pm13$\\
SMM 4a & 0.21 & $0.04\pm0.01$ & $5.2\pm1.1$ & $0.5\pm0.4$ & $289\pm234$\\
SMM 6a & 0.16 & $0.11\pm0.03$ & $1.4\pm0.3$ & $2.7\pm2.1$ & $10\pm9$\\
SMM 7 & 0.09 & $0.02\pm0.003$ & $5.9\pm1.3$ & $0.2\pm0.1$ & $1\,095\pm982$\\
IRAS 13039-6108a & 0.17 & $0.03\pm0.006$ & $6.4\pm1.4$ & $0.3\pm0.2$ & $1\,307 \pm 1\,158$\\
\hline
\end{tabular} 
\tablefoot{The listed parameters are the observed, projected separation between the core fragments ($d_{\rm sep}$), thermal Jeans length ($\lambda_{\rm J}$), ratio between the latter two separations ($d_{\rm sep}/\lambda_{\rm J}$), thermal Jeans mass ($M_{\rm J}$), and the Jeans number defined as $N_{\rm J}=M/M_{\rm J}$.}
\end{minipage} 
\end{table*}

\section{Summary and conclusions}

We used data from the \textit{WISE}, \textit{IRAS}, and \textit{Herschel} satellites together with our previous submillimetre 
dust continuum observations with the APEX telescope to construct the far-IR to submillimetre SEDs of the SABOCA 350~$\mu$m selected 
cores in the Seahorse IRDC G304.74+01.32. The SEDs were fitted using single and two-temperature modified blackbody models. Our main results are summarised as follows:

\begin{enumerate}
\item For the 12 analysed cores, out of which two are IR dark (no \textit{WISE} detection), the mean values of the derived dust temperatures for the cold (warm) component, masses, luminosities, H$_2$ number densities, and surface densities were found to be $13.3\pm1.4$~K ($47.0\pm5.0$~K), $113\pm29$~M$_{\sun}$, $192\pm94$~L$_{\sun}$, $(4.3\pm1.2)\times10^5$~cm$^{-3}$, and $0.77\pm0.19$~g~cm$^{-2}$. All the cores in our sample were found to be gravitationally bound ($\alpha_{\rm vir}<2$).
\item The most luminous source ($L=(1.1\pm0.4)\times10^3$~L$_{\sun}$) in our sample was found to be IRAS~13039-6108a, which is known to be in the \ion{H}{ii} region stage of evolution.
\item Two out of the nine analysed IR bright cores (22\%) were found to have luminosities that are consistent with the accretion luminosity where the mass accretion rate was assumed to be $10^{-5}$~M$_{\sun}$~yr$^{-1}$, the stellar mass was fixed at 10\% of the parent core mass, and the radius of the central star was assumed to be $5$~R$_{\sun}$. Most of the remaining cores (6 out of 10) were found to lie within 1~dex below the aforementioned accretion luminosity value.
\item Seven out of the 12 analysed cores (58\%) were found to lie above the mass-radius thresholds of high-mass star formation presented by Kauffmann \& Pillai (2010) and Baldeschi et al. (2017). The same seven cores were derived to have mass surface densities of $>0.4$~g~cm$^{-3}$ that also make them potential high-mass star-forming cores. Hence, in addition to IRAS~13039-6108a, the Seahorse IRDC is potentially hosting substructures capable of forming high-mass stars.   
\item The average dust temperatures and luminosities of dense cores in the Seahorse IRDC were found to be fairly similar (within a factor of 
$\sim1.8$) to those in the well-studied Snake IRDC G11.11-0.12, which is also known to host a high-mass star-forming object (the P1 region) and a number of lower mass cores. The Snake cores were found to be about 2.5 times less massive on average than the Seahorse cores, but at least part of the aforementioned differences can be explained by the fact that we also included the available submillimetre wavelength data in the SED fits of the Seahorse cores. 
\item Five out of the 12 analysed cores (42\%; SMM~1a, 4a, 6a, 7, and IRAS~13039-6108a) show evidence of fragmentation in the SABOCA 350~$\mu$m image, and the fragment separation in SMM~6a is consistent with thermal Jeans fragmentation (i.e. $d_{\rm sep}=(1.4\pm0.3)\times \lambda_{\rm J}$), while the other four cores appear to require non-thermal fragmentation processes. 
\end{enumerate}

Although the Seahorse IRDC lies about $1\fdg3$ or $\sim60$~pc above the Galactic plane, it appears to have comparable star-forming properties to the \textit{Spitzer}-selected filamentary IRDCs in or closer to the Galactic plane. More detailed studies of the Seahorse core fragments and implications for the cores' ability to form massive stars require high-resolution follow-up observations. In particular, (sub-)millimetre dust continuum and molecular spectral line imaging of the cores with instruments like ALMA would be the next natural step in studying the Seahorse IRDC filament.

\begin{acknowledgements}

I would like to thank the anonymous referee for providing comments and suggestions. This research has made use of NASA's Astrophysics Data System Bibliographic Services, the NASA/IPAC Infrared Science Archive, which is operated by the Jet Propulsion Laboratory, California Institute of Technology, under contract with the National Aeronautics and Space Administration, and {\tt Astropy}\footnote{\url{http://www.astropy.org}}, a community-developed core Python package for Astronomy (\cite{astropy2013}, \cite{astropy2018}). This publication makes use of data products from the \textit{Wide-field Infrared Survey Explorer}, which is a joint project of the University of California, Los Angeles, and the Jet Propulsion Laboratory/California Institute of Technology, and NEOWISE, which is a project of the Jet Propulsion Laboratory/California Institute of Technology. \textit{WISE} and NEOWISE are funded by the National Aeronautics and Space Administration.

\end{acknowledgements}

\appendix

\section{Multiwavelength images of the analysed cores and the source appearances}

In Fig.~\ref{figure:images}, we show the \textit{WISE} 12~$\mu$m and 22~$\mu$m and \textit{Herschel}/PACS 70~$\mu$m and 100~$\mu$m images 
towards the analysed cores. Each image is overlaid with contours showing the SABOCA 350~$\mu$m and LABOCA 870~$\mu$m dust continuum emission. Selected sources are discussed below. 

The \textit{WISE} source J130622.27-613014.7 seen towards SMM~1 appears to lie in between the SABOCA cores SMM~1a and SMM~1b. 
The \textit{WISE} source is interpreted to be associated with SMM~1b, where the projected separation between the two is $8\farcs9$. Also, the colours of the \textit{WISE} source suggest that it could be related to shock emission (\cite{miettinen2018}), which could explain the fairly large distance from SMM~1b. Nevertheless, this would still be an indication that SMM~1b is a star-forming core. On the other hand, 
the \textit{WISE} source could also be associated with the SMM~1a's subcondensation ($5\farcs95$ separation) $15\farcs3$ to the east of the stronger SABOCA 350~$\mu$m peak. The core SMM~1a itself is classified as IR dark because it has no \textit{WISE} counterpart. However, SMM~1a is not 70~$\mu$m dark as illustrated in Fig.~\ref{figure:images} although the PACS Point Source Catalogue did not contain a corresponding counterpart.

The \textit{WISE} source J130637.23-612848.8 seen towards SMM~3 appears weak in the 12~$\mu$m and 22~$\mu$m images, but it can be found in the AllWISE catalogue. The PACS 70~$\mu$m source is well coincident with the \textit{WISE} source's catalogue position as shown in Fig.~\ref{figure:images}. There are also two other PACS sources to the east and south-west of SMM~3, both of which are visible in the \textit{WISE} images, and there is weak ($3\sigma$) SABOCA 350~$\mu$m emission seen towards the former source.

The source WISE J130646.35-612855.3, which is assigned to SMM~4a, is not detected in the W3 (12~$\mu$m) band of \textit{WISE} ($>9.344$~mag), and it is also not obviously seen in the 22~$\mu$m image although the AllWISE catalogue contains such detection (${\rm W4}=6.001\pm0.233$~mag). The core SMM~4a also appears to contain a subcondensation $16\farcs8$ to the south-west of the brighter component. This subcondensation also appears in the PACS images and is hence a real source, and an indication that SMM~4a is fragmented. The parent clump SMM~4 appears to contain multiple SABOCA sources and the system is likely a formation site of a stellar group or cluster. 

We note that the \textit{WISE} source seen $29\farcs4$ to the west of WISE J130651.93-612822.2 associated with IRAS~13037-6112a, is the  source WISE J130647.85-612819.5 associated with SMM~4b. Also, the images towards IRAS~13037-6112 shown in Fig.~\ref{figure:images} contain the SABOCA cores SMM~6a and 6b in the northern part of the images (cf.~the SMM~6 panels). 

The core SMM~6a shows an additional condensation $12\farcs7$ to the south-west of the stronger peak position. The \textit{WISE} source J130654.93-612725.9 is associated with the latter position (see the 12~$\mu$m image in Fig.~\ref{figure:images}). The \textit{WISE} source J130652.20-612735.0 associated with SMM~6b is not clearly visible in the \textit{WISE} 12~$\mu$m and 22~$\mu$m images shown in Fig.~\ref{figure:images}, but the AllWISE catalogue reported a detection in all the four bands W1--W4 (3.4~$\mu$m--22~$\mu$m). The source WISE J130652.20-612735.0 has \textit{WISE} colours in between an embedded YSO and a shock feature (\cite{miettinen2018}), but there is a PACS 100~$\mu$m source $5\farcs4$ away from the \textit{WISE} source, which suggests that it could be a YSO.

In our previous studies of the Seahorse IRDC (\cite{miettinen2018}, 2020), the \textit{WISE} source J130704.50-612620.6 seen towards SMM~7 ($6\arcsec$ offset) was considered a background extragalactic object, namely a star-forming galaxy or an active galactic nucleus, because the colours of this weak \textit{WISE} source were suggestive of its extragalactic nature and the source is only detected at 22~$\mu$m. However, SMM~7 shows a hint of an additional SABOCA 350~$\mu$m peak $7\farcs5$ to the south of the brighter SABOCA peak, and WISE J130704.50-612620.6 could in fact be an embedded YSO associated with this secondary SABOCA peak ($2\farcs8$ projected offset). The latter source position is coincident with the PACS far-IR emission. Hence, in the present work we take SMM~7 to be an IR bright core. 

IRAS~13039-6108a, which is associated with the \textit{WISE} source J130707.22-612438.5 ($7\farcs7$ separation), 
shows another SABOCA 350~$\mu$m fragment $13\farcs4$ to the north of the emission maximum. There is also an extension of SABOCA 350~$\mu$m emission to the west of IRAS~13039-6108a. This \ion{H}{ii} region is extended in the \textit{WISE} 12~$\mu$m and 22~$\mu$m images, which suggests the presence of a photodissociation region.

There is a \textit{WISE} source in the AllWISE catalogue that is seen towards SMM~9. However, it is not identifiable in the \textit{WISE} 12~$\mu$m and 22~$\mu$m images shown in Fig.~\ref{figure:images}, and indeed it is reported to be undetected at 12~$\mu$m in the AllWISE catalogue (${\rm W3}>10.159$~mag). This source, WISE J130713.09-612238.2, lies $9\farcs1$ from the SABOCA 350~$\mu$m peak position of SMM~9, and its \textit{WISE} colours suggest that it is a foreground star rather than an embedded YSO (\cite{miettinen2018}). Hence, SMM~9 is taken to be IR dark but the source is detected with PACS at 70~$\mu$m.

\begin{figure*}[!htb]
\begin{center}
\includegraphics[width=0.2425\textwidth]{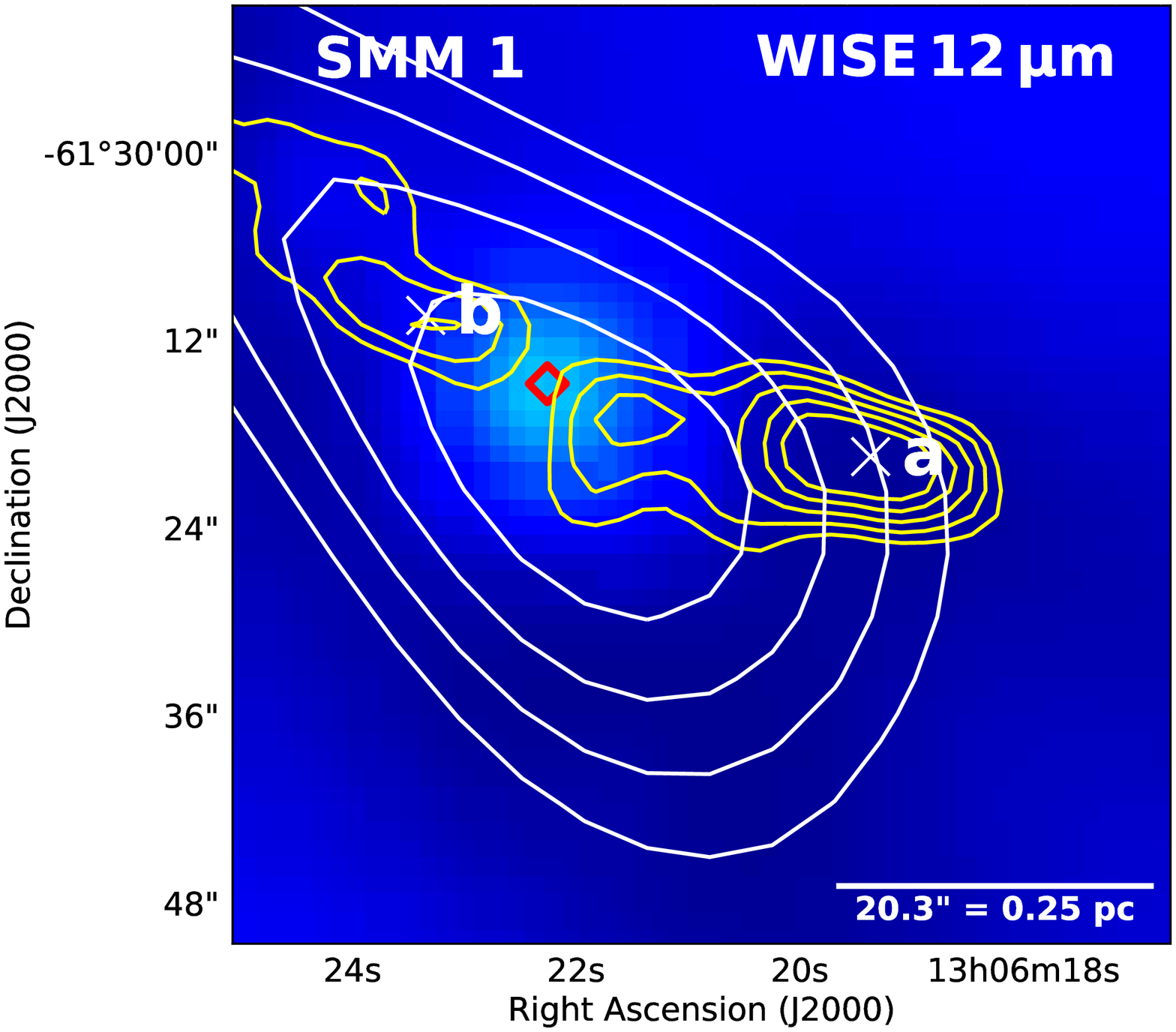}
\includegraphics[width=0.21\textwidth]{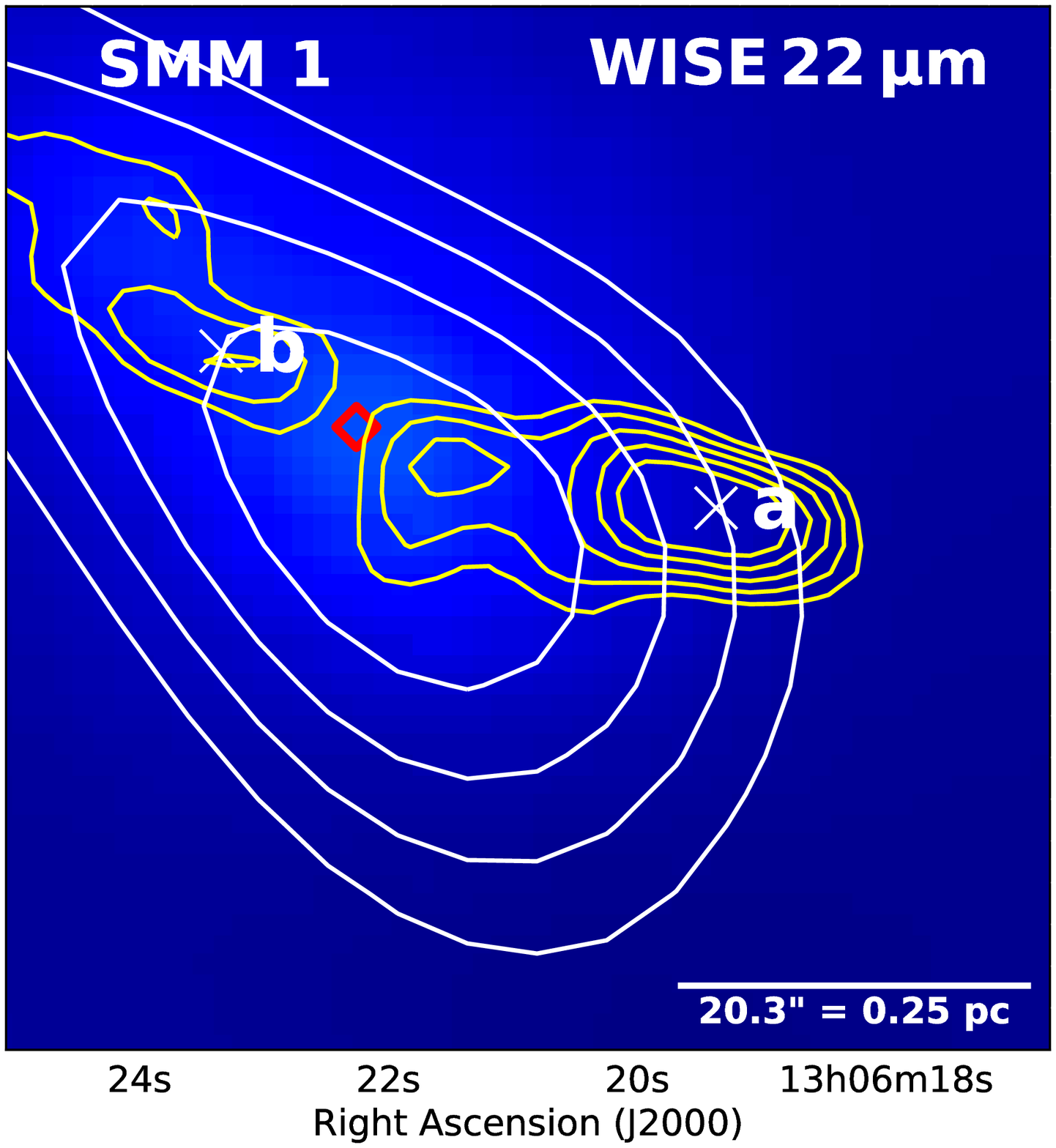}
\includegraphics[width=0.21\textwidth]{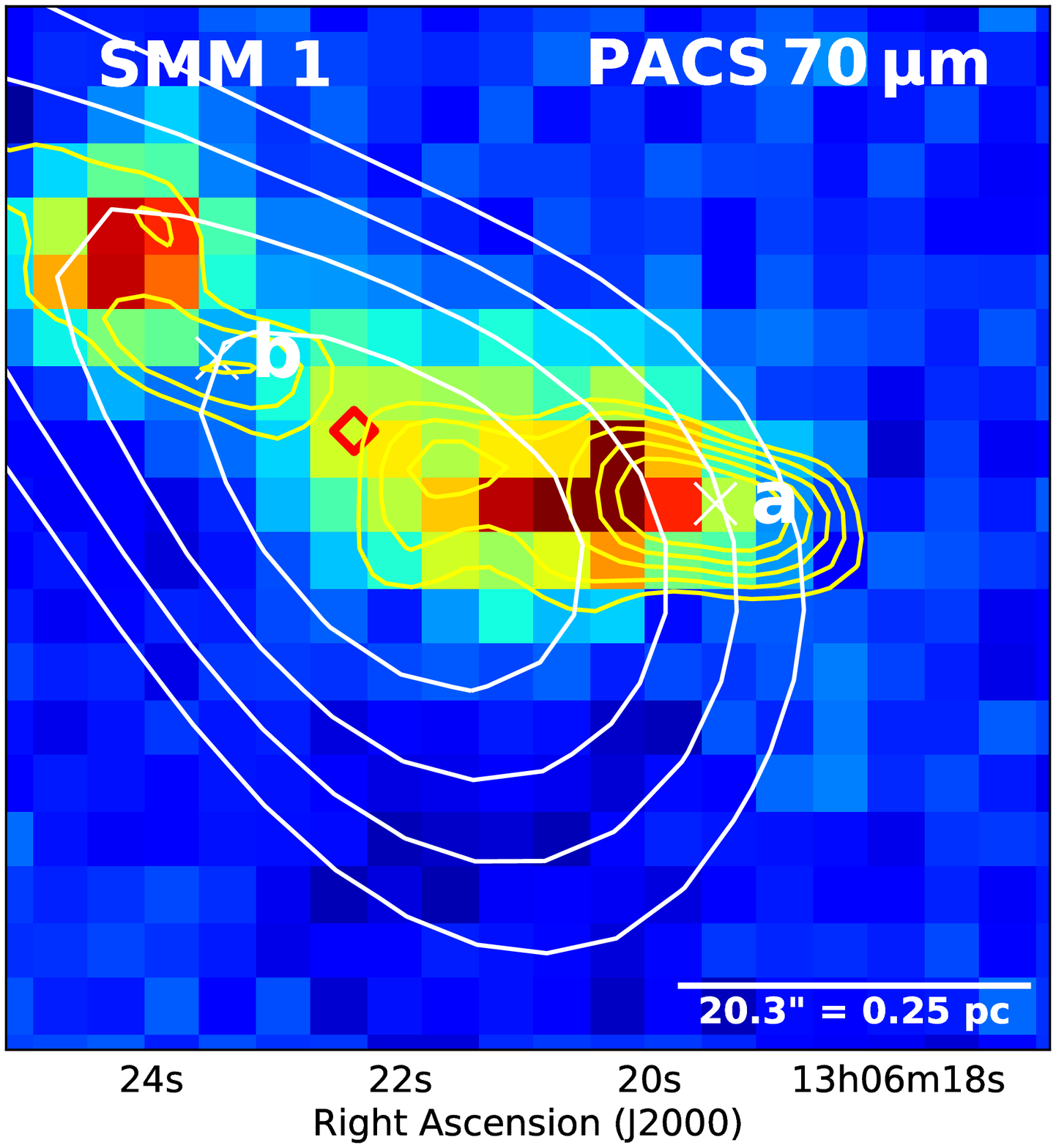}
\includegraphics[width=0.21\textwidth]{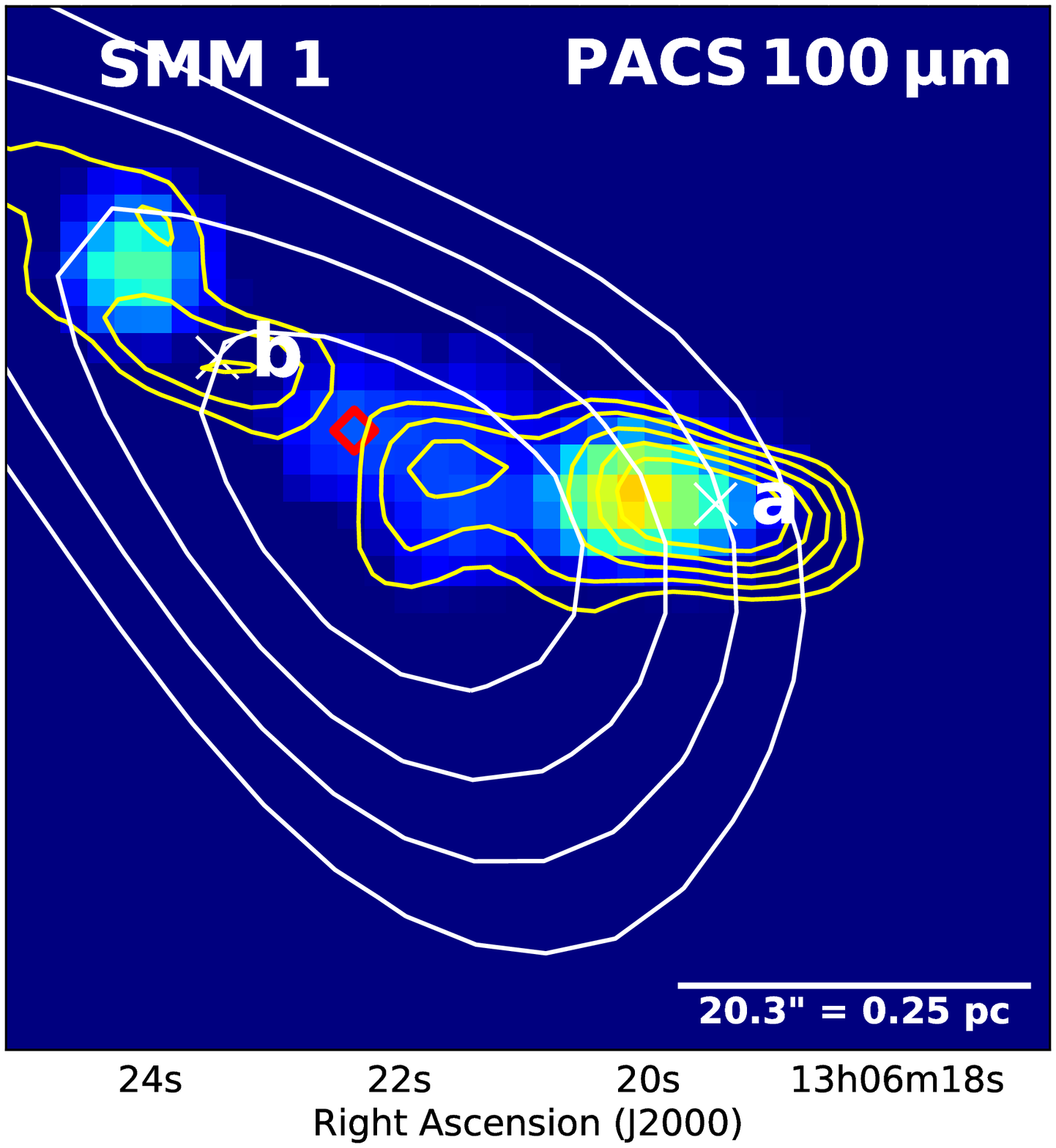}
\includegraphics[width=0.2425\textwidth]{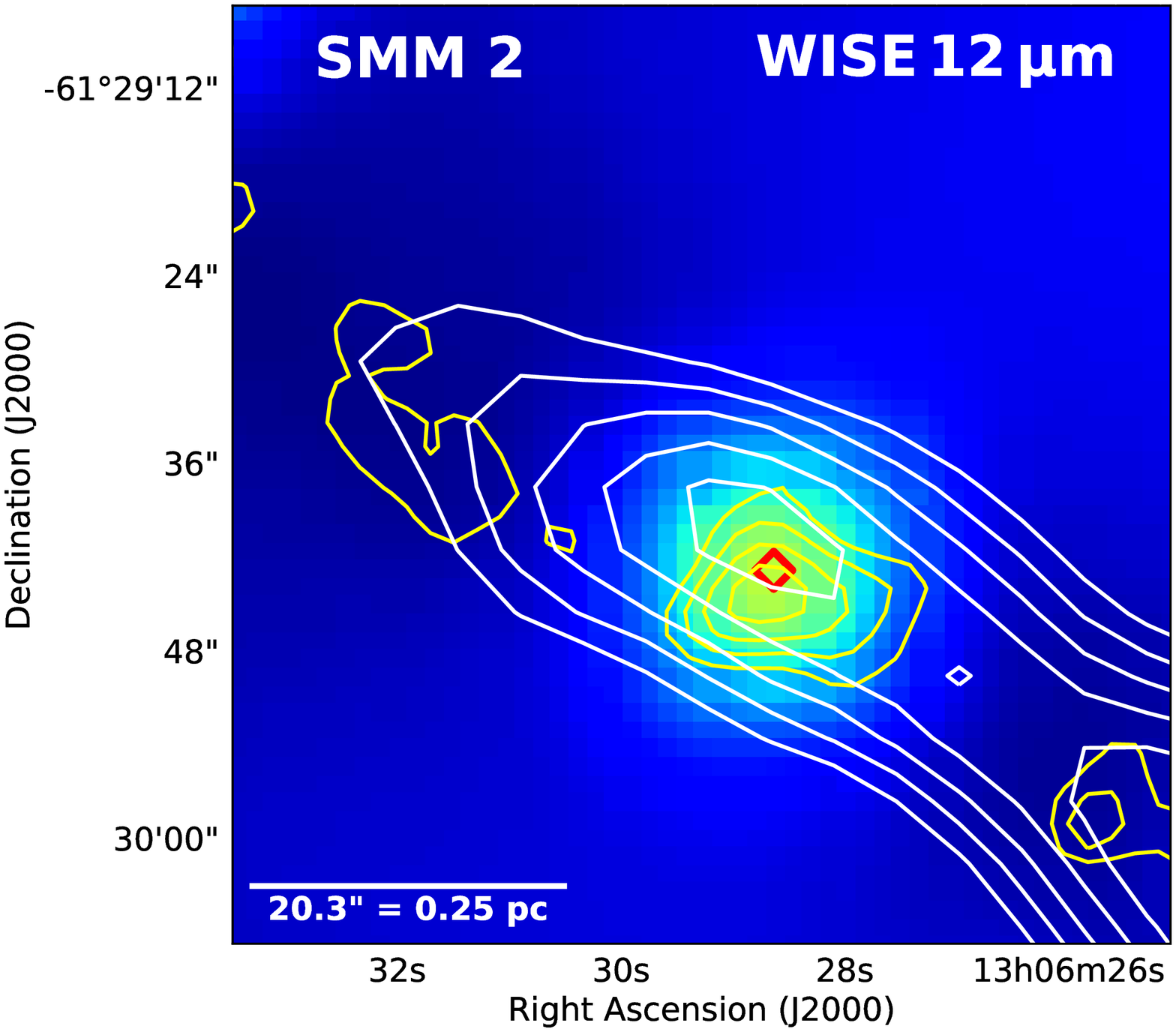}
\includegraphics[width=0.21\textwidth]{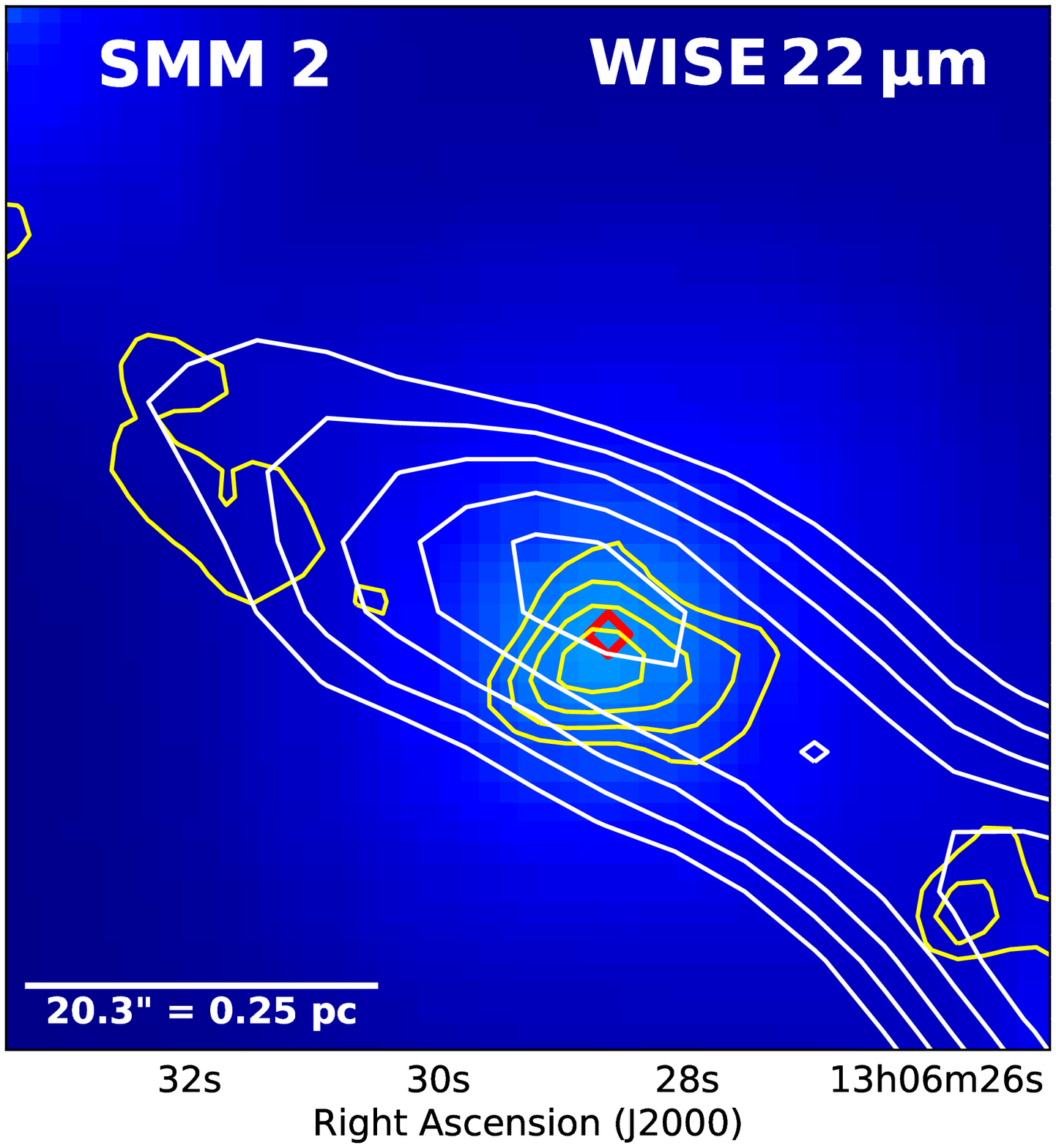}
\includegraphics[width=0.21\textwidth]{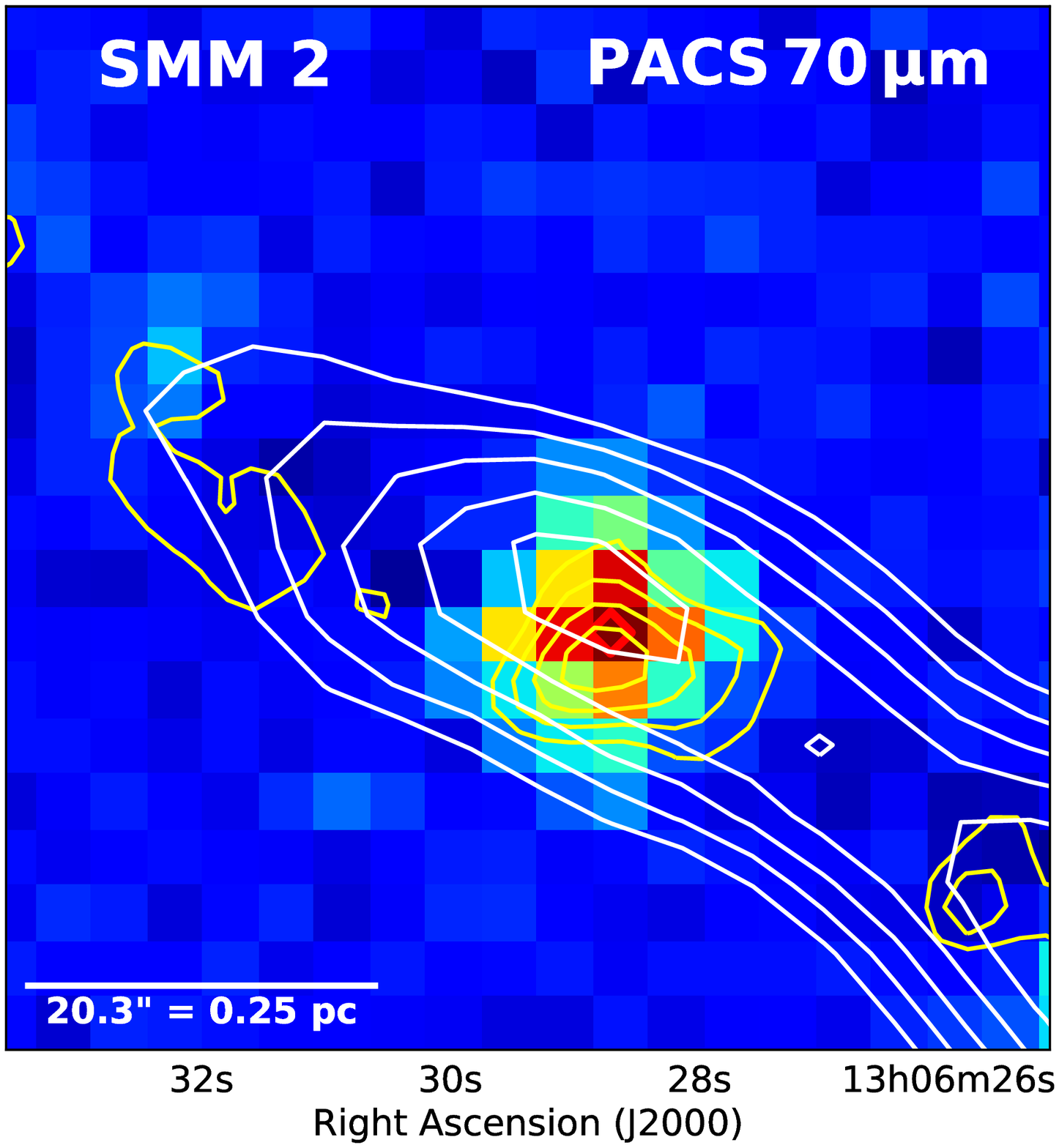}
\includegraphics[width=0.21\textwidth]{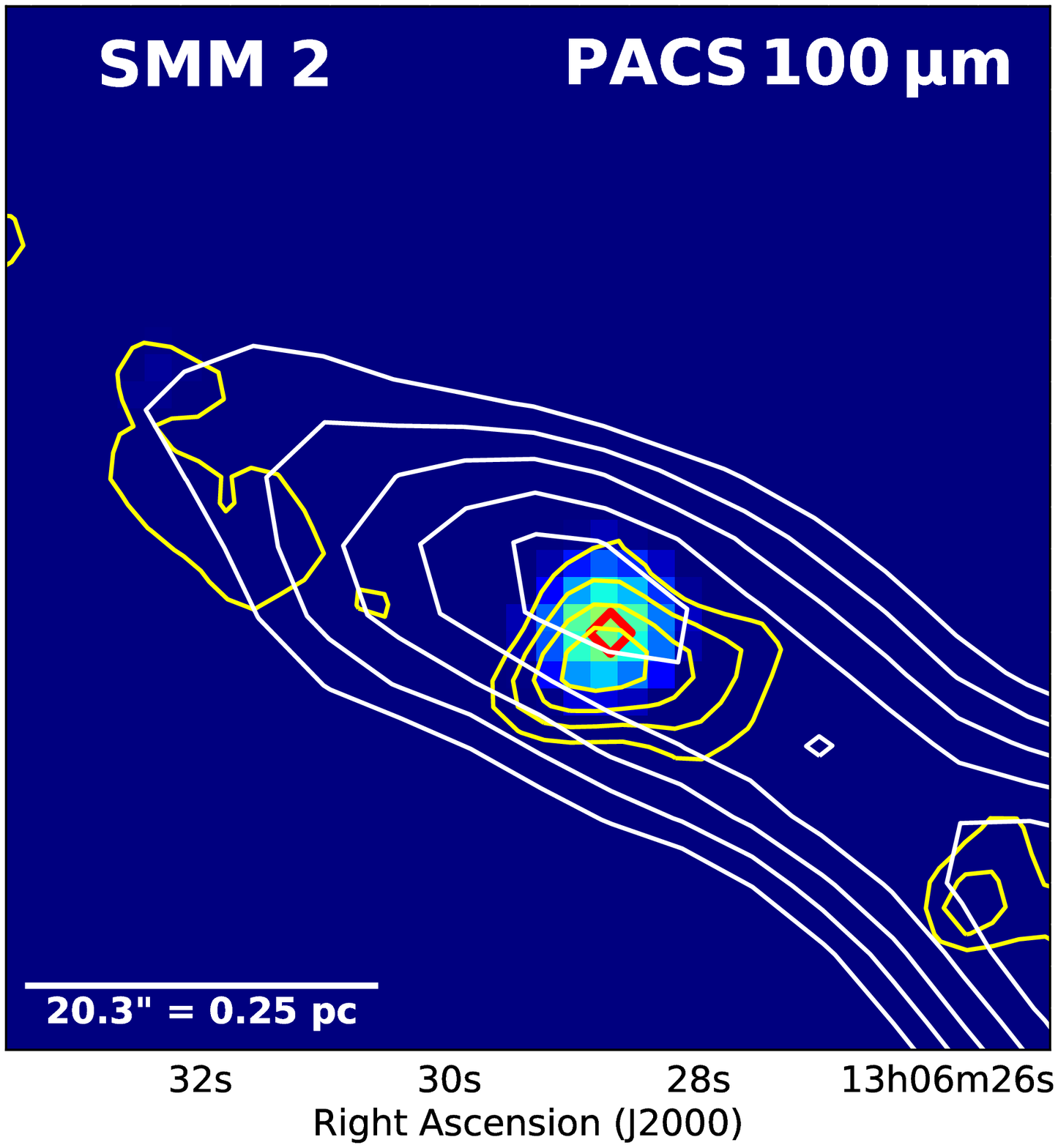}
\includegraphics[width=0.2425\textwidth]{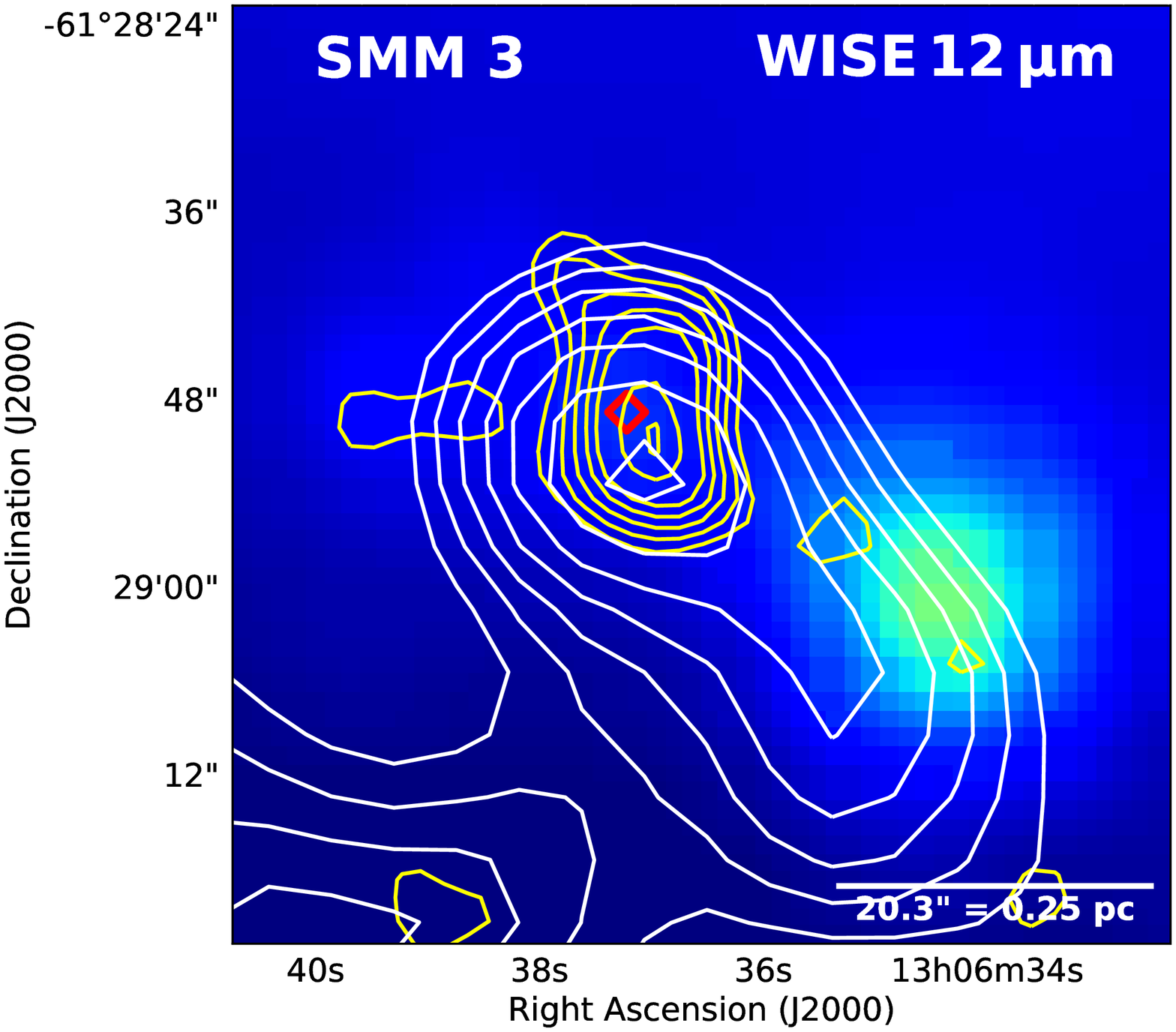}
\includegraphics[width=0.21\textwidth]{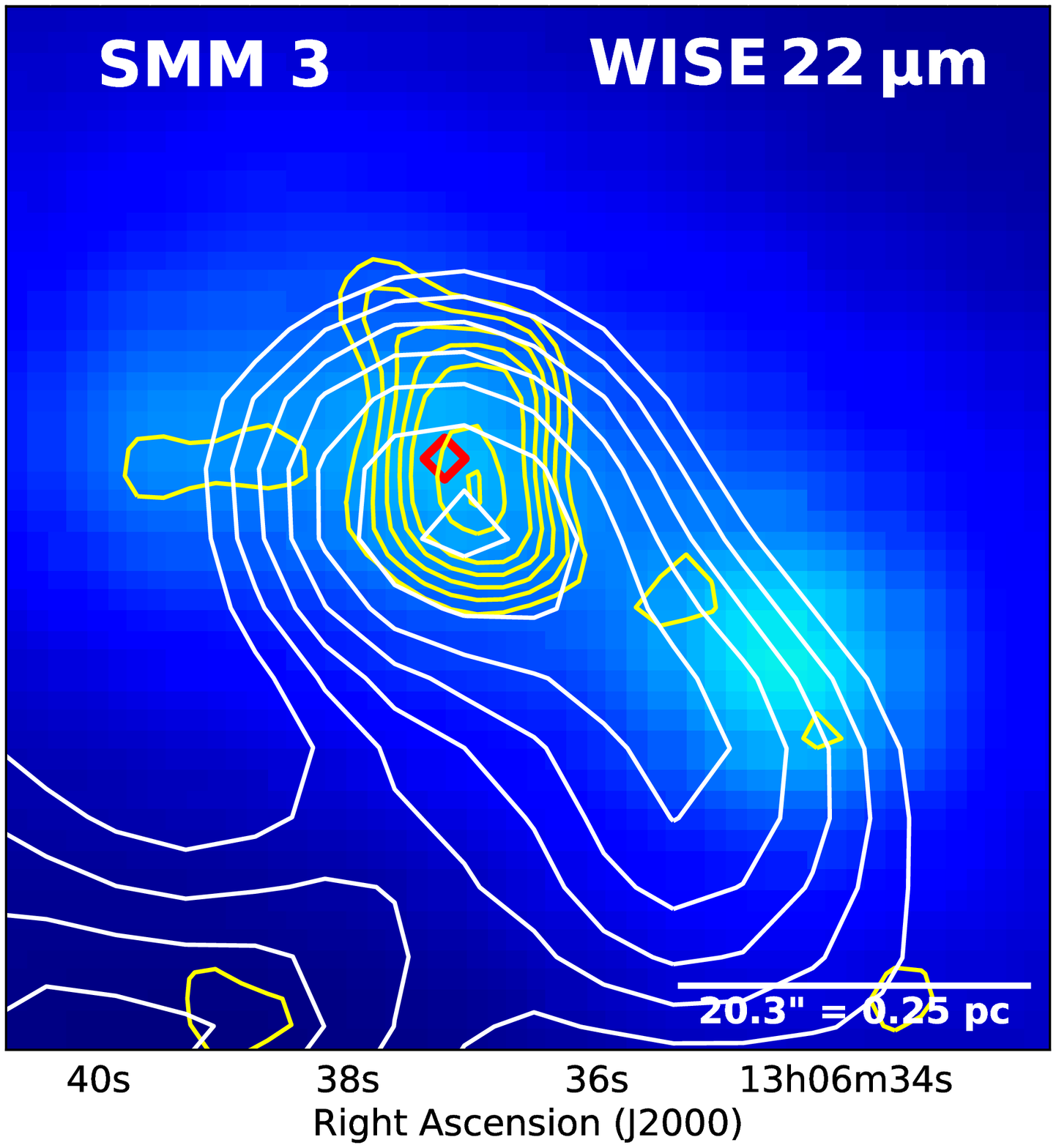}
\includegraphics[width=0.21\textwidth]{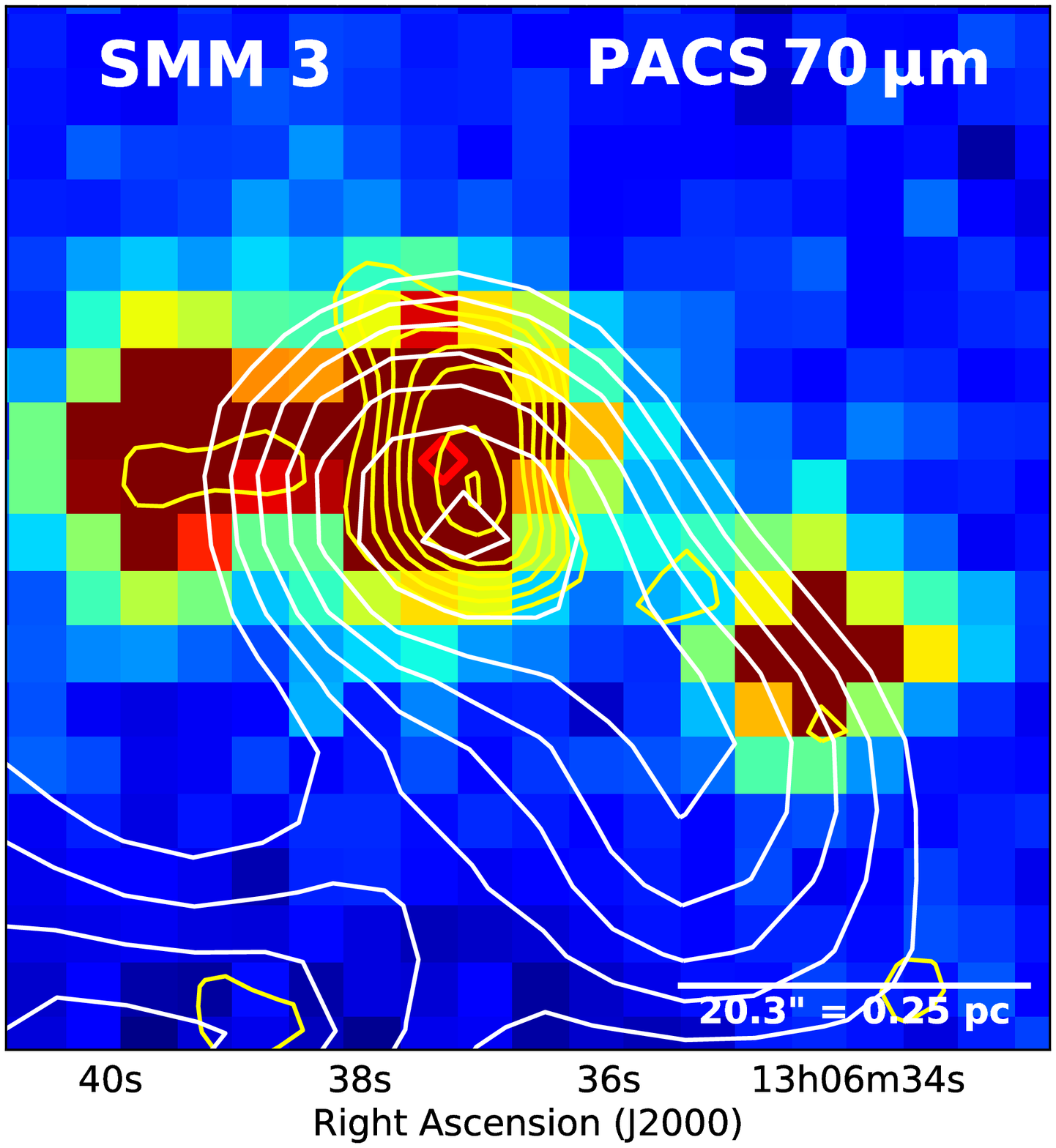}
\includegraphics[width=0.21\textwidth]{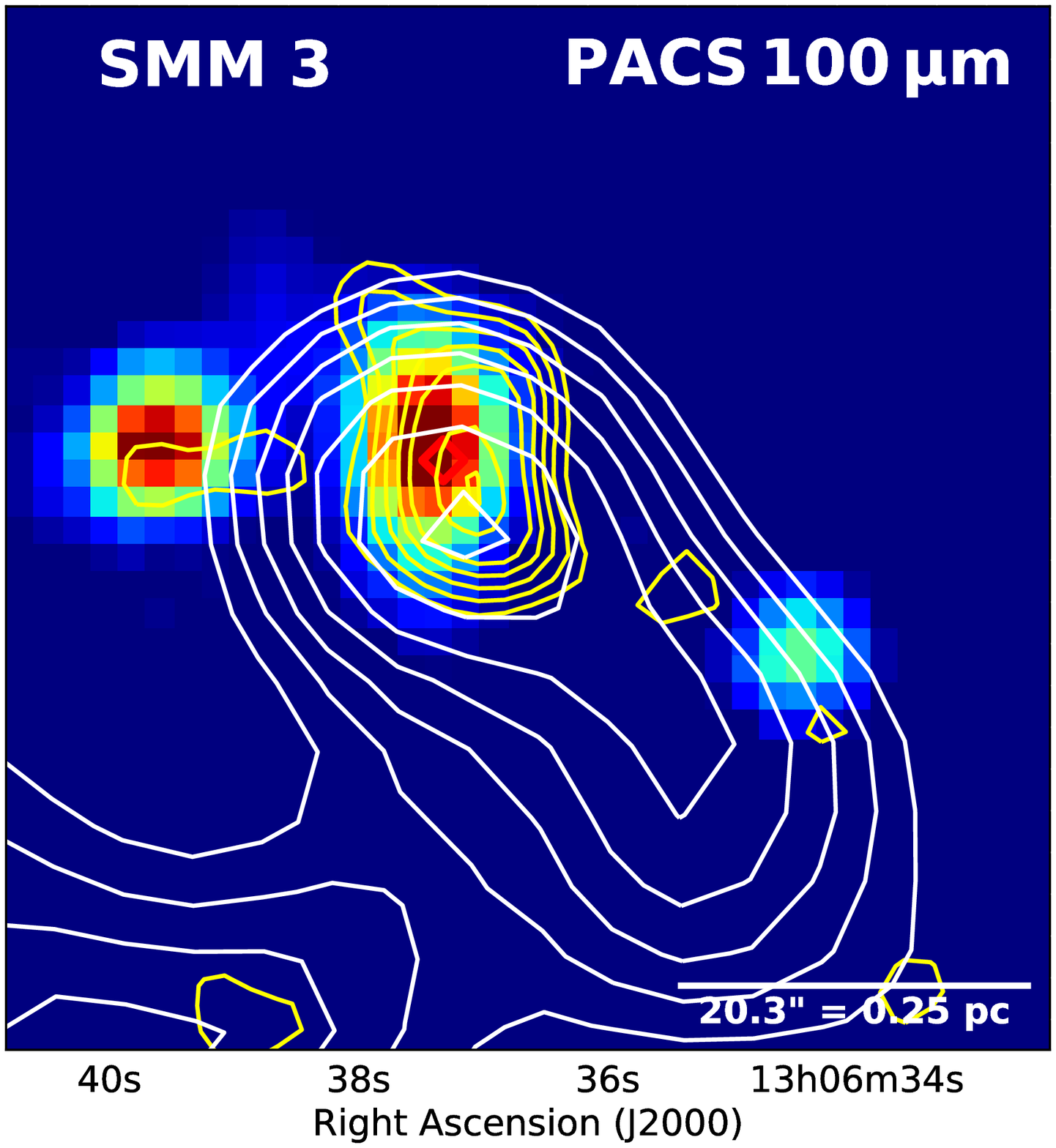}
\includegraphics[width=0.2425\textwidth]{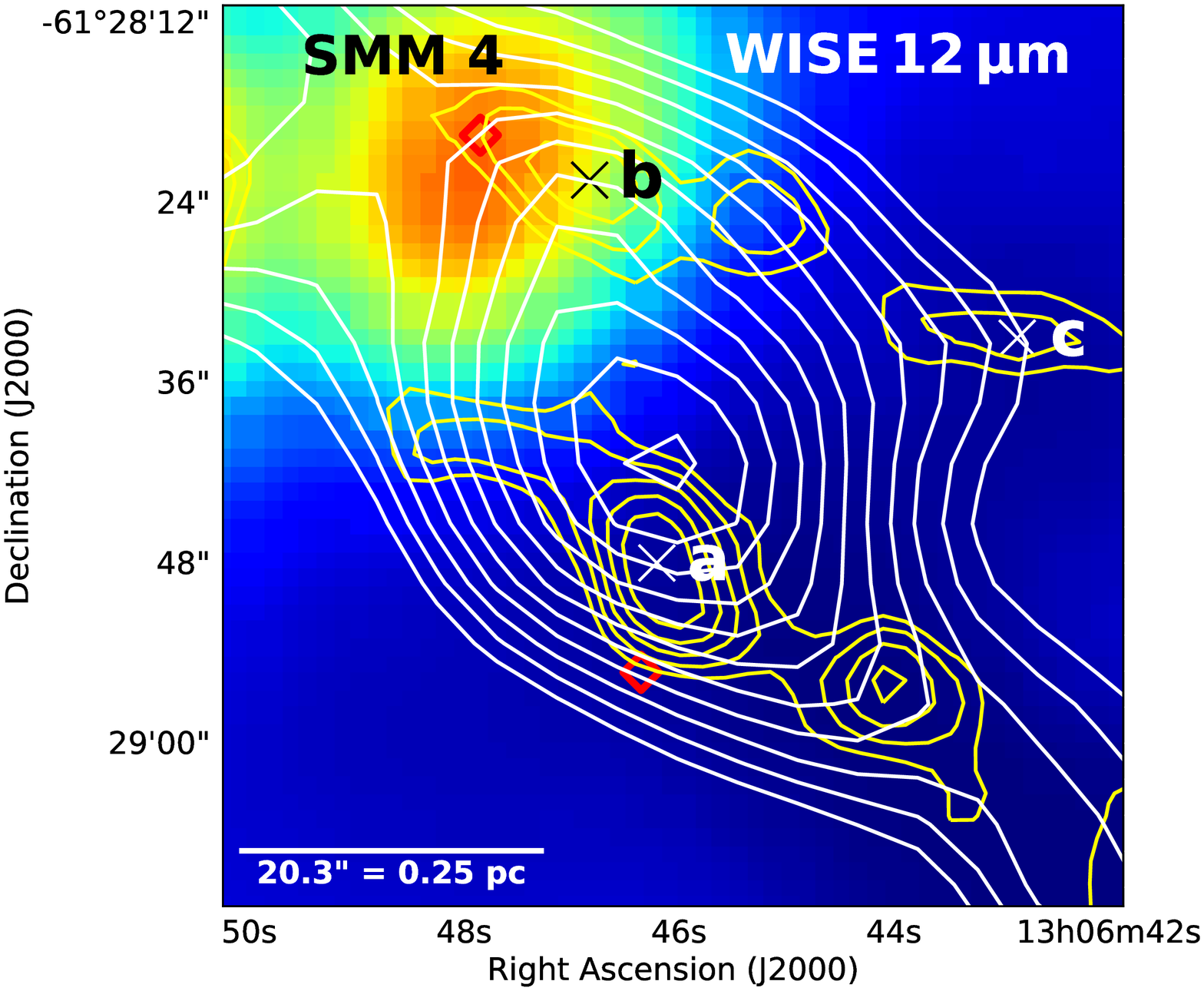}
\includegraphics[width=0.21\textwidth]{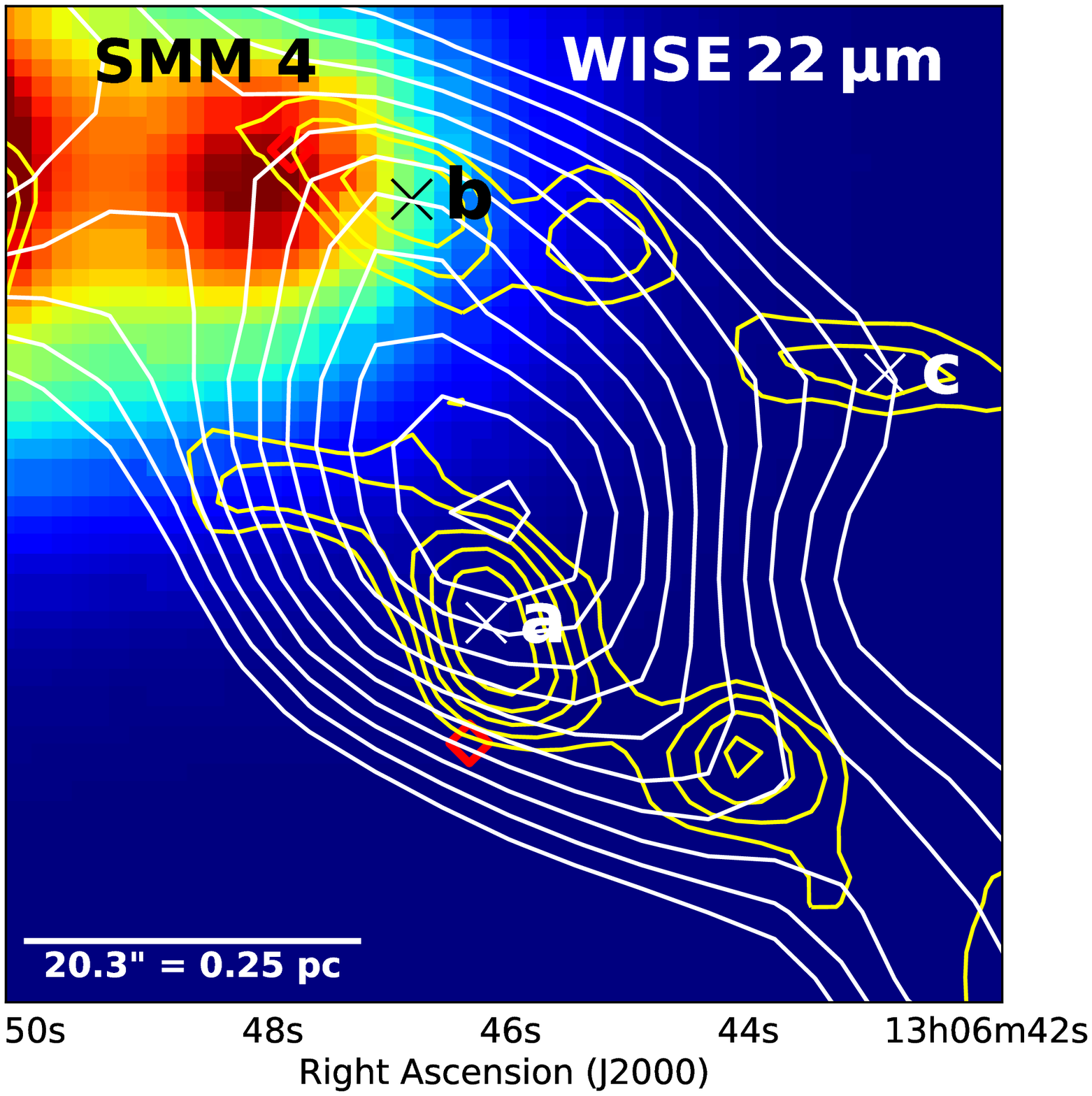}
\includegraphics[width=0.21\textwidth]{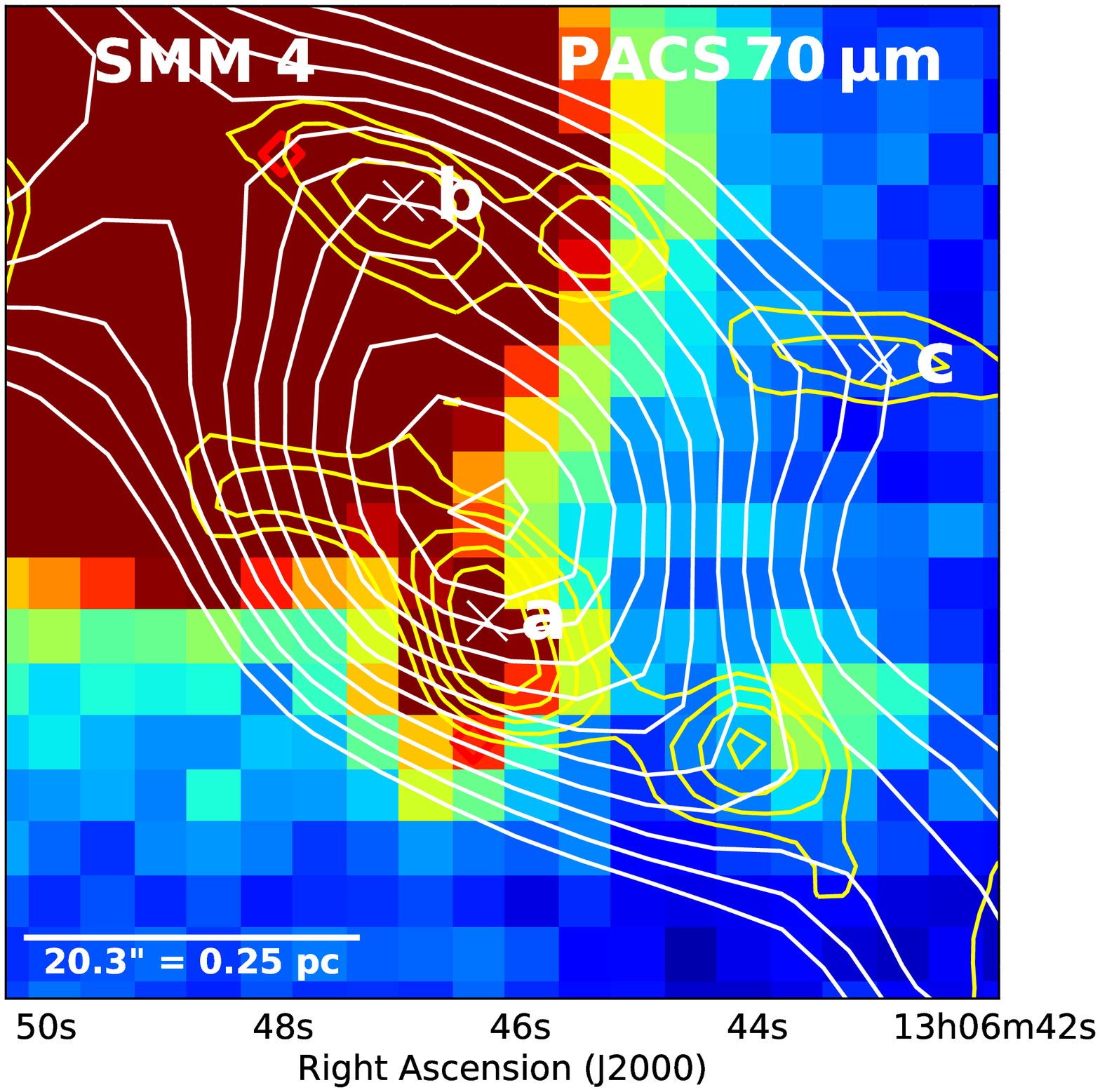}
\includegraphics[width=0.21\textwidth]{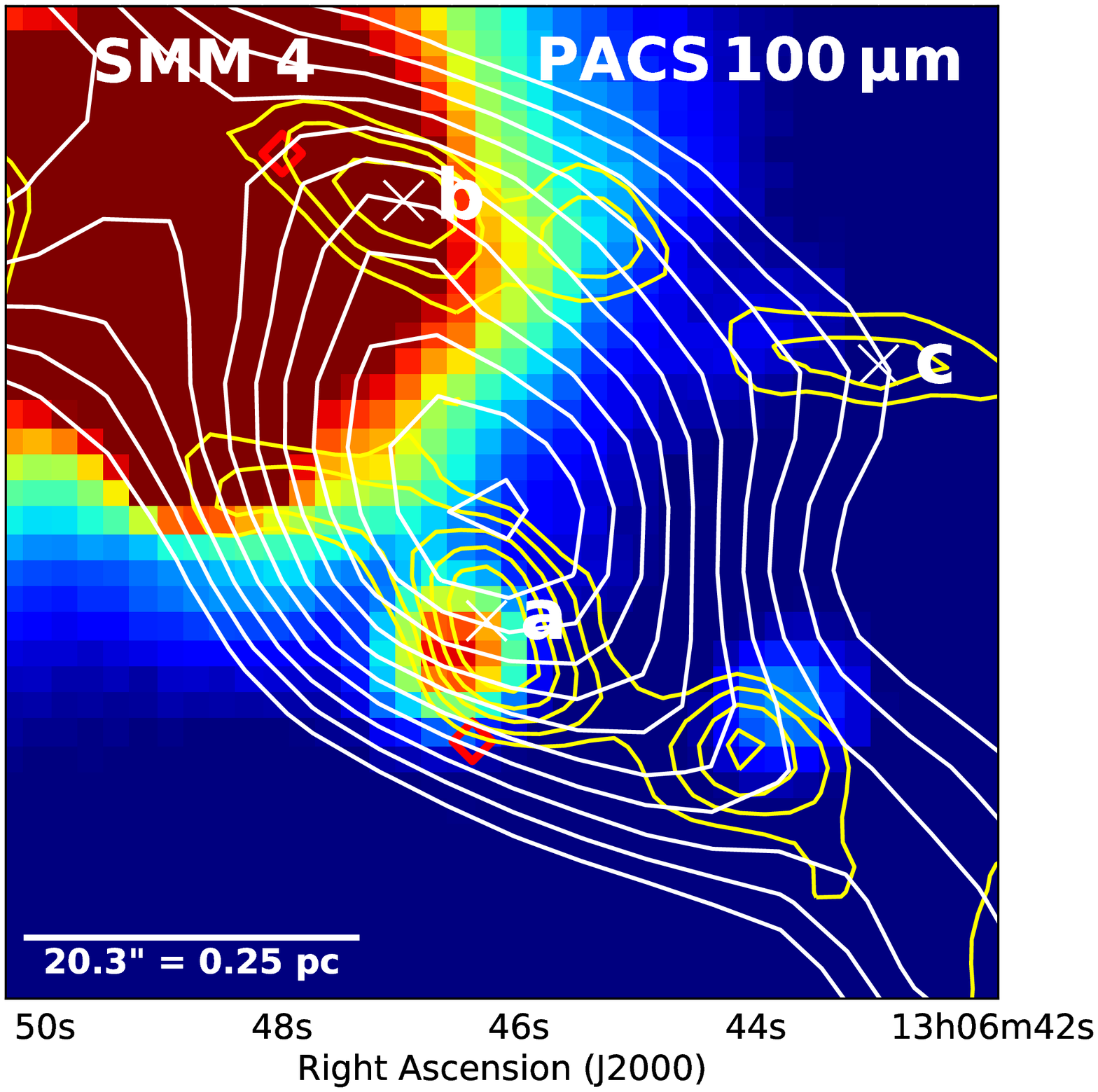}
\includegraphics[width=0.2425\textwidth]{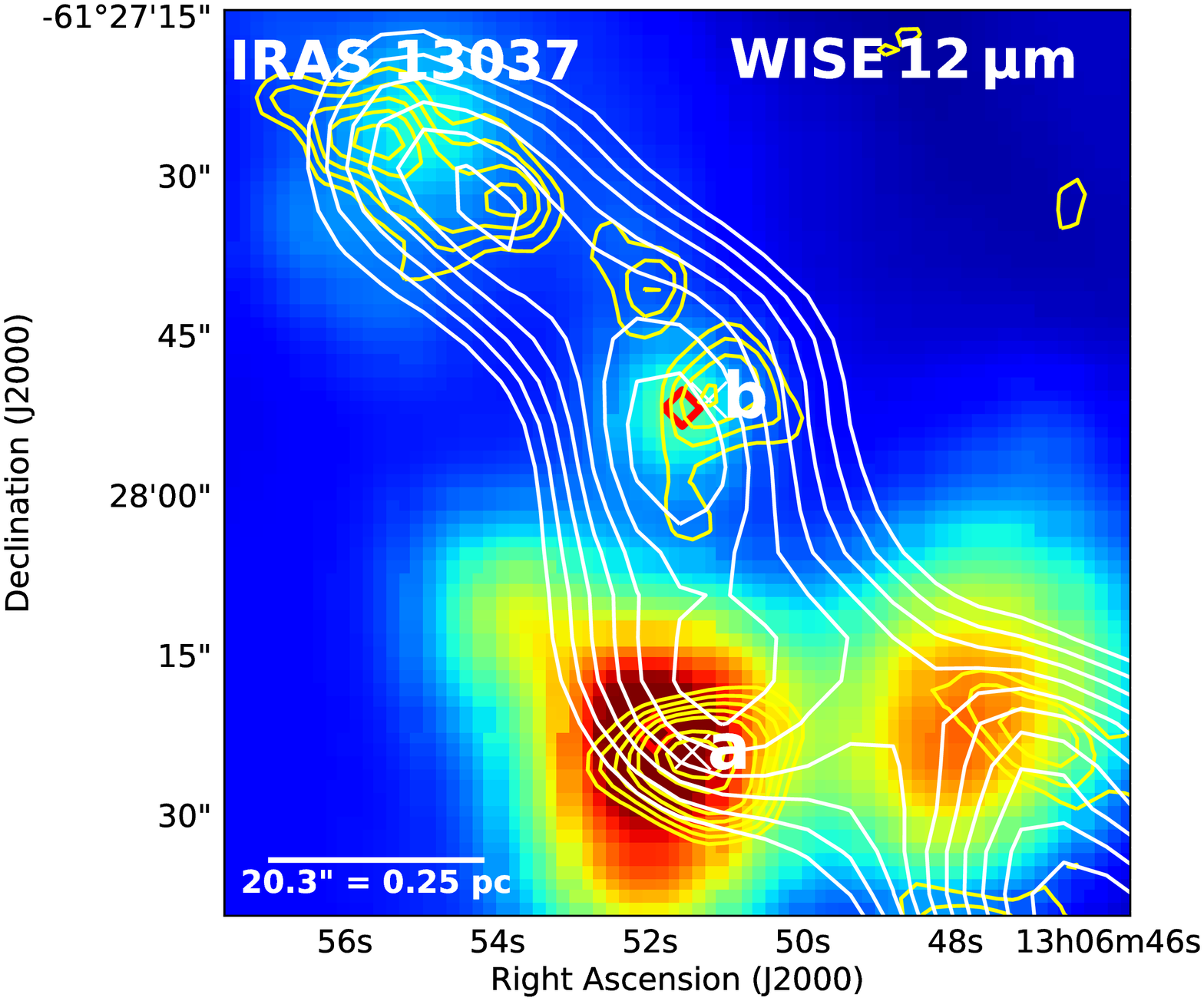}
\includegraphics[width=0.21\textwidth]{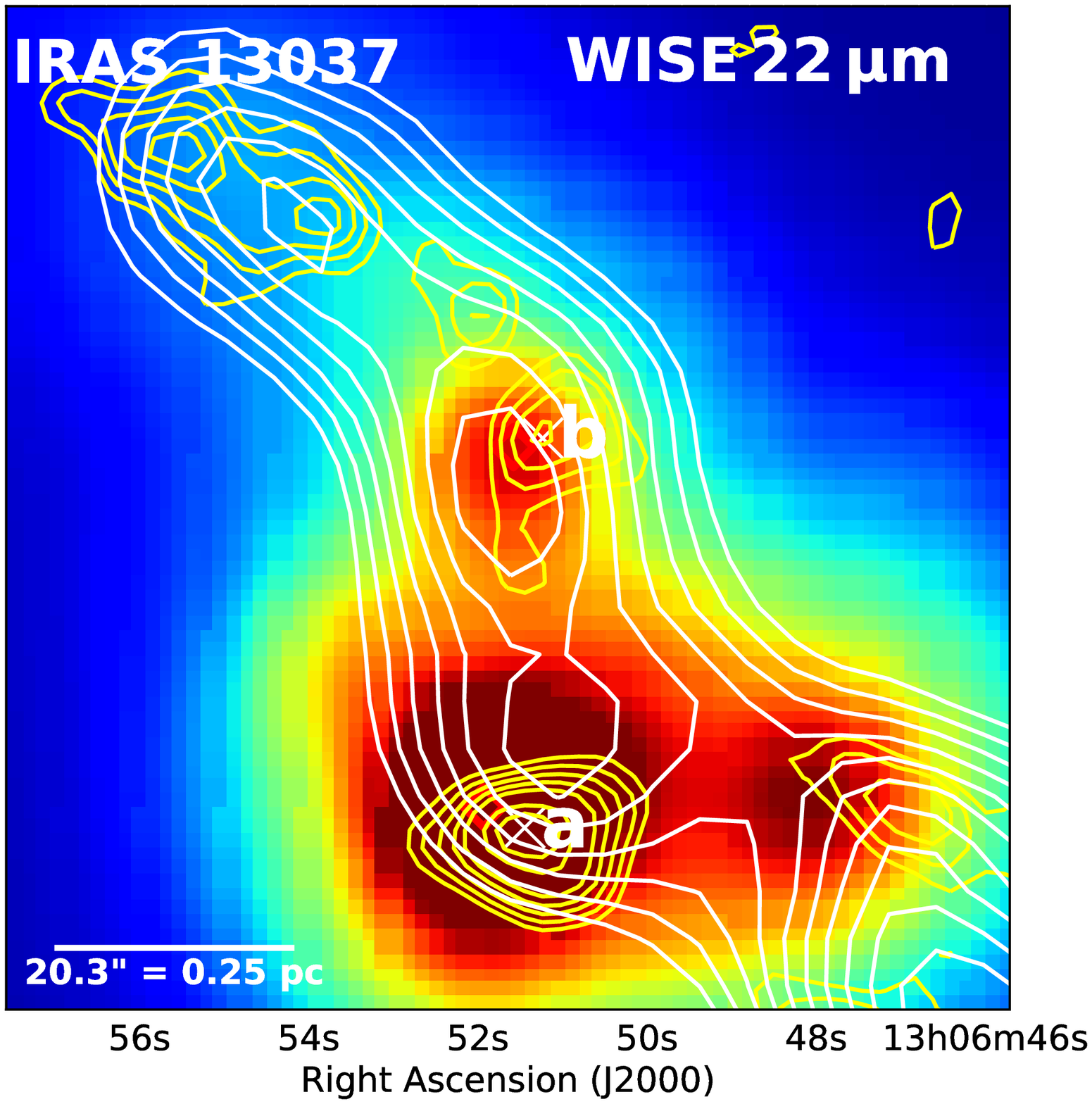}
\includegraphics[width=0.21\textwidth]{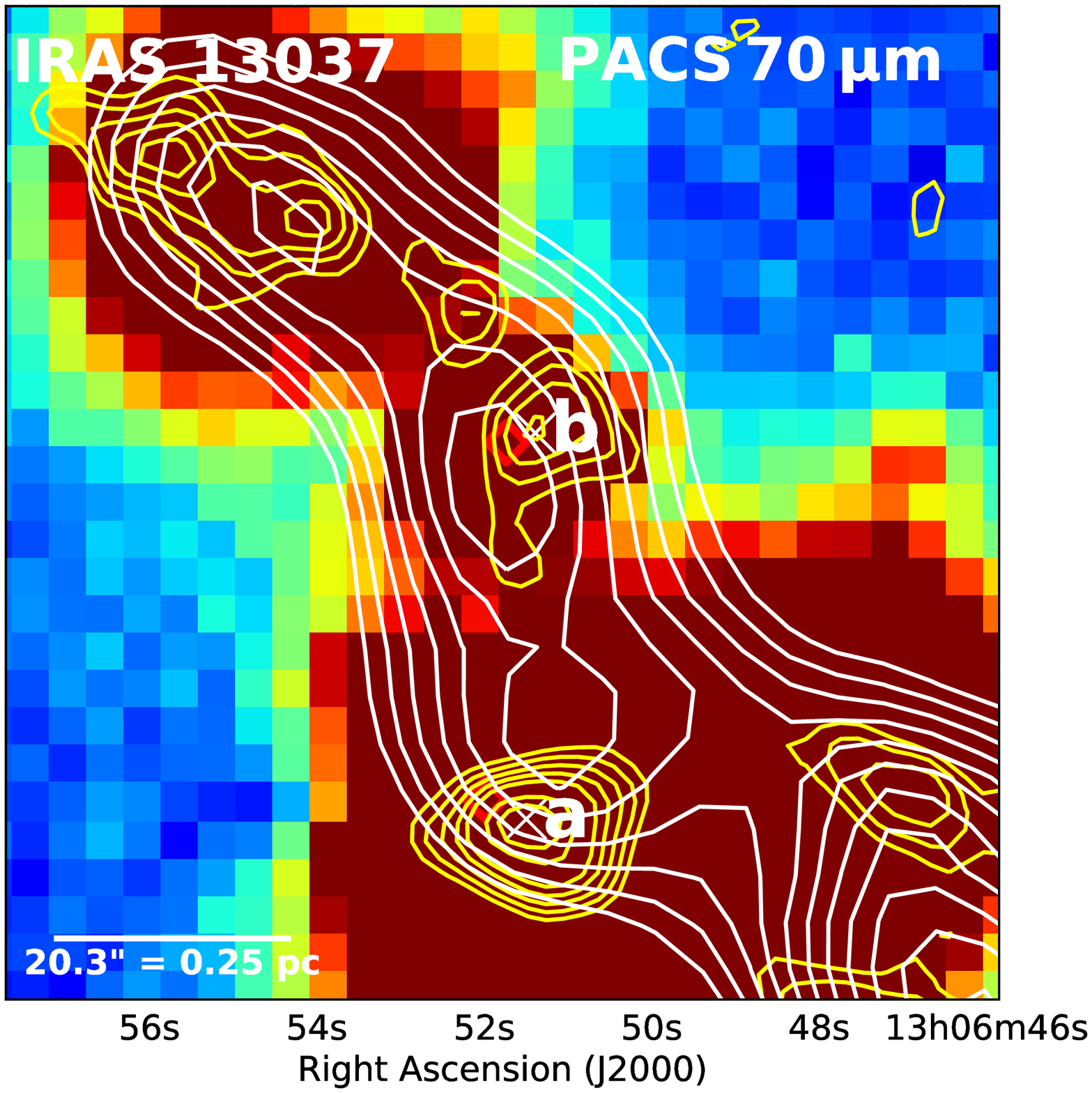}
\includegraphics[width=0.21\textwidth]{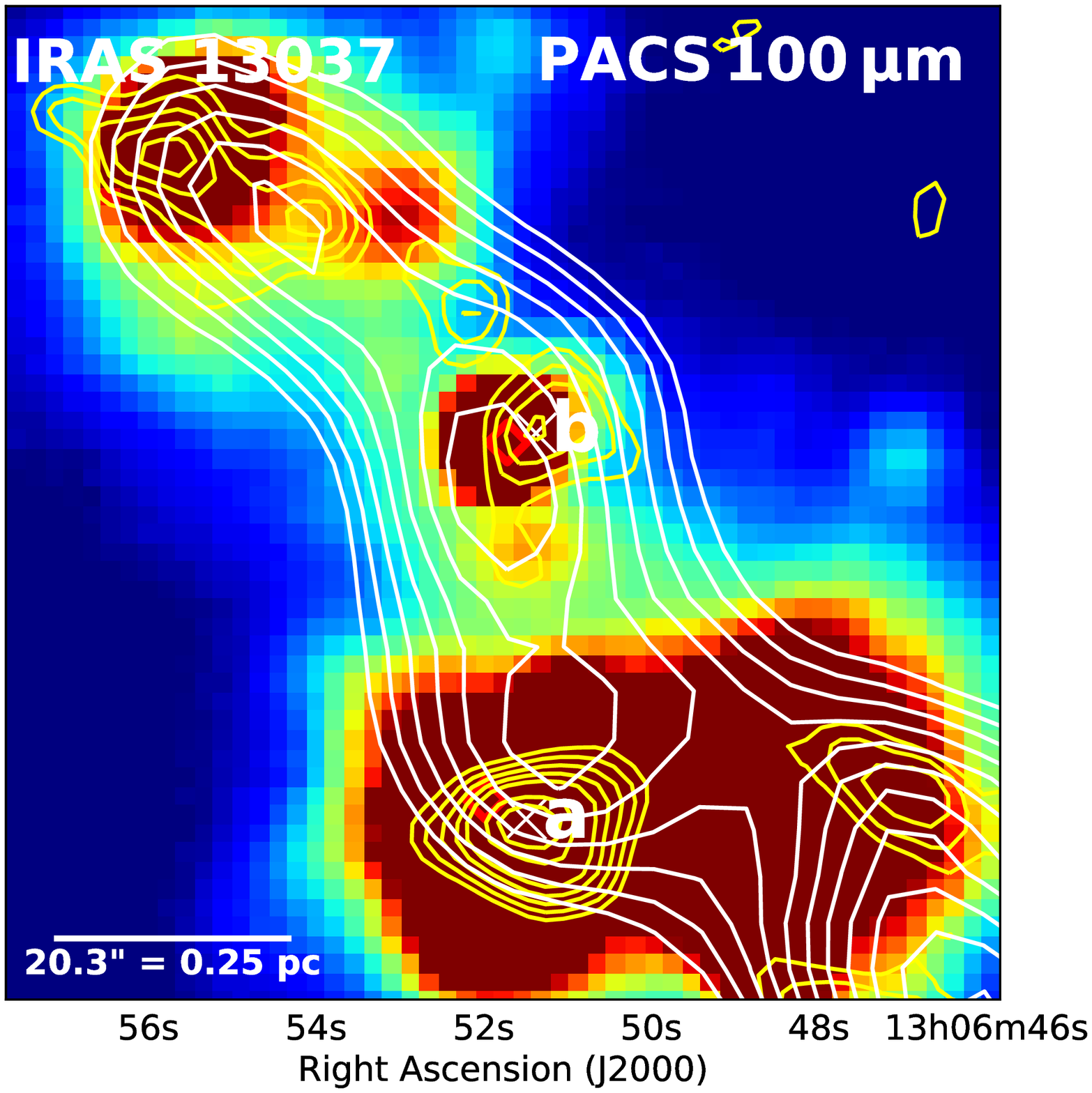}
\caption{Multiwavelength images towards the target cores (\textit{WISE} 12~$\mu$m and 22~$\mu$m and \textit{Herschel}/PACS 70~$\mu$m and 100~$\mu$m). The \textit{WISE} images are shown with logarithmic scaling except the 22~$\mu$m images of SMM~7 and SMM~9 that are shown with power-law scaling to enhance the colour contrast. The PACS images are displayed using a linear stretch. In each panel, the overlaid yellow contours show the SABOCA 350~$\mu$m dust continuum emission and the white contours show the LABOCA 870~$\mu$m emission. The contour levels start from $3\sigma$, and progress in steps of $1\sigma$, where $1\sigma=200$~mJy~beam$^{-1}$ for the SABOCA data and 
$1\sigma=40$~mJy~beam$^{-1}$ for the LABOCA data (\cite{miettinen2018}). Each image is centred
on the LABOCA peak position of the parent clump, and is $60\arcsec \times 60\arcsec$ in size with the exception of 
the IRAS~13037 and IRAS~13039 images, which are $85\arcsec$ on a side. In those cases where the LABOCA clump contains more than one SABOCA core (e.g. components a and b), the crosses indicate the 350~$\mu$m peak positions of the cores. The red diamond symbols show the \textit{WISE} source positions (see \cite{miettinen2018}, Table~3 therein). The plus sign in the IRAS~13039 image indicates the position of the 18~GHz and 22.8~GHz radio continuum source found by S{\'a}nchez-Monge et al. (2013). A scale bar of 0.25 pc is shown in each panel.}
\label{figure:images}
\end{center}
\end{figure*}

\begin{figure*}[!htb]
\addtocounter{figure}{-1}
\begin{center}
\includegraphics[width=0.2425\textwidth]{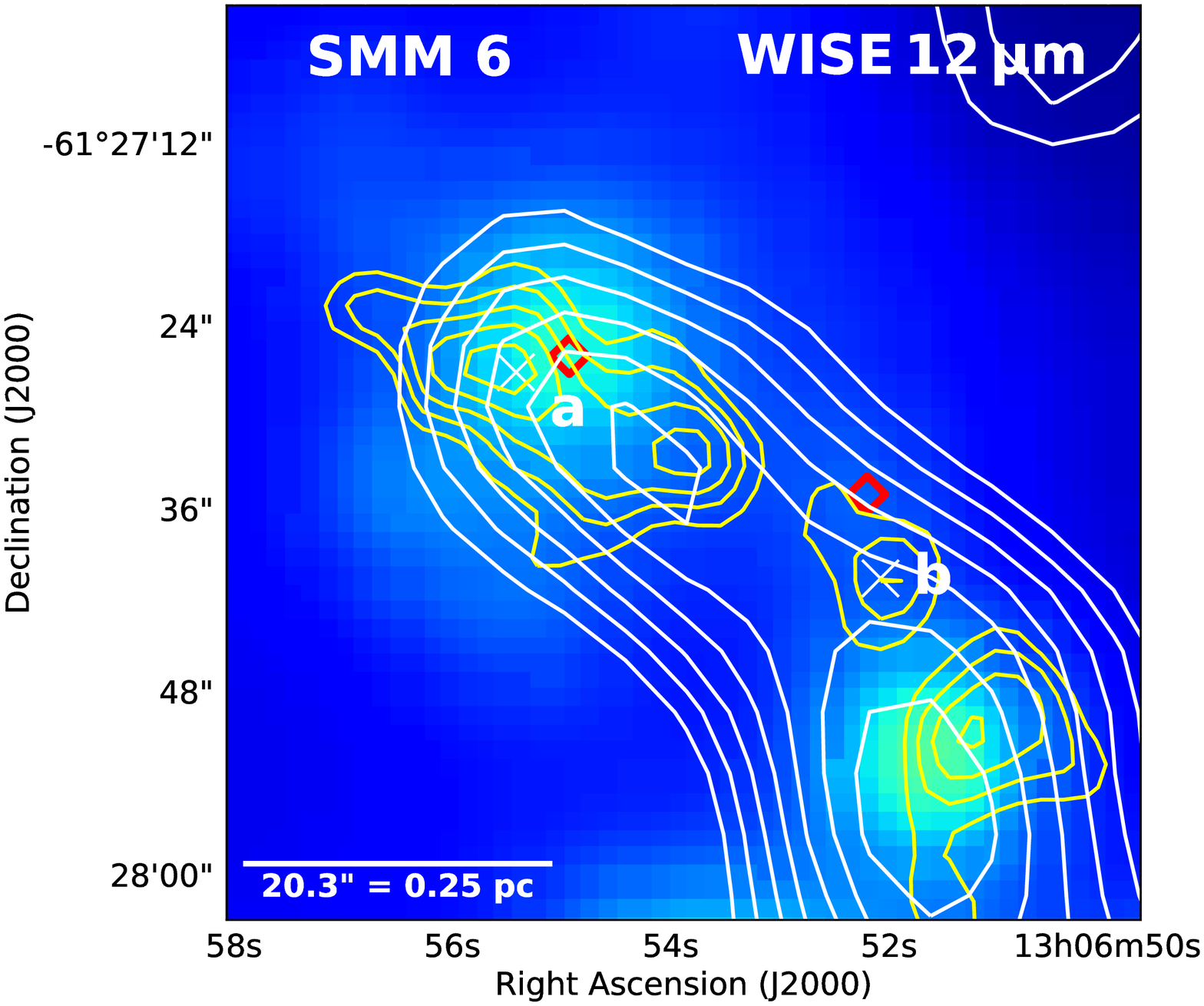}
\includegraphics[width=0.21\textwidth]{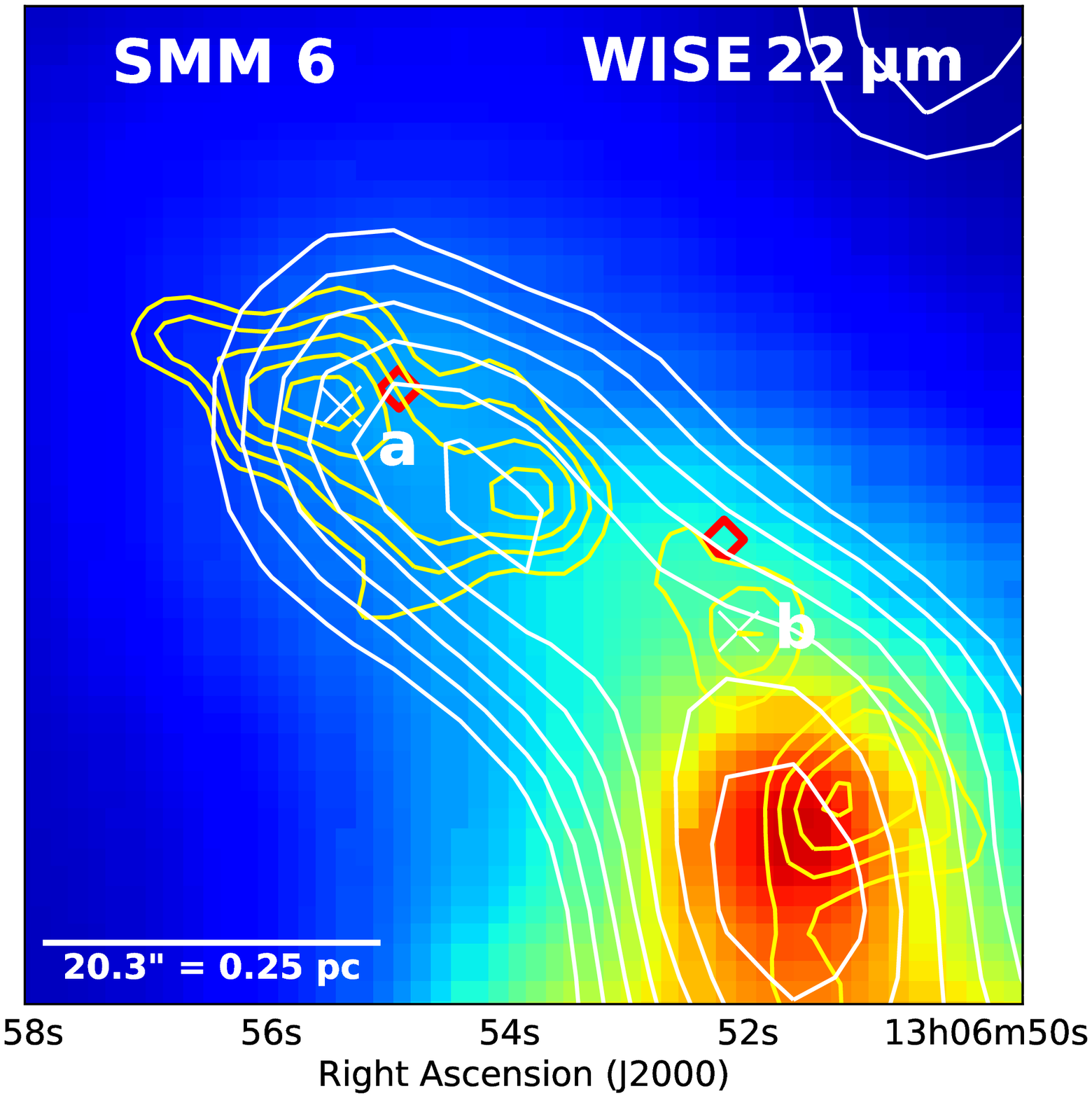}
\includegraphics[width=0.21\textwidth]{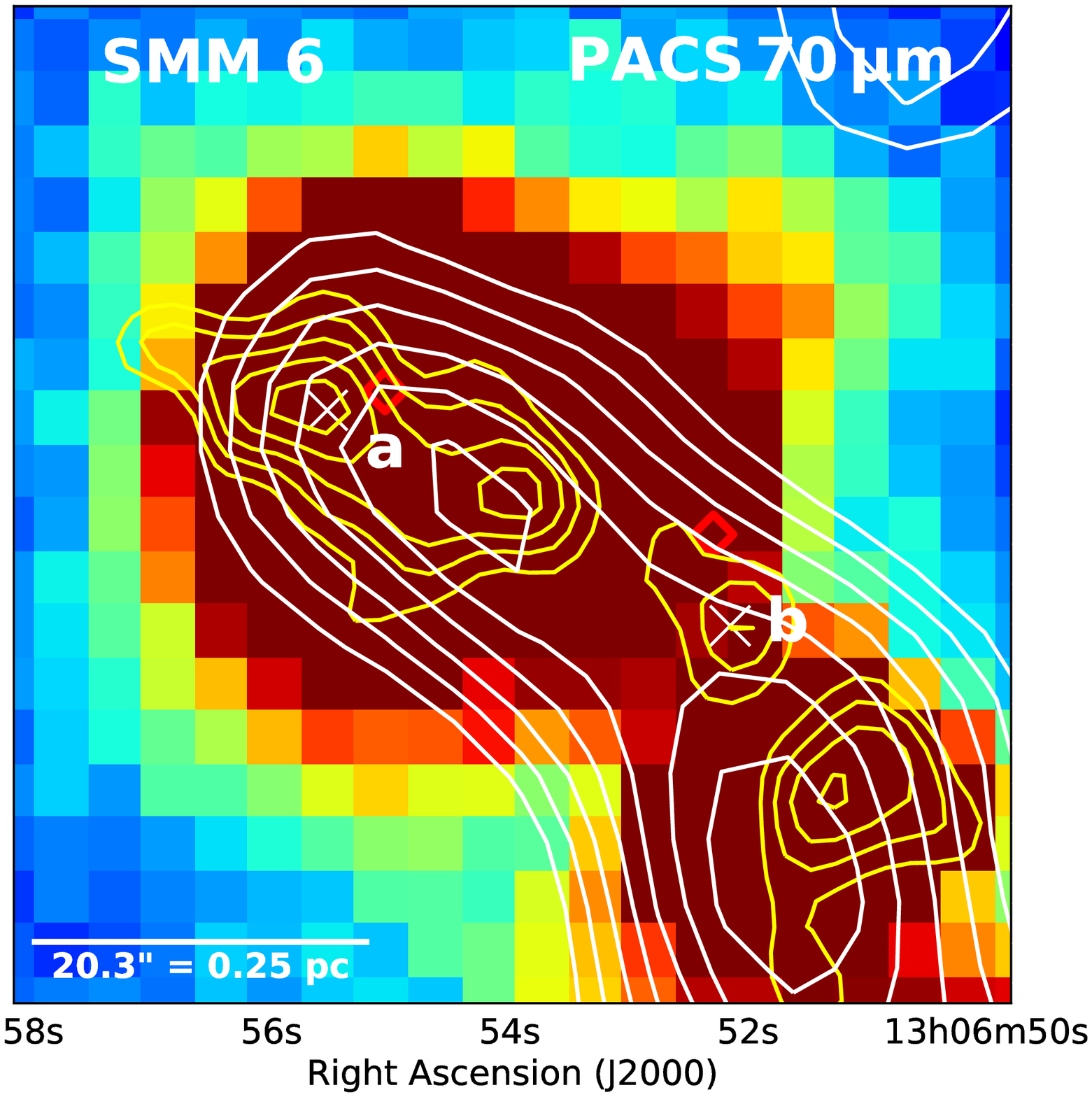}
\includegraphics[width=0.21\textwidth]{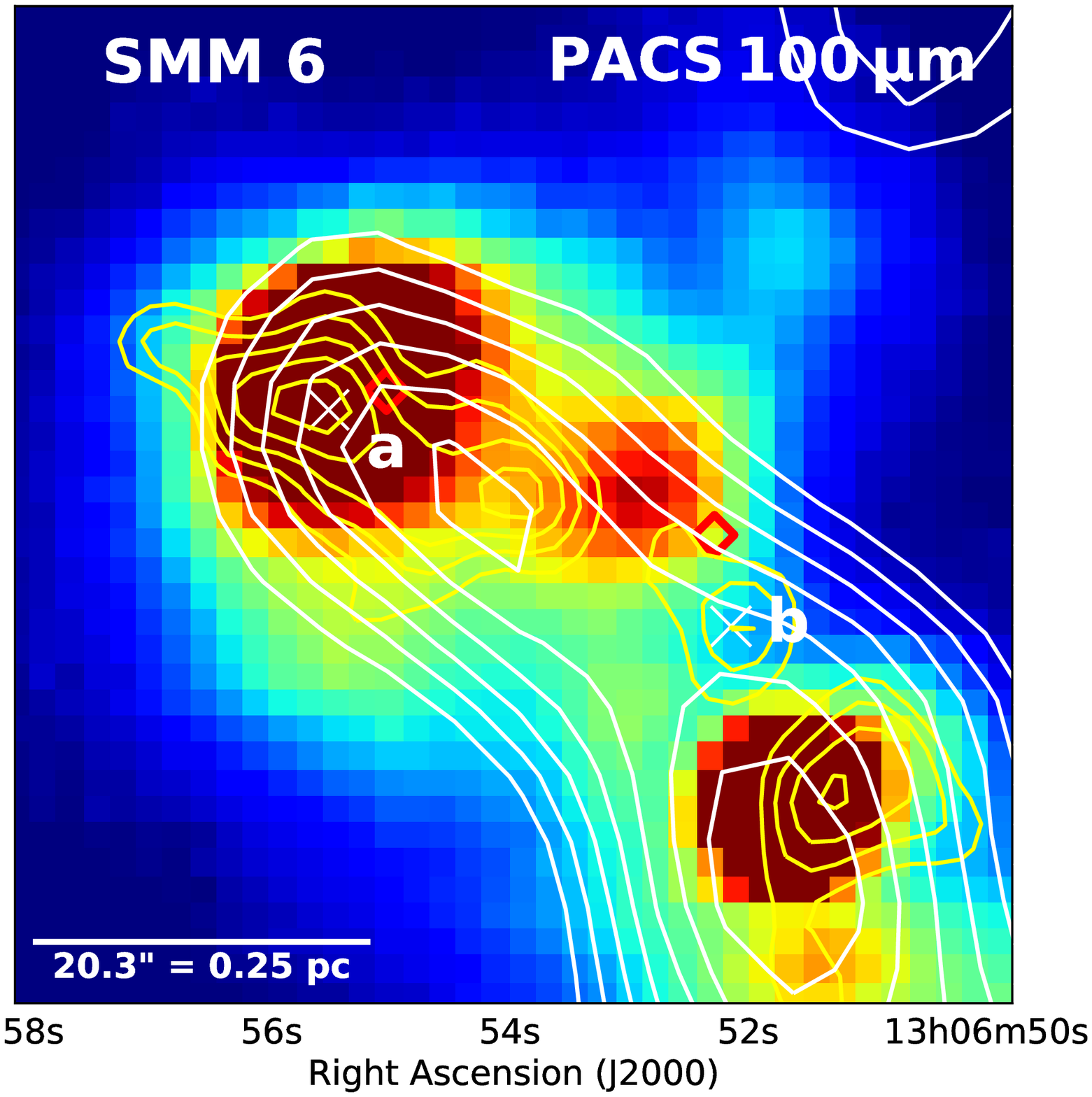}
\includegraphics[width=0.2425\textwidth]{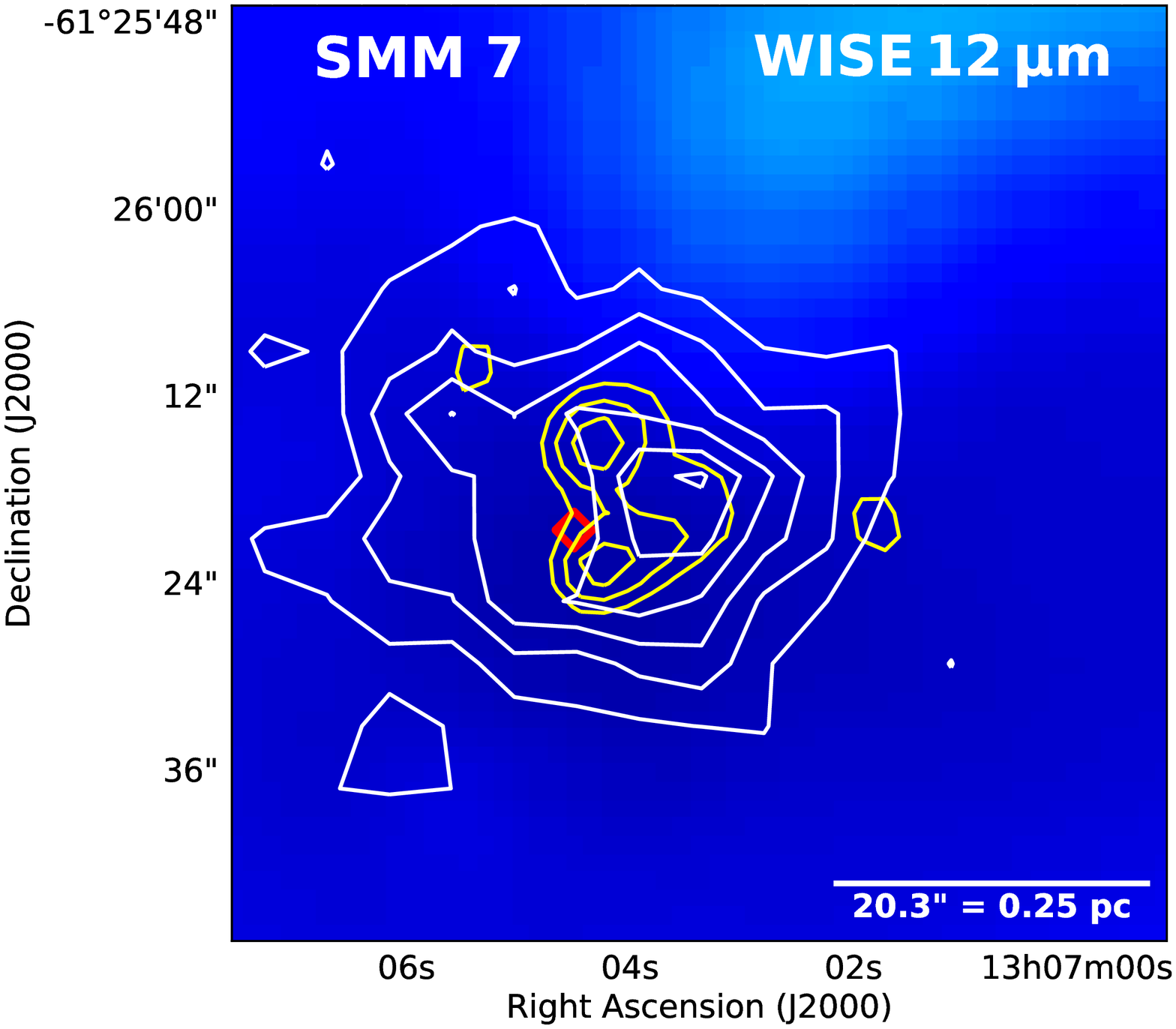}
\includegraphics[width=0.21\textwidth]{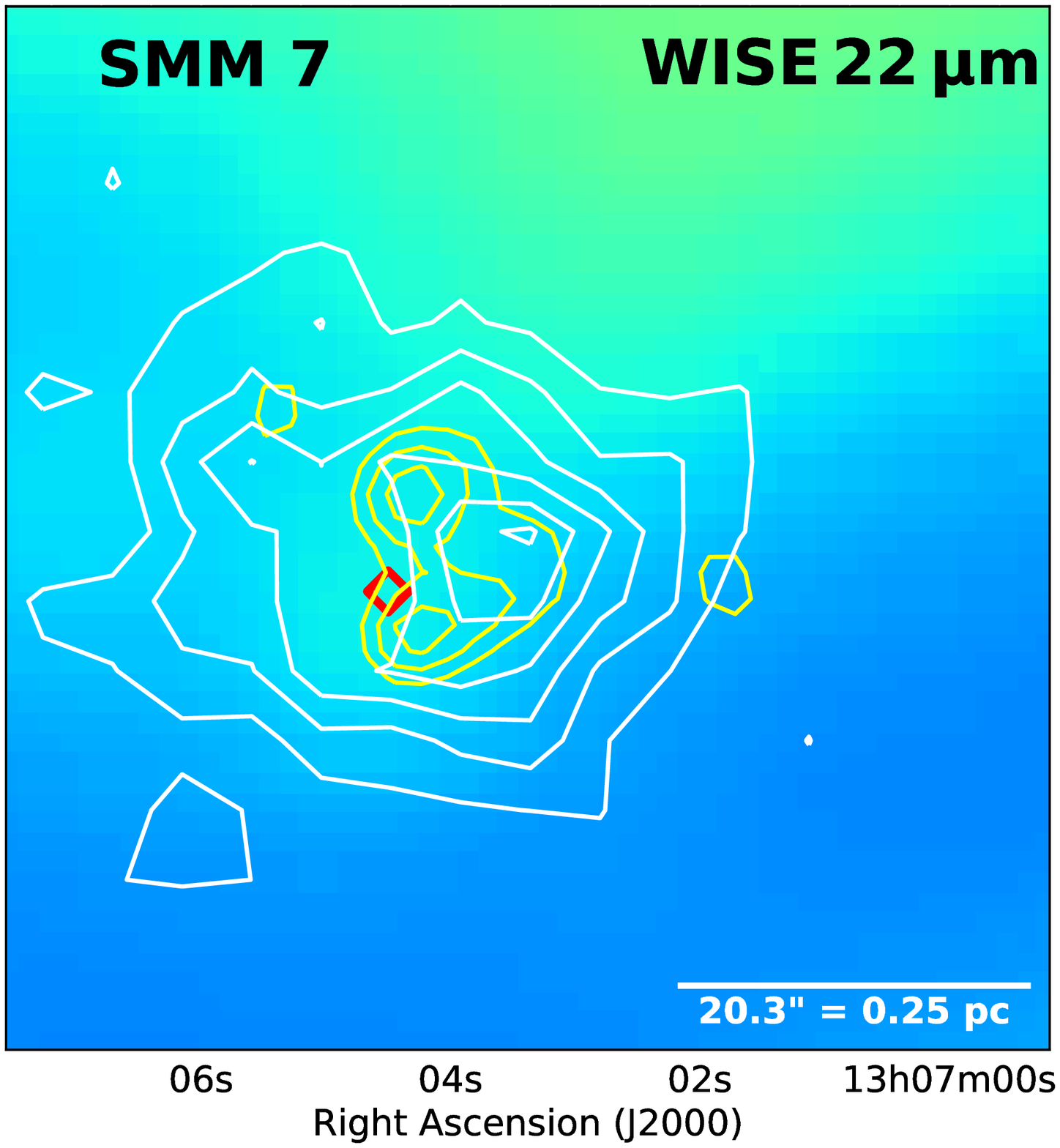}
\includegraphics[width=0.21\textwidth]{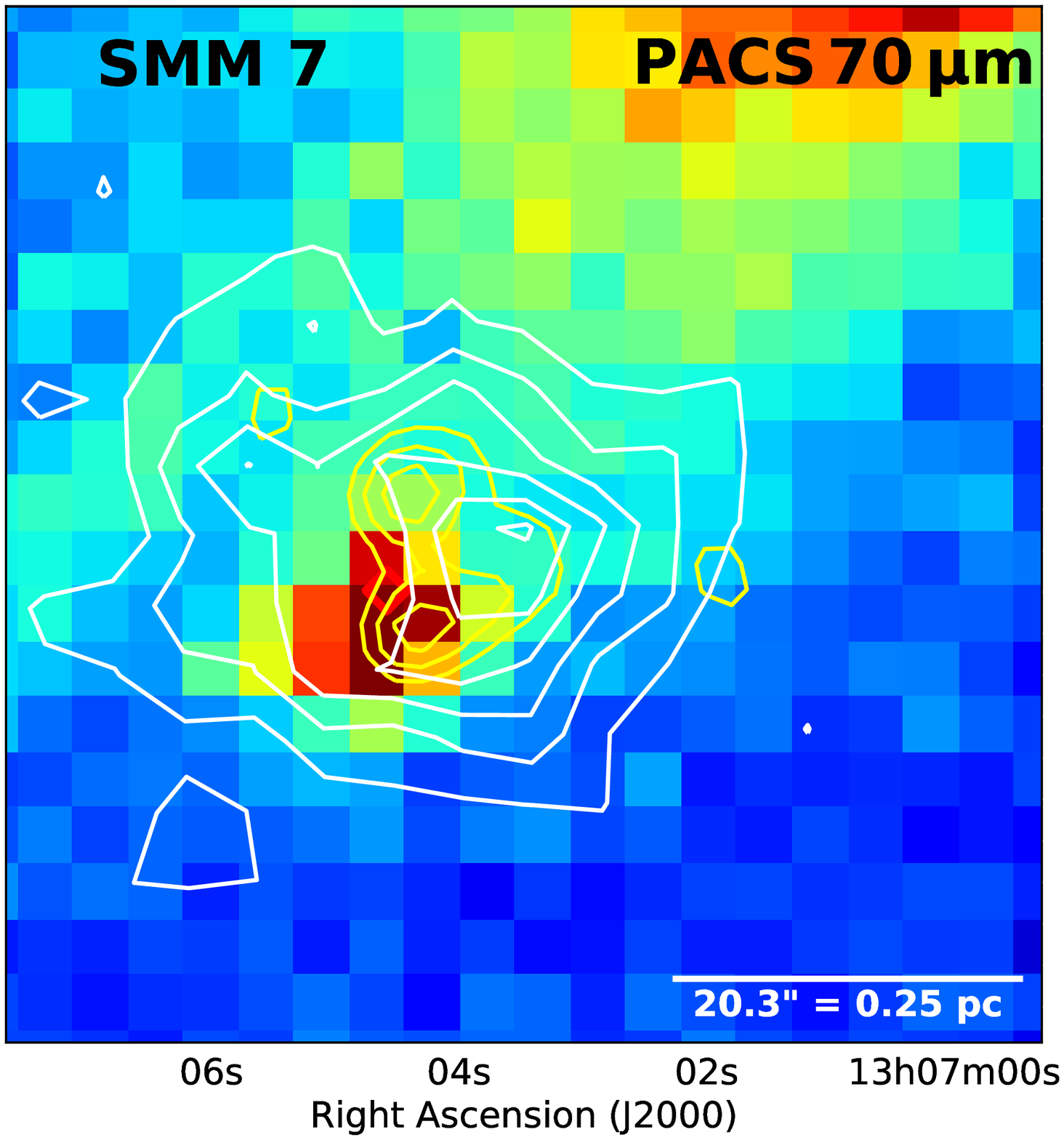}
\includegraphics[width=0.21\textwidth]{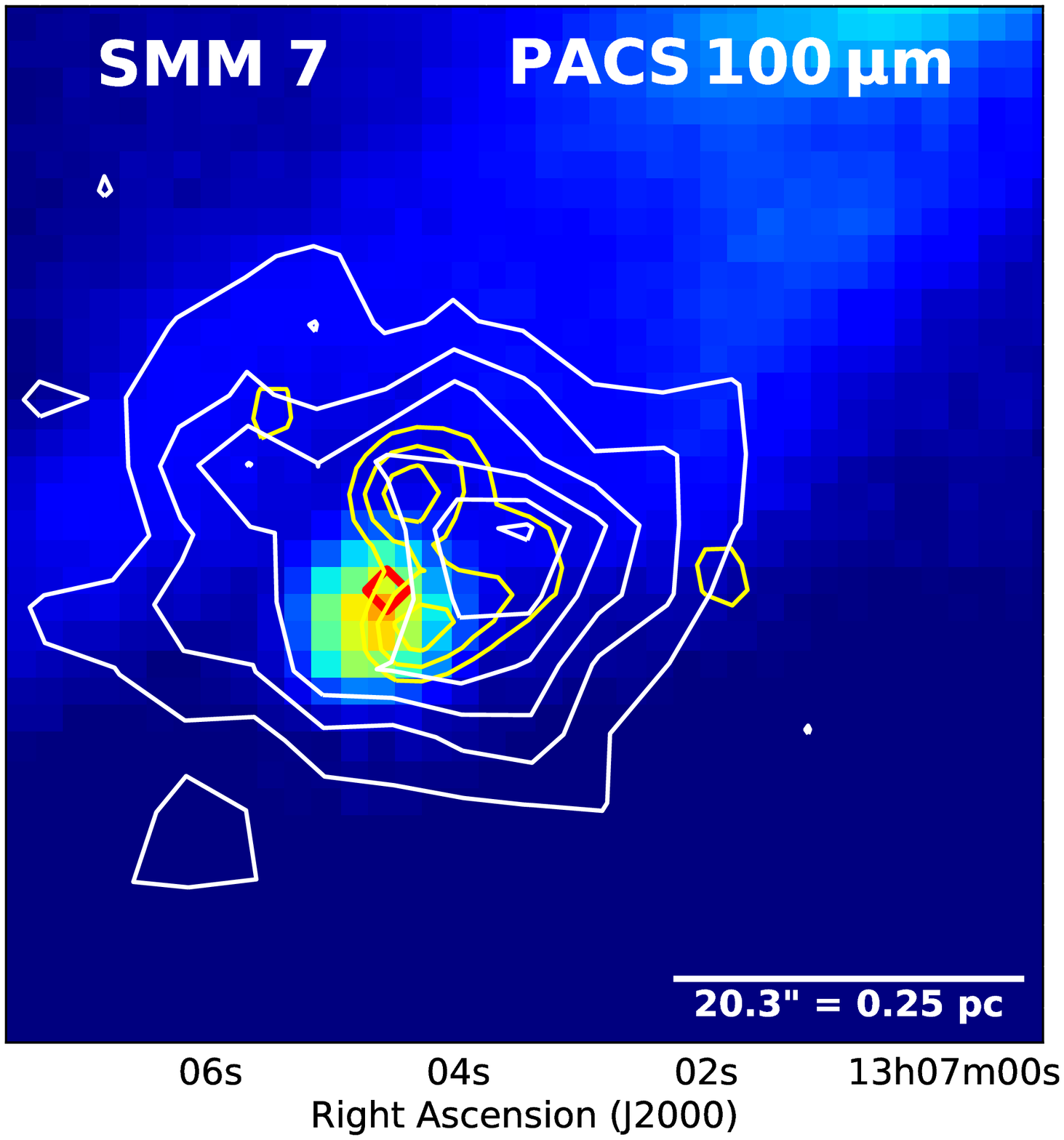}
\includegraphics[width=0.2425\textwidth]{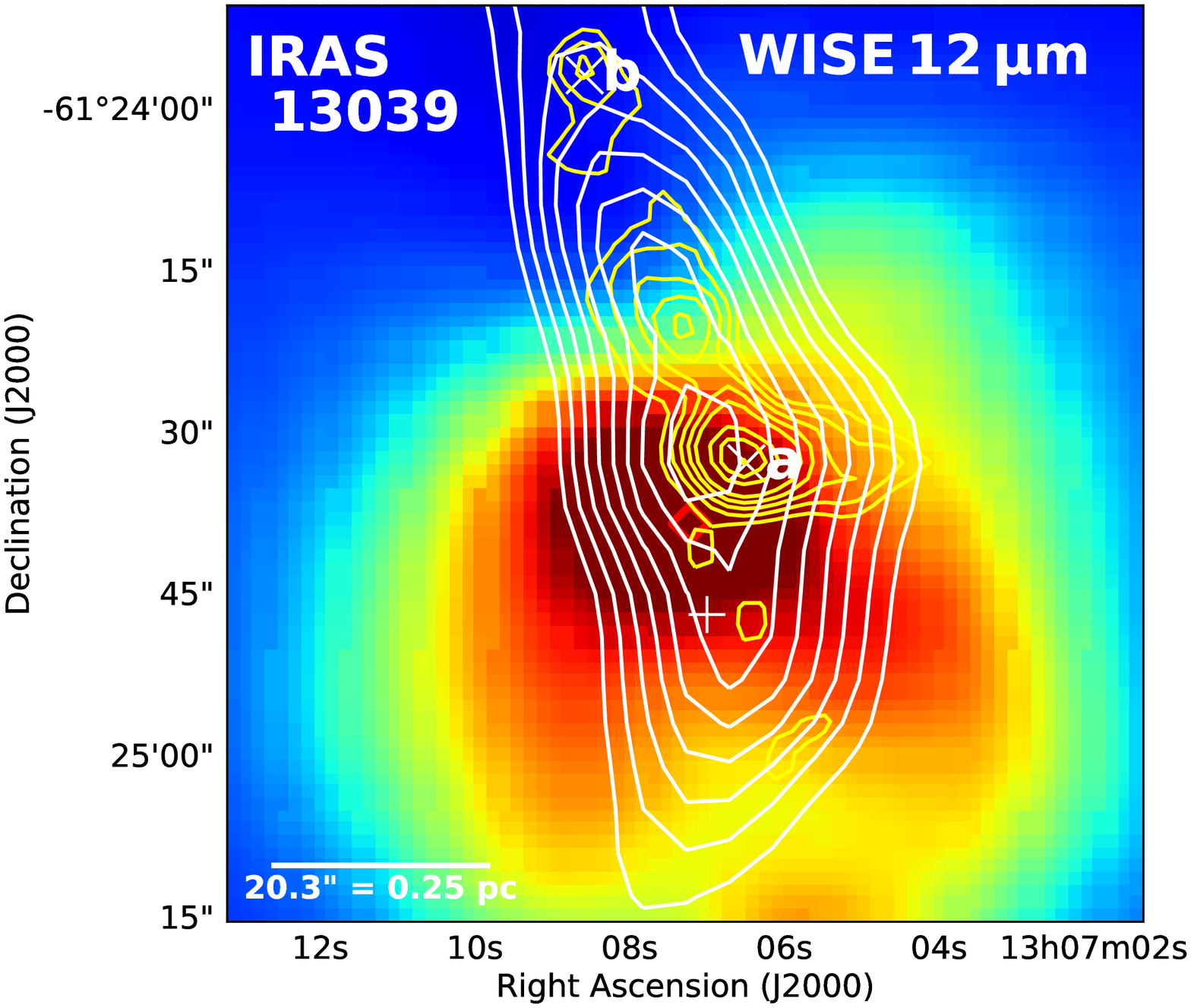}
\includegraphics[width=0.21\textwidth]{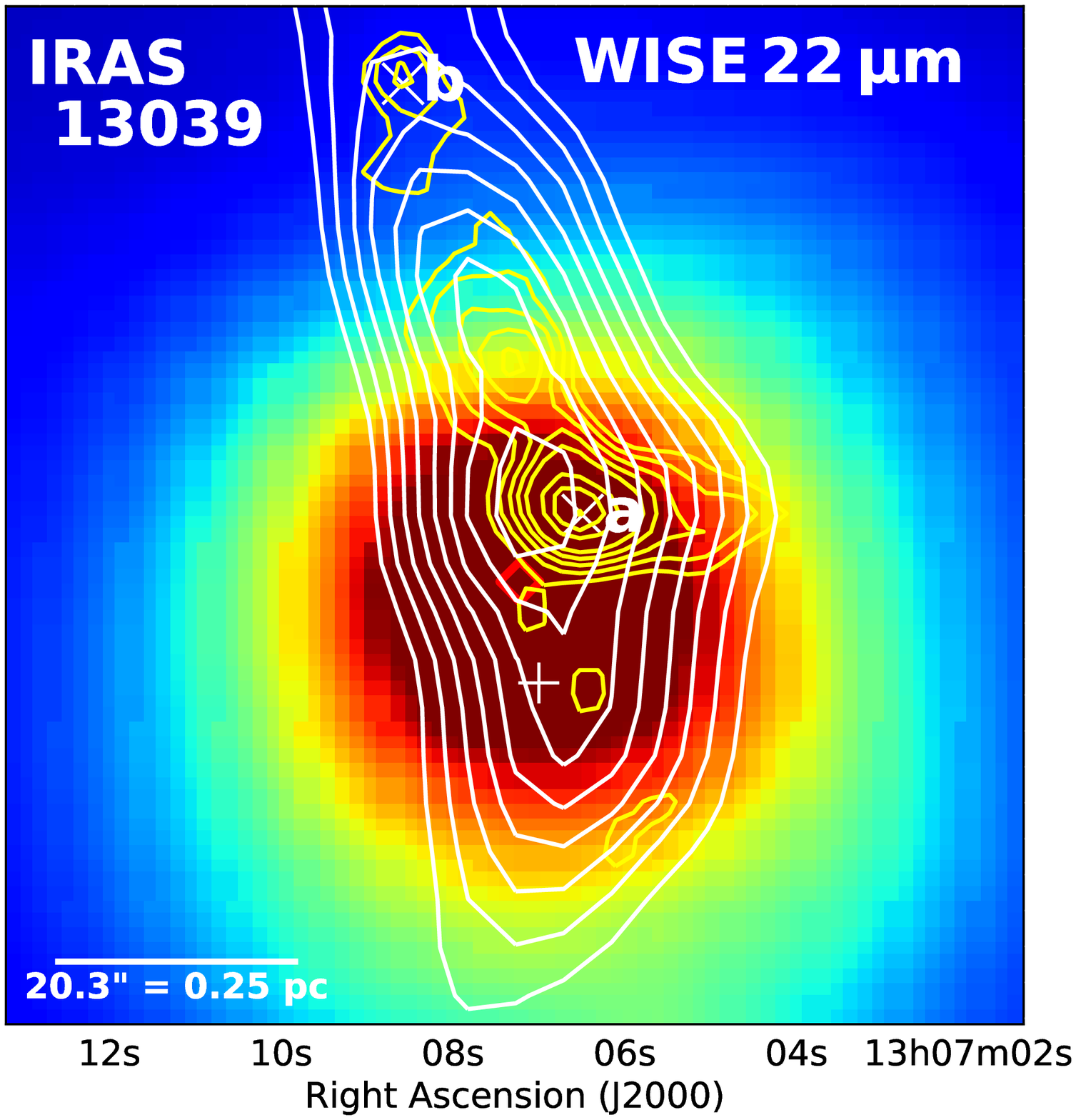}
\includegraphics[width=0.21\textwidth]{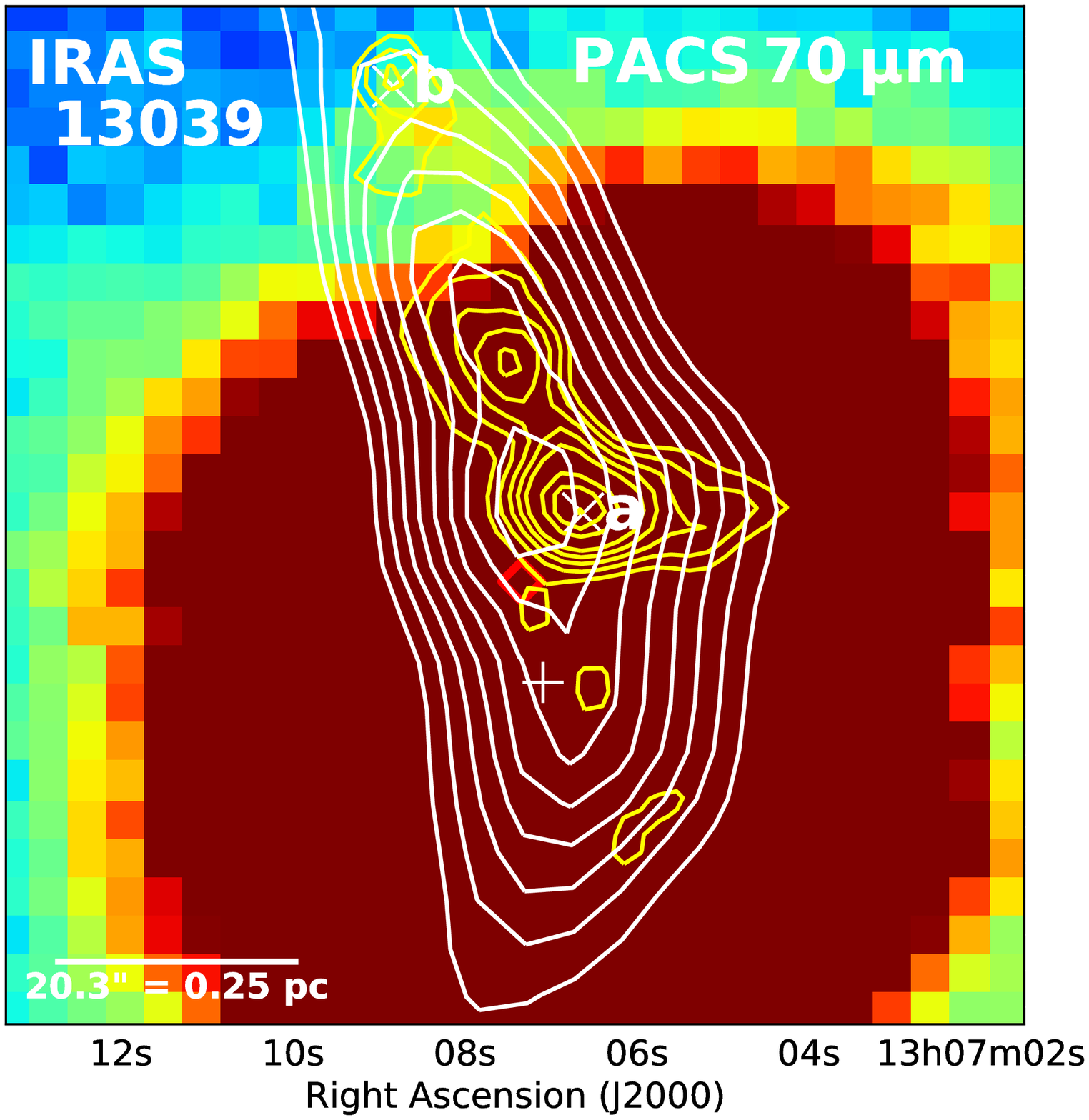}
\includegraphics[width=0.21\textwidth]{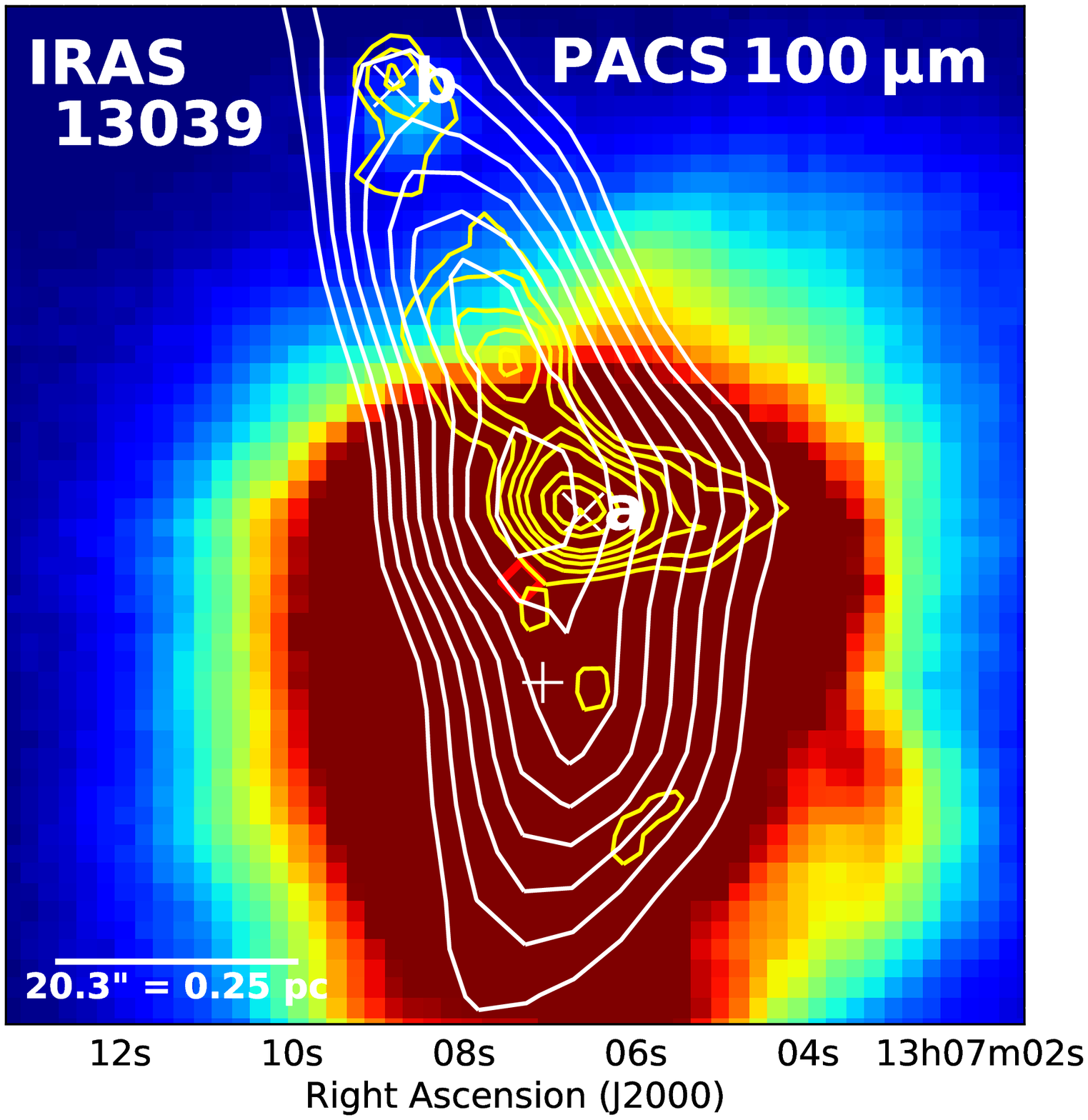}
\includegraphics[width=0.2425\textwidth]{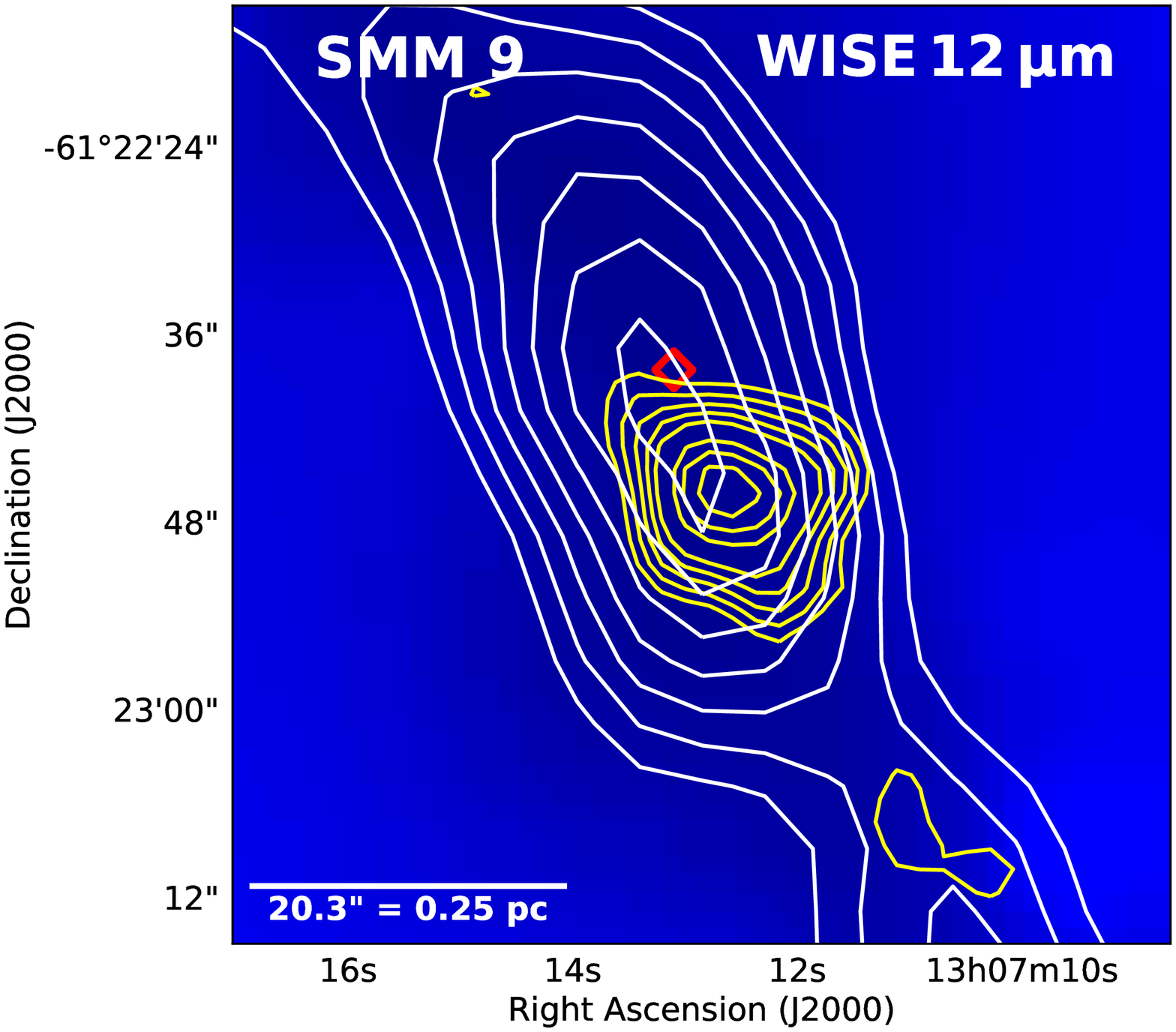}
\includegraphics[width=0.21\textwidth]{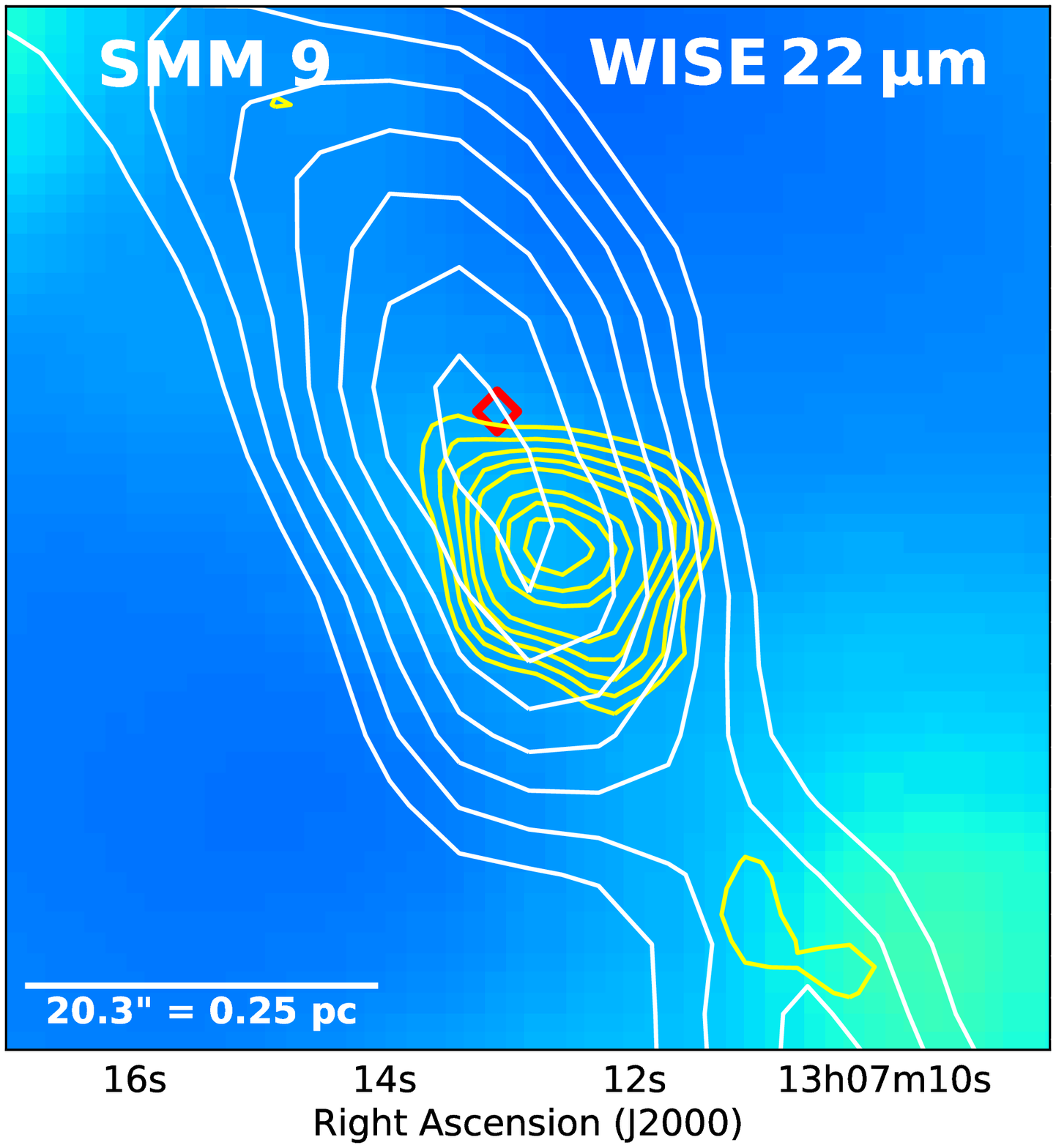}
\includegraphics[width=0.21\textwidth]{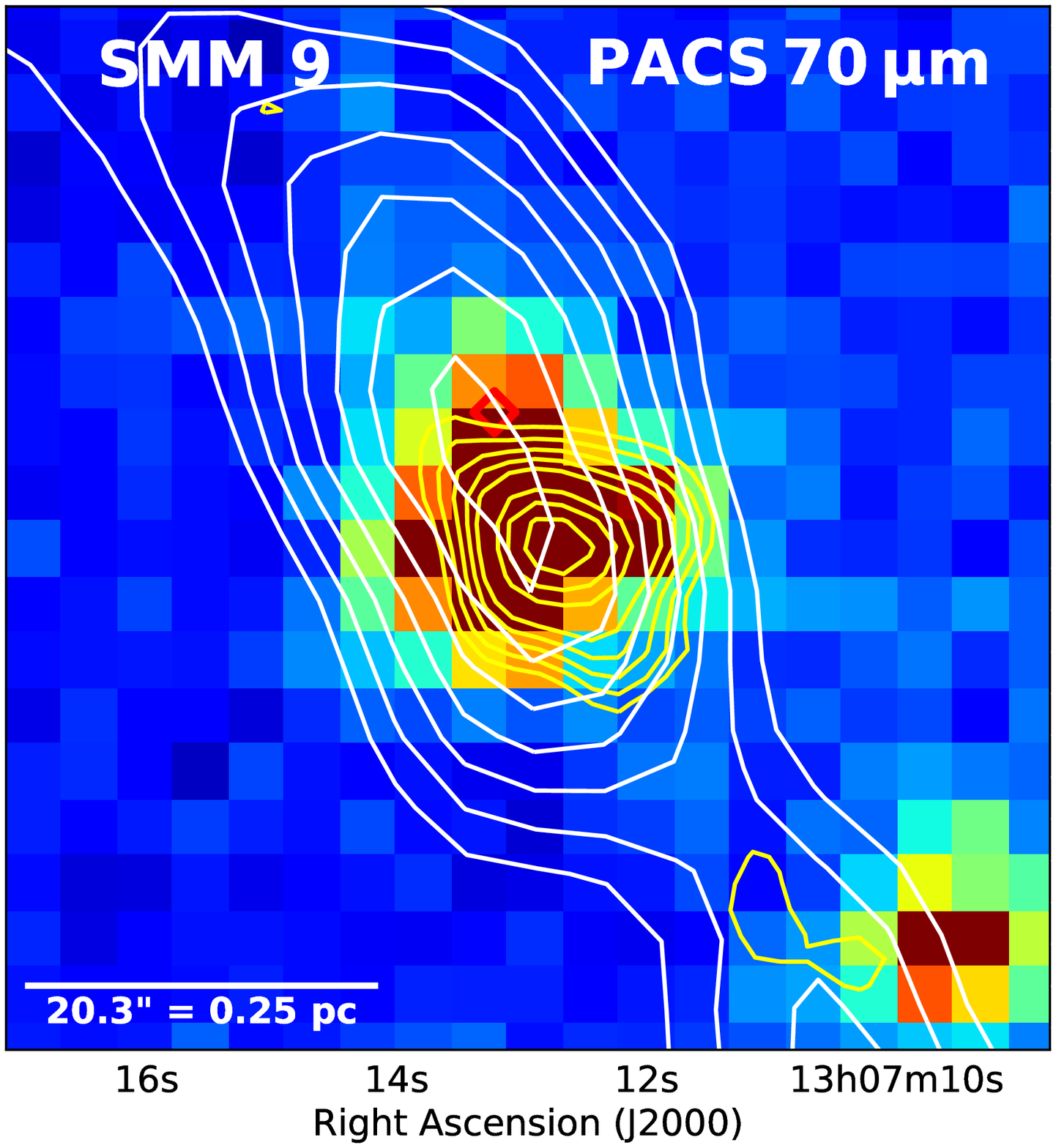}
\includegraphics[width=0.21\textwidth]{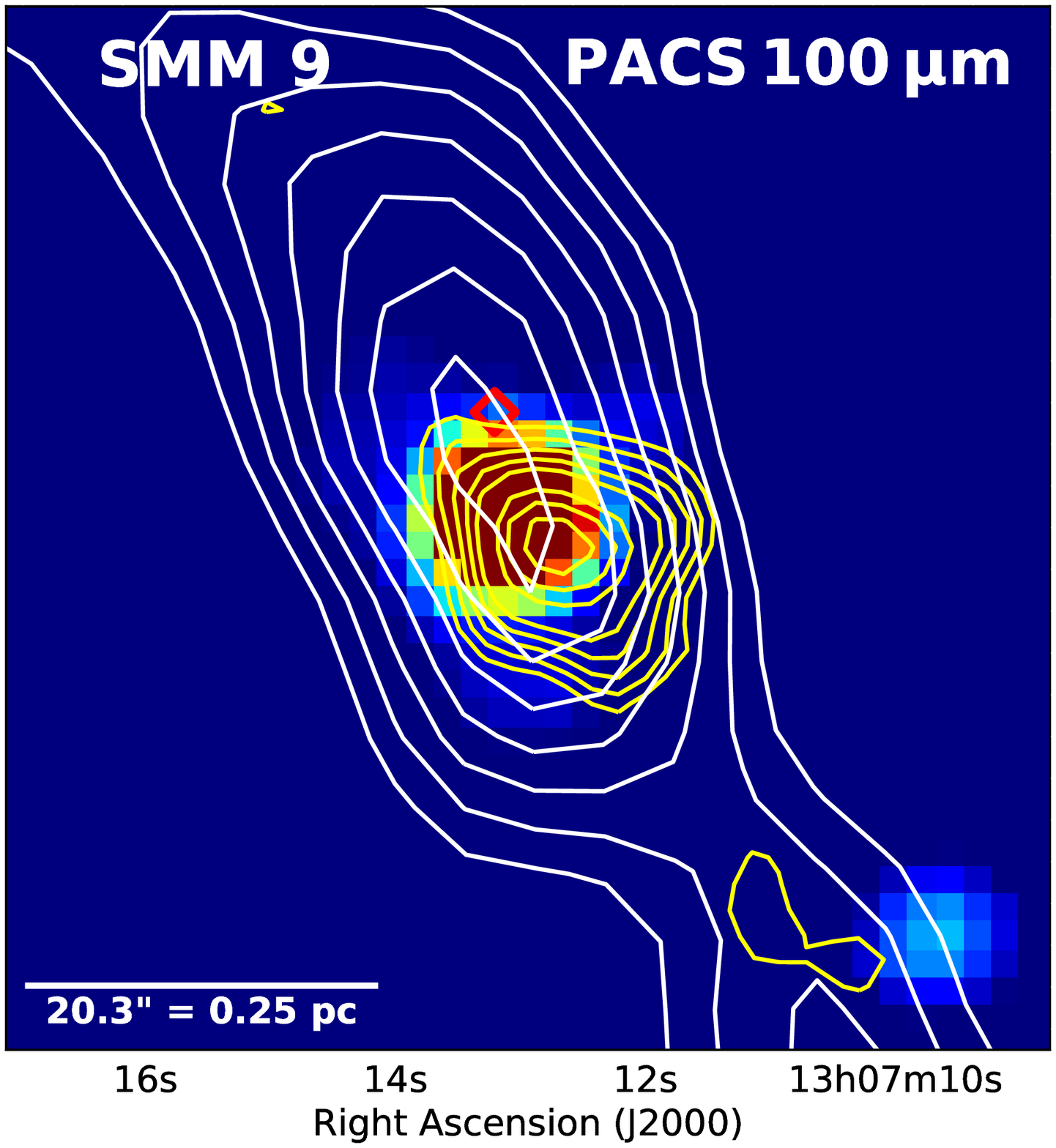}
\includegraphics[width=0.2425\textwidth]{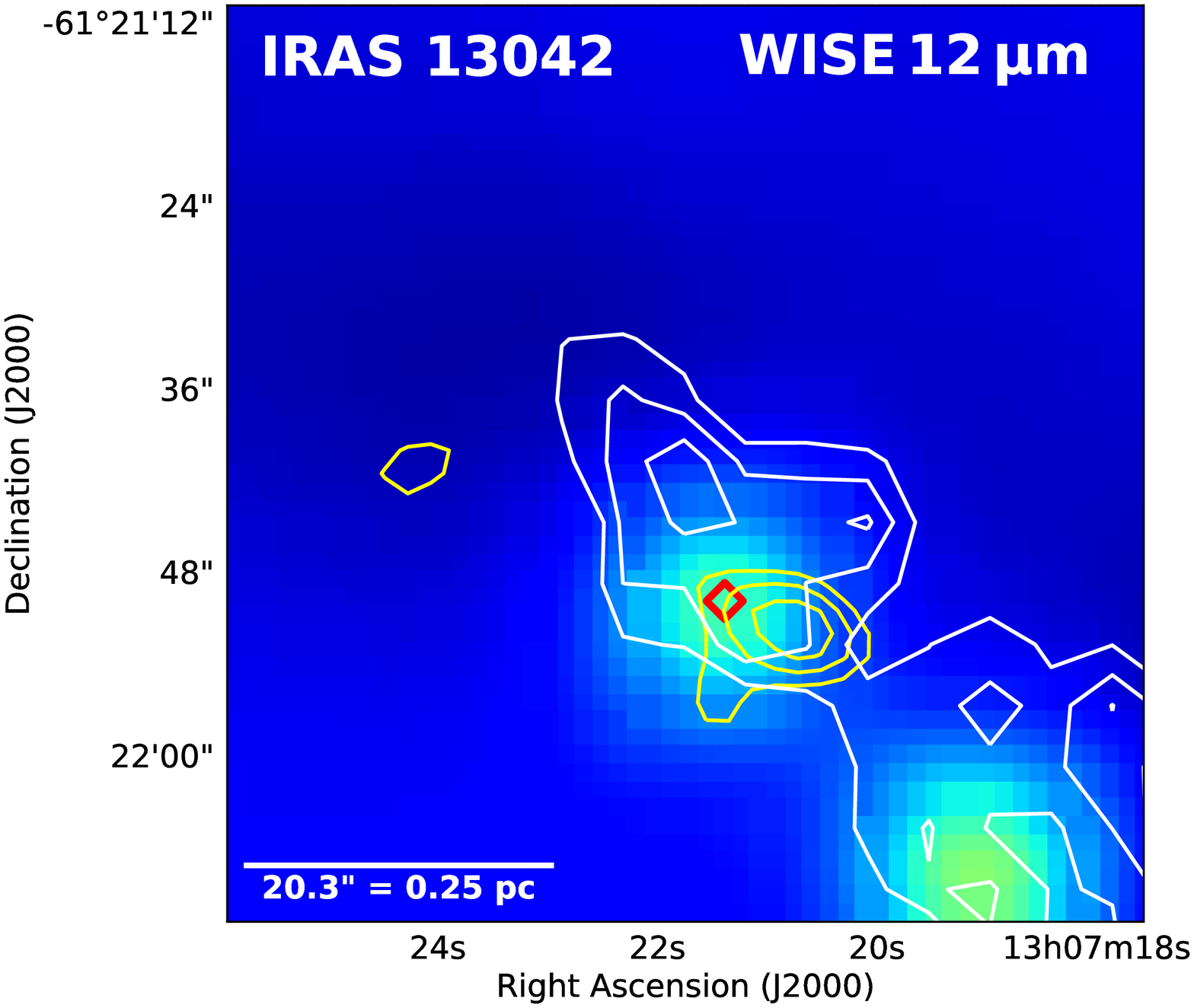}
\includegraphics[width=0.21\textwidth]{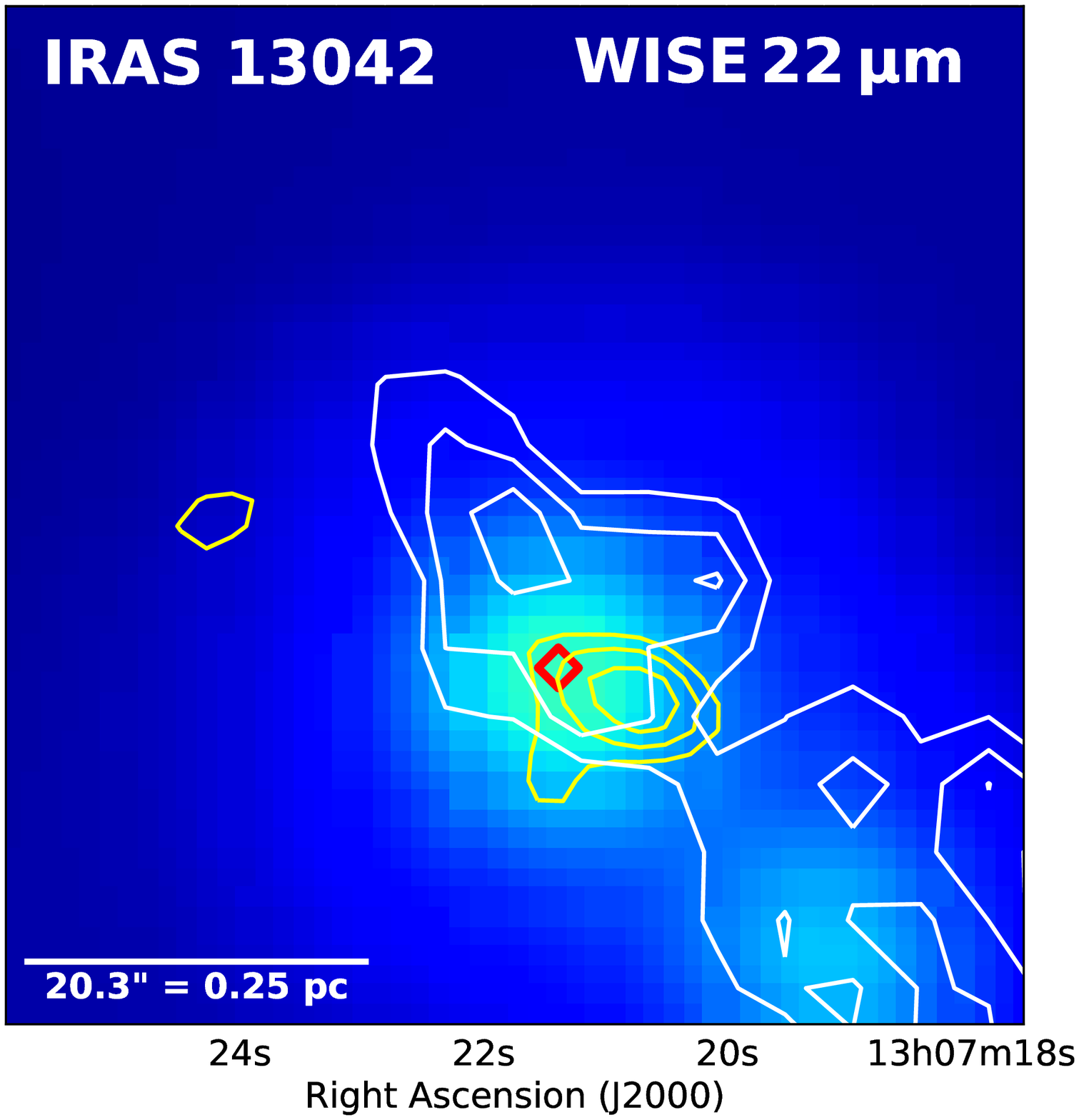}
\includegraphics[width=0.21\textwidth]{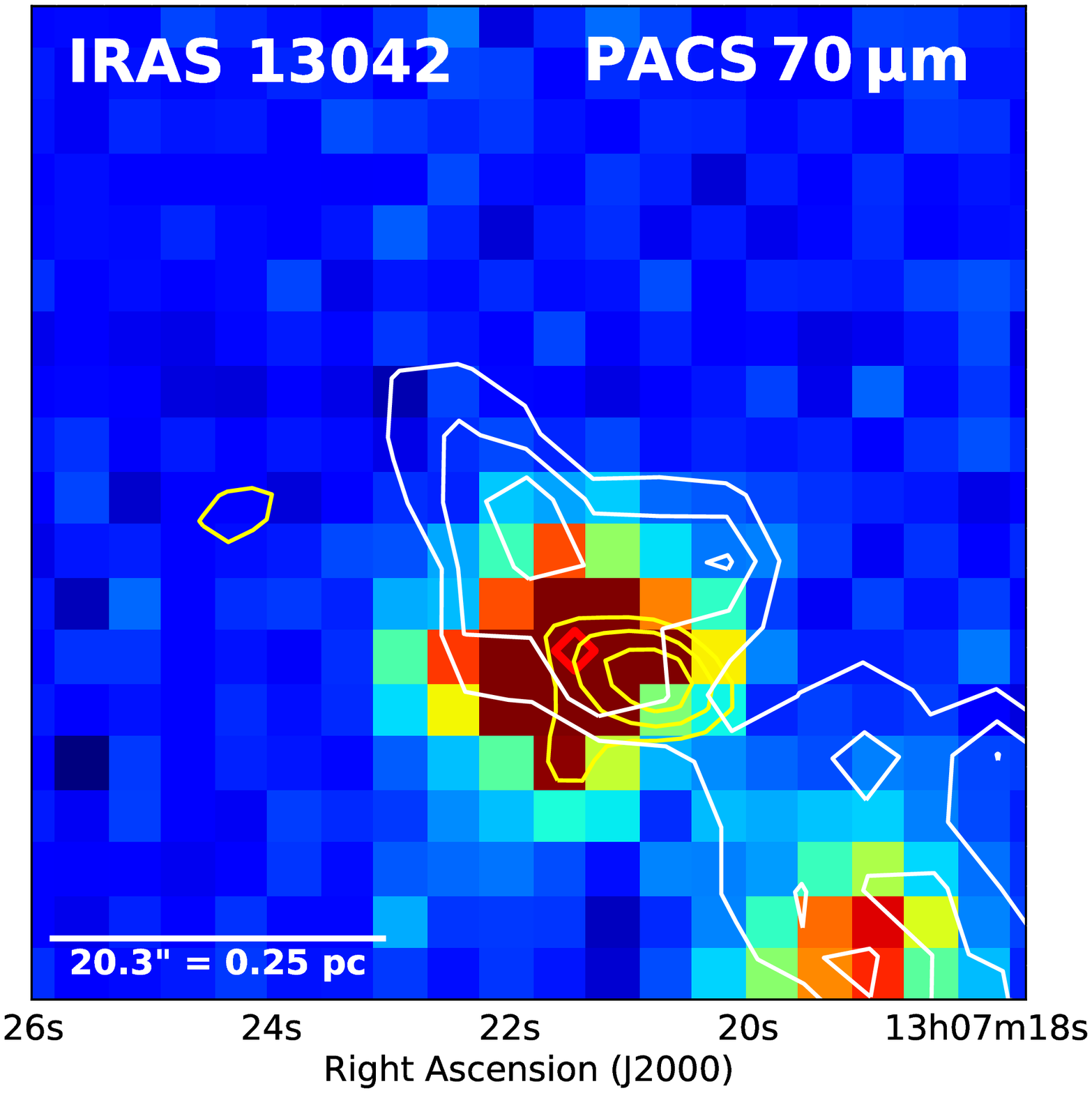}
\includegraphics[width=0.21\textwidth]{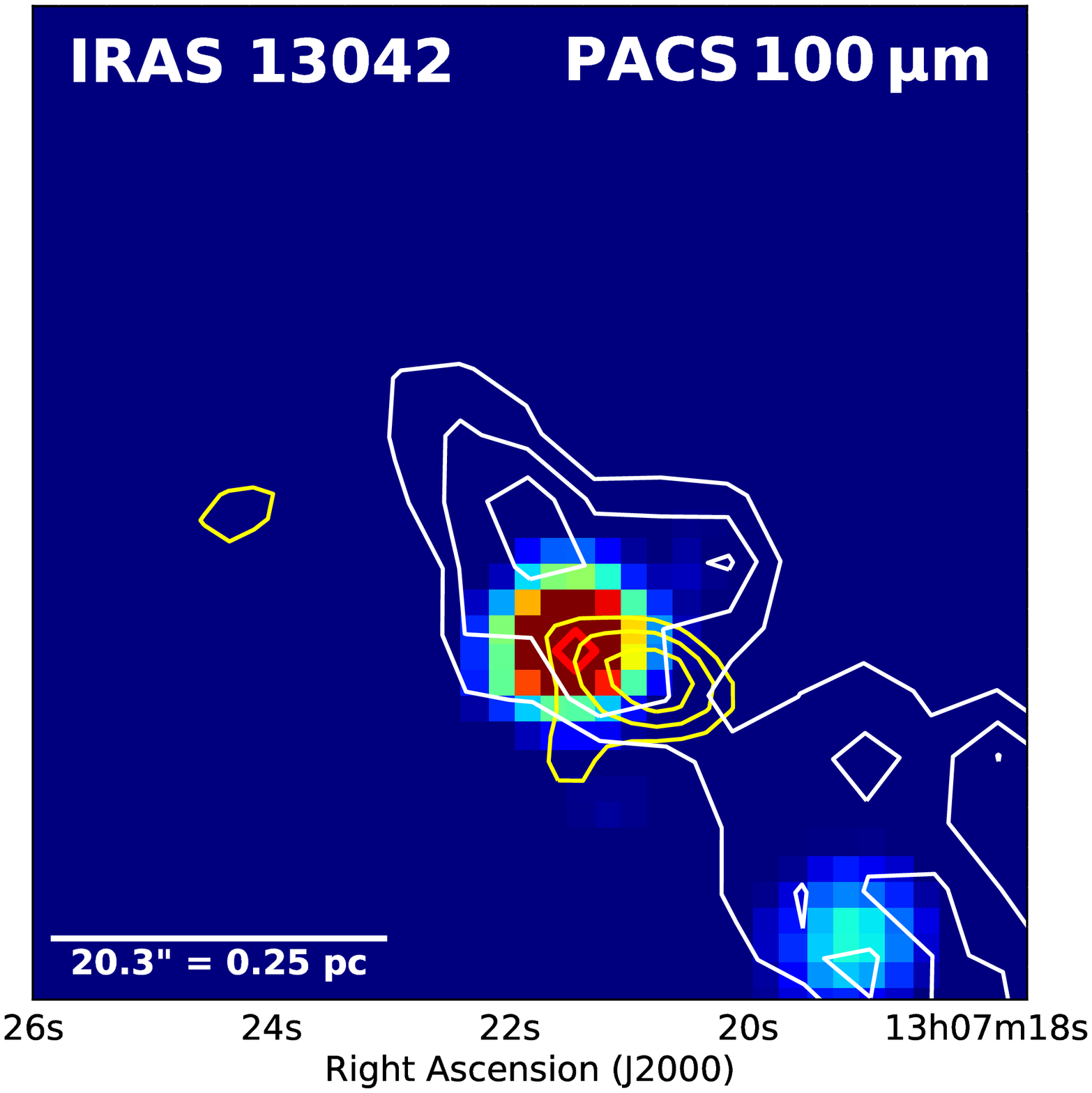}
\caption{continued.}
\label{figure:images}
\end{center}
\end{figure*}

\section{SED analysis of dense cores in the Snake IRDC}

Similarly to the present study, Henning et al. (2010) also used modifed blackbody SED fitting technique to derive the Snake IRDC’s core properties. However, the difference is that while we used two-temperature modifed blackbodies to fit the \textit{WISE} 22~$\mu$m data point, Henning et al. (2010) did not inlucde the \textit{Spitzer}/MIPS (Multiband Imaging Photometer for \textit{Spitzer}; \cite{rieke2004}) 
24~$\mu$m data in the SED fitting because the corresponding emission might be optically thick. Instead, the authors fit the \textit{Herschel}/PACS data only and also neglected the longer wavelength data, for example those measured with \textit{Herschel}/SPIRE owing to its coarser resolution, which prevented the identification of the PACS cores. 

For better comparison with the present results, we fit the SEDs of the cores in the Snake IRDC filament following the same method as outlined in Sect.~3.1. We note that the \textit{Spitzer} and \textit{Herschel}/PACS photometric data used by Henning et al. (2010) were not reported by the authors, and hence we serched for the photometric data from the MIPSGAL (MIPS Galactic Plane Survey; \cite{carey2009}) Catalogue\footnote{{\tt https://irsa.ipac.caltech.edu/cgi-bin/Gator/ \\nph-scan?submit$=$Select\&projshort$=$SPITZER}} and the PACS Point Source Catalogues. As a search radius, we used $10\arcsec$ with respect to the coordinates given in Table~1 in Henning et al. (2010). When the PACS 70~$\mu$m or 100~$\mu$m photometric data could not be found from the PACS Point Source Catalogues, we searched for the data from the Hi-GAL (the \textit{Herschel} infrared Galactic Plane Survey; \cite{molinari2016}) 70~$\mu$m and 100~$\mu$m Photometric Catalogues\footnote{{\tt https://irsa.ipac.caltech.edu/cgi-bin/Gator/ \\nph-scan?submit$=$Select\&projshort$=$HERSCHEL}}. 

For the Snake core~3, no photometric data were found at all from the aforementioned catalogues, and for the cores~6, 7, 11, 14, and 17 only two photometric data points were found, and hence their SEDs were not analysed. We note that there is a MIPS 24~$\mu$m source, namely MG011.1545-00.1688, which lies $5\farcs8$ away from core~7. However, Henning et al. (2010) reported the core to be 24~$\mu$m dark (the two photometric data points we found from the \textit{Herschel} catalogues were those at 70~$\mu$m and 100~$\mu$m). On the other hand, no MIPS 24~$\mu$m counterpart was found for core~13 (within $10\arcsec$) although it was reported to be 24~$\mu$m bright by Henning et al. (2010). Also, the PACS data of this core could not be successfully fit with our modifed blackbody model. For core~12, three photometric data points were found, but one of them was at 24~$\mu$m (the other two being 70~$\mu$m and 160~$\mu$m), and hence the wavelength coverage was not sufficient for the present SED analysis. Hence, among the cores~1-18 in the Snake IRDC filament (\cite{henning2010}, Table~1 therein), we could analyse ten cores. Of these ten cores, only one (core~1) is 24~$\mu$m dark.

\begin{figure}[!htb]
\centering
\resizebox{.24\textwidth}{!}{\includegraphics{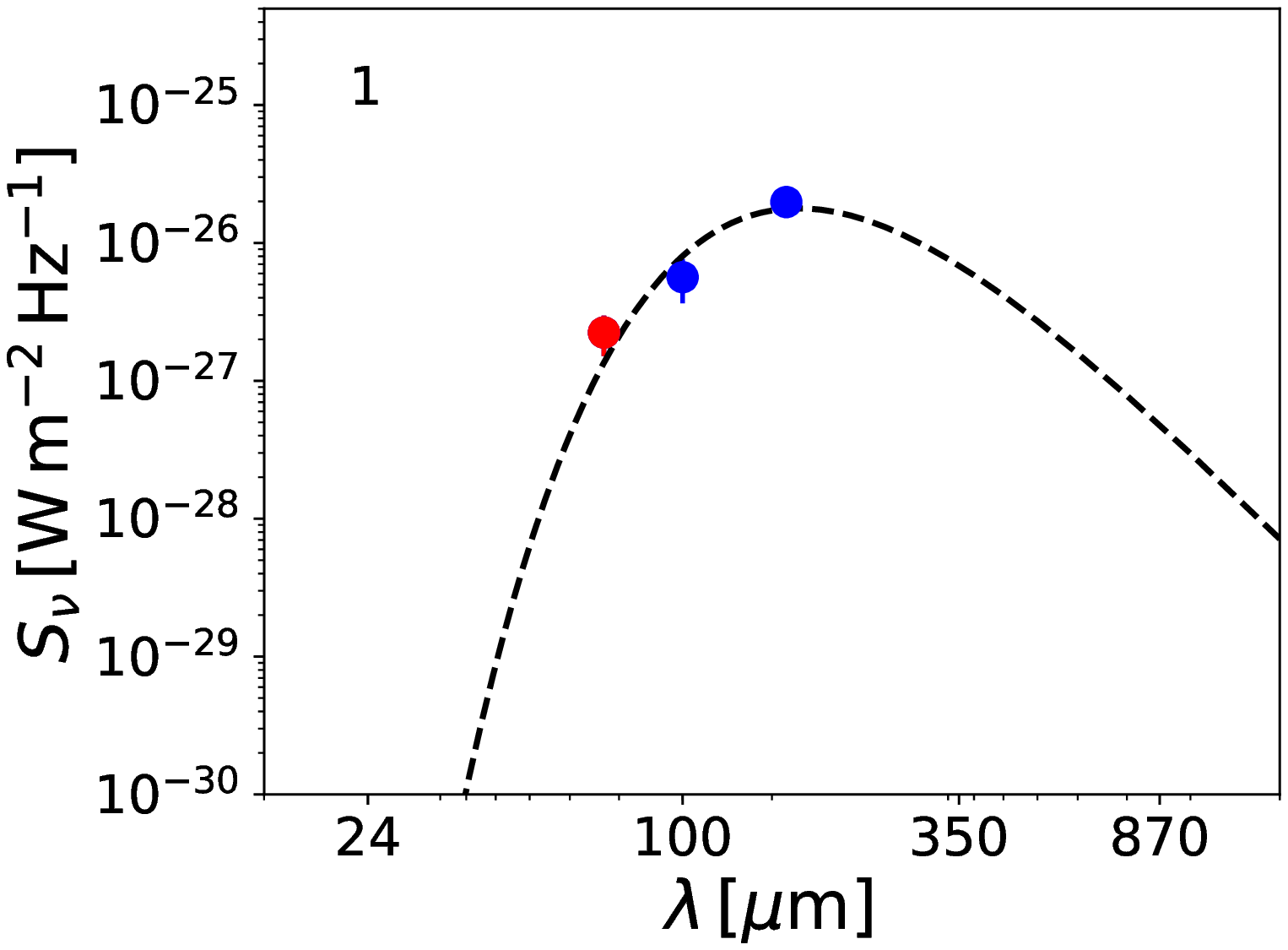}}
\resizebox{.24\textwidth}{!}{\includegraphics{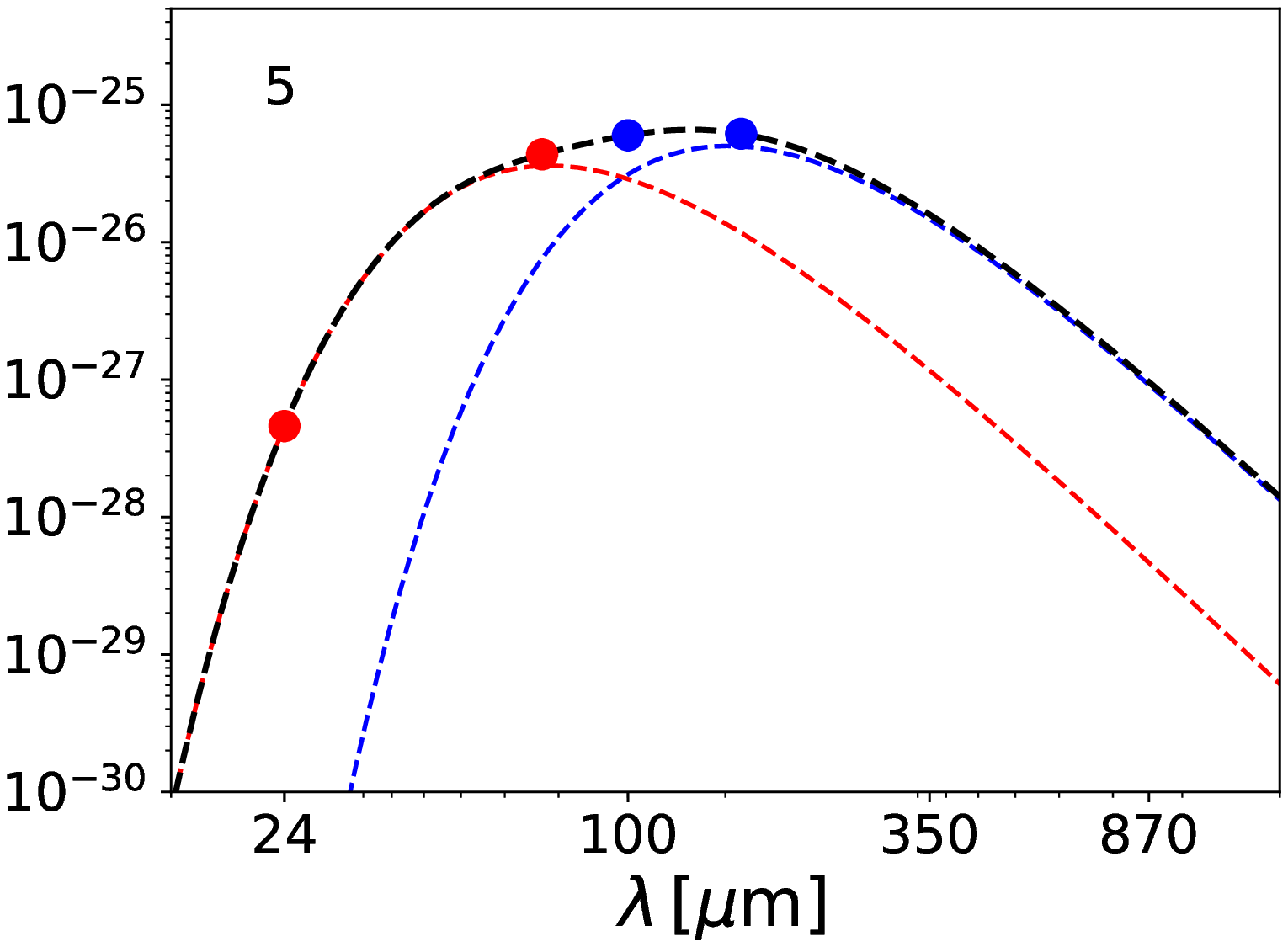}}
\caption{Two examples of SEDs of dense cores in the Snake IRDC. Core~1 (left panel) is 24~$\mu$m dark, while core~5 (right panel) is associated with a 24~$\mu$m source. Data points at $\lambda \leq 70$~$\mu$m are highlighted in red, while the longer wavelength data are shown in blue. The black, dashed lines represent the best modified blackbody fits to the data. For core~5, the latter represents the sum of the cold and warm components, where the individual fits are shown by the blue and red dashed lines, respectively.}
\label{figure:snake_seds}
\end{figure}

The derived SED parameters for the ten analysed cores in the Snake IRDC are listed in Table~\ref{table:snake}. Two example SEDs are plotted in Fig.~\ref{figure:snake_seds}. For three of the cores (1, 2, and 4), the data were fitted with a single-temperature model. Of these, core~1 is 24~$\mu$m dark and had only the three PACS data points to fit. For core~2, our two-temperature model could not fit the 24~$\mu$m data point, and hence we fit the PACS data only with a single-temperature model. This is a similar case as SMM~3 in the Seahore IRDC (see Fig.~\ref{figure:seds}). For core~4, we did not find a MIPS 24~$\mu$m counterpart but Henning et al. (2010) reported the core to be associated with a 24~$\mu$m source. The PACS data of core~4 were fitted with a single-temperature model. The SEDs of the remaining seven cores listed in Table~\ref{table:snake} were fitted with a two-temperature model. 

The dust temperatures (of the cold component), core masses, and luminosities we derived for the Snake cores are about 0.7--1.2, 0.1--19, and 1.0--3.1 times the values reported by Henning et al. (2010). On average, our analysis yielded about 0.8, 3.2, and 2.2 times the temperature, mass, and luminosity values from Henning et al. (2010). For example, the difference in the luminosity values is caused by the inclusion of the \textit{Spitzer} 24~$\mu$m data in our SED fitting. We note that Henning et al. (2010) adopted a dust opacity of 1~cm$^2$~g$^{-1}$ at 850~$\mu$m with $R_{\rm gd}=1/100$. In the Ossenkopf \& Henning (1994) dust model adopted in the present work, the 850~$\mu$m dust opacity would be about 1.44~cm$^2$~g$^{-1}$, which together with our 1.41 times lower dust-to-gas mass ratio yield a factor of 0.98 to scale the core masses from Henning et al. (2010) to be consistent with our assumptions. Because this factor is so close to unity, we used the original core masses reported by Henning et al. (2010) in the above comparison.

\begin{table}[H]
\renewcommand{\footnoterule}{}
\caption{SED parameters of dense cores in the Snake IRDC.}
{\scriptsize
\begin{minipage}{1\columnwidth}
\centering
\label{table:snake}
\begin{tabular}{c c c c c c c}
\hline\hline
Source & $n_{\rm SED}^{\rm fit}$ & $\chi_{\rm red}^2$ & $\lambda_{\rm peak}$ & $T_{\rm dust}$ & $M$ & $L$ \\
no.    & & & [$\mu$m] & [K] & [M$_{\sun}$] & [L$_{\sun}$] \\
\hline 
1 & 3 & 3.23 & 166.7 & $17.5\pm1.3$ & $8\pm4$ & $15^{+21}_{-11}$ \\ [1ex]
2 & 3 & 1.69 & 111.1 & $27.0\pm2.0$ & $1\pm0.5$ & $27^{+30}_{-17}$ \\ [1ex]
4 &	3 & 11.51 & 187.5 & $16.2\pm1.3$ & $19\pm12$ & $22^{+35}_{-17}$ \\ [1ex]
5 &	4 & \ldots\tablefootmark{a} & 130.5 & $19.5\pm2.8$ & $13\pm11$ & $118^{+172}_{-66}$ \\ [1ex]
     &   &      &       & $41.0\pm0.7$ & $0.24\pm0.05$ & \\ [1ex]
8 & 4 & \ldots\tablefootmark{a} & 166.7 & $16.9\pm3.1$ & $24\pm30$\tablefootmark{b} & $64^{+227}_{-62}$ \\ [1ex]
      &   &      &       & $51.7\pm3.2$ & $0.023\pm0.017$ & \\ [1ex]
9 & 4 & \ldots\tablefootmark{a} & 130.5 & $20.3\pm0.7$ & $285\pm45$ & $2\,739^{+792}_{-686}$ \\ [1ex]
      &   &      &       & $45.0\pm0.4$ & $3.0\pm0.4$ & \\ [1ex]
10 & 4 & \ldots\tablefootmark{a} & 157.9 & $18.0\pm1.9$ & $13\pm7$ & $50^{+79}_{-38}$ \\ [1ex]
      &   &      &       & $47.4\pm2.5$ & $0.03\pm0.02$ & \\ [1ex]
15 & 4 & \ldots\tablefootmark{a} & 142.9 & $19.8\pm4.6$ & $14\pm20$\tablefootmark{b} & $115^{+515}_{-114}$ \\ [1ex]
      &   &      &       & $56.0\pm4.4$ & $0.033\pm0.027$ & \\ [1ex]
16 & 4 & \ldots\tablefootmark{a} & 125.0 & $23.0\pm14.1$ & $4\pm12$\tablefootmark{b} & $78^{+5\,119}_{-80}$\tablefootmark{b} \\ [1ex]
      &   &      &       & $65.0\pm33.3$ & $0.009\pm0.045$\tablefootmark{b} & \\ [1ex]
18 & 4 & \ldots\tablefootmark{a} & 166.7 & $17.3\pm1.1$ & $102\pm43$ & $265^{+239}_{-150}$  \\ [1ex]
     &   &      &       & $50.0\pm1.6$ & $0.09\pm0.03$ & \\ [1ex]
\hline
\end{tabular} 
\tablefoot{The parameters given in the table are the number of flux density data points used in the SED fit, reduced $\chi^2$ value defined in Eq.~(\ref{eqn:chi}), wavelength of the peak position of the fitted SED, dust temperature, total (gas+dust) mass, and luminosity (Eq.~(\ref{eqn:luminosity})). For the sources whose SED was fitted with a two-temperature model, the dust temperature and mass of the warm component are reported beneath the cold component values.\tablefoottext{a}{The number of flux density data points is equal to the number of free parameters of the model, and hence there are zero degrees of freedom. Therefore, the value of $\chi_{\rm red}^2$ becomes infinite.}\tablefoottext{b}{The uncertainty in the derived parameter is larger than the nominal value.}}
\end{minipage} }
\end{table}

\end{document}